\documentclass[%
 reprint,
superscriptaddress,
 amsmath,amssymb,
prm,
]{revtex4-2}

\usepackage{graphicx}
\usepackage{dcolumn}
\usepackage{bm}
\usepackage{xcolor}

\usepackage{setspace} 
\begin{document}

\preprint{PRM}

\title{Magnetic interlayer coupling between ferromagnetic SrRuO$_3$ layers \\ through a SrIrO$_3$ spacer}

\author{Lena Wysocki}
\thanks{These two authors contributed equally.}
 \affiliation{%
Institute of Physics II, University of Cologne, Germany
} 
\author{Sven Erik Ilse}%
\thanks{These two authors contributed equally.}
\affiliation{%
 Max Planck Institute for Intelligent Systems, Stuttgart, Germany
}%
\author{Lin Yang}%
\affiliation{%
 Institute of Physics II, University of Cologne, Germany
}%
\author{Eberhard Goering}
 \affiliation{%
Max Planck Institute for Intelligent Systems, Stuttgart, Germany
}%
\author{Felix Gunkel}%
\affiliation{%
Peter Grünberg Institut (PGI-7), Forschungszentrum Jülich GmbH, Jülich, Germany
}%
\author{Regina Dittmann}%
\affiliation{%
Peter Grünberg Institut (PGI-7), Forschungszentrum Jülich GmbH, Jülich, Germany
}%
\author{Paul H.M. van Loosdrecht}
\affiliation{%
  Institute of Physics II, University of Cologne, Germany
}%
\author{Ionela Lindfors-Vrejoiu}
\affiliation{%
  Institute of Physics II, University of Cologne, Germany
}%
\date{\today}
\begin{abstract}
A key element to tailor the properties of magnetic multilayers is the coupling between the individual magnetic layers. In case of skyrmion hosting multilayers, coupling of skyrmions across the magnetic layers is highly desirable.
Here the magnetic interlayer coupling was studied in epitaxial all-oxide heterostructures of ferromagnetic perovskite SrRuO$_3$ layers separated by spacers of the strong spin-orbit coupling oxide SrIrO$_3$. This combination of oxide layers is being discussed as a potential candidate system to host N\'{e}el skyrmions. First order reversal curve (FORC) measurements were performed in order to distinguish between magnetic switching processes of the individual layers and to disentangle the signal of soft magnetic impurities from the samples´ signal. Additionally, FORC investigations enabled to determine whether the coupling between the magnetic layers is ferromagnetic or antiferromagnetic. The observed interlayer coupling strength was weak for all the heterostructures, with SrIrO$_3$ spacers between 2 monolayers and 12 monolayers thick.
\end{abstract}
\maketitle
\section{\label{sec:level1}Introduction}
The ferromagnetic perovskite oxide SrRuO$_3$ attracted attention recently due to the proposal of the formation of N\'{e}el-type skyrmions when it is interfaced with the large spin-orbit coupling 5\emph{d} oxide SrIrO$_3$ \cite{Matsuno2016,Ohuchi2018,Meng2019}. In a multilayer, if skyrmions can form, their ferromagnetic coupling across the stack has to be achieved, as it was realized in metallic superlattices \cite{Nandy2016, Moreau-Luchaire2016,Pollard2017}. Although several studies focused on the discussion of the origin of unconventional features in the magneto-transport of SrRuO$_3$/SrIrO$_3$ heterostructures and the existence of topologically non-trivial textures in SrRuO$_3$ thin films is still under debate \cite{Matsuno2016,Ohuchi2018,Meng2019,Groenendijk,Wysocki2020,Groenendijk,Kan2018,Kan2018b, Kan2020,Qin2019,Wu2019,Wysocki2020,Yang2020,Wysocki2020b}, the interlayer coupling in SrRuO$_3$-based multilayers was only little investigated.
Experimental studies of the magnetic interlayer coupling between SrRuO$_3$ layers were performed only in multilayers with spacers that are not expected to induce interfacial Dzyaloshinskii-Moriya interaction (DMI), such as LaNiO$_3$ or SrTiO$_3$ \cite{Yang2021, Herranz2003}. Yang \textit{et al.} achieved strong ferromagnetic coupling of the SrRuO$_3$ layers by introducing a 4 monolayers (MLs) thick metallic LaNiO$_3$ spacer, while weak ferromagnetic coupling was observed for the separation of the SrRuO$_3$ layers by 2 MLs of LaNiO$_3$ \cite{Yang2021}. Insulating SrTiO$_3$ spacers, $1.6$ nm to $2.5$ nm thick, were found to result also in weak coupling or in magnetic decoupling of two epitaxial SrRuO$_3$ layers  in the study by Herranz \textit{et al.} \cite{Herranz2003}. In our previous study \cite{Wysocki2018}, the interlayer coupling between SrRuO$_3$ layers separated by an asymmetric spacer of the strong spin-orbit coupling oxide SrIrO$_3$ and the large band gap insulator SrZrO$_3$ was addressed. Weak ferromagnetic coupling was observed with enhanced coupling strength for the reduction of the total spacer thickness from 0.8 nm to 0.4 nm \cite{Wysocki2018}. For symmetric  SrRuO$_3$/SrIrO$_3$ multilayers with 2 MLs thick SrIrO$_3$, where the SrIrO$_3$ is discussed to induce interfacial DMI, only theoretical calculations by Esser \textit{et al.} exist, which predict that ferromagnetic coupling between the SrRuO$_3$ layers is more favorable than an antiferromagnetic type of coupling \cite{Esser2021}.\\
Our present study addresses the magnetic interlayer coupling in such symmetric SrRuO$_3$-SrIrO$_3$ multilayers experimentally. Here the magnetic interlayer coupling was investigated for heterostructures in which the SrRuO$_3$ layers were separated by SrIrO$_3$ spacers of various thickness by means of superconducting quantum interference device (SQUID) magnetometry (full and minor hysteresis loops) and first order reversal curve measurements (FORC).
The FORC method has proven to provide valuable information in many different systems that is inaccessible for conventional magnetometry measurements. For example, microstructural information without actual lateral resolution in microstructured and model magnetic systems \cite{Gross2019b,Beron2006,Graefe2016,Graefe2016b}, information about coercive and interaction field distribution in permanent hard magnetic systems \cite{Dobrota2013,Ilse2021,Muralidhar2017}, as well as interaction strength and interaction type between different magnetic components in systems \cite{Gross2019b,Muralidhar2017} can be achieved.
Performing minor hysteresis loops and FORC measurements enabled us to quantify the sign and strength of the magnetic interlayer coupling between the SrRuO$_3$ layers for various SrIrO$_3$ spacer thicknesses. For the heterostructure with only 2 MLs SrIrO$_3$ spacer, the minor loops showed a small positive shift with respect to the major hysteresis loops above 30 K, indicating that the coupling turned weakly antiferromagnetic. However, the estimated coupling strength of about -7 $\mu$J/m$^2$ at 40 K led to the conclusion that the two SrRuO$_3$ layers switch their magnetization almost independently.
In its bulk form and thick films, the spacer material SrIrO$_3$ is a paramagnetic semimetal with a Fermi surface that consists of electron- and holelike pockets  \cite{Guo2020}. The transition to an insulating state can be induced in SrIrO$_3$ thin films by the reduction of the film thickness in the ultrathin limit \cite{Groenendijk2017, Manca2019,Guo2020} and by tailoring of epitaxial strain \cite{Biswas2014,Kim2017}. Manca \textit{et al.} reported the (semi-)metal-to-insulator transition to take place in SrIrO$_3$ films between 3 and 4 MLs \cite{Manca2019}. A resistivity increase was observed upon temperature enhancement in these SrIrO$_3$ layers of minimum 4 MLs thickness that indicated the metallic properties \cite{Manca2019}. In contrast, a 20 nm thick SrIrO$_3$ film showed only weakly temperature-dependent resistivity in the study by Gruenewald \textit{et al.} \cite{Gruenewald2014}. In our current study, it was therefore expected that the 2 MLs SrIrO$_3$ spacer is insulating and might undergo a transition from the insulating to the (semi-)metallic state, upon thickness increase. In case of a transition to the semimetallic state with clear temperature dependent resisitivity, the influence of the SrIrO$_3$ electronic transport properties on the interlayer coupling could be adressed in our study, in addition to the commonly observed thickness dependence of the interlayer coupling mediated by exchange or magnetostatic interactions. It turned out that the coupling strength did not increase upon the increase of the spacer thickness to 12 MLs and the two SrRuO$_3$ layers stayed decoupled. Resistivity investigations of SrIrO$_3$ reference films show that they are semimetallic with very weakly temperature-dependent behavior. Thus SrIrO$_3$ layers may be unsuitable as spacers for achieving a strong magnetic coupling between ferromagnetic SrRuO$_3$ and other oxide layers ought to be considered for realizing this end.

\section{Sample Design and Experimental Methods}
For investigating the type and strength of the magnetic coupling of the ferromagnetic SrRuO$_3$ layers, a set of heterostructures with two ferromagnetic SrRuO$_3$ layers of distinct thicknesses was designed. To make use of the thickness dependence of the coercive field $H_\text{c}$ and ferromagnetic transition temperature $T_\text{c}$ of SrRuO$_3$ thin films \cite{Xia2009}, each multilayer was composed of two separated SrRuO$_3$ layers with 6 MLs and 18 MLs thickness.  The 18 MLs thick SrRuO$_3$ was deposited directly on the SrTiO$_3$ (100) substrate, while the top 6 MLs SrRuO$_3$ layer was grown on top of the spacer layer, as illustrated in the scheme of the heterostructure design in \textbf{Figure S1a} of the supplemental material (Ref. \cite{supplemental}). For heterostructure RIR2, with the thinnest (2 MLs) SrIrO$_3$ spacer of this study, the 6 MLs SrRuO$_3$ layer was additionally capped by 2 MLs SrIrO$_3$.\\
The heterostructures were fabricated by pulsed-laser deposition (PLD), using a KrF excimer laser with 248 nm wavelength. The multilayers were grown on  SrTiO$_3$ (100) substrates. The substrates were etched in NH$_4$F - buffered HF solution and annealed in air at 1000$^{\circ}$C for 2 hours to achieve uniform TiO$_2$- termination of the surface.
During the growth, the deposition temperature was  650$^{\circ}$C, the oxygen pressure was kept at 0.133 mbar and the laser fluence was set to about 2 J/cm$^2$. We used 5 Hz repetition rate for the SrRuO$_3$ and 1 Hz for SrIrO$_3$. In order to ensure a smooth epitaxial growth for enhanced thicknesses of the SrIrO$_3$ spacer, the deposition temperature was increased for the heterostructure RIR12 to 700$^{\circ}$C .
Employing \emph{in situ} high-energy electron diffraction (RHEED) enabled the precise control of the SrIrO$_3$ layer thickness, which grew in a layer-by-layer mode (see \textbf{Figure S1b} in the supplemental information (Ref. \cite{supplemental})). Atomic force microscopy (AFM) investigations confirmed the smooth topography of the heterostructure surface resembling the stepped terrace structure of the SrTiO$_3$ (100) substrates, which indicates the pseudomorphic, crystalline growth. Further details on the thin film deposition and structural characterization can be found in the SI (Ref.\cite{supplemental}).\\
The magnetic interlayer coupling was investigated by a combination of conventional SQUID magnetometry (temperature-dependent and magnetic field-dependent magnetic moment measurements) and FORC investigations. The study was complemented by polar magneto-optical Kerr effect (p-MOKE) and Hall voltage measurements for selected samples. All Hall measurements were performed in the van der Pauw geometry in a custom-built set up.\\
SQUID magnetometry was performed by a commercially available SQUID magnetometer (MPMS-XL , Quantum Design inc.). In order to extract the magnetic response of the ferromagnetic SrRuO$_3$ layers, the linear contribution of the diamagnetic SrTiO$_3$ substrate was subtracted by linear fitting in the high magnetic field range. Furthermore, the nonlinear magnetic moment measured above the Curie temperature of the SrRuO$_3$ layers was subtracted to correct the additional background response originating from magnetic impurities introduced most likely during the required sample cutting (see section 2 of the supplemental material \cite{supplemental}). \\
The FORC measurements were performed with a SQUID magnetometer (MPMS 3 , Quantum Design inc.). Processing of raw data was done with LeXtender \cite{Graefe2021}, and the FORC densities were calculated using the gFORC algorithm \cite{Gross2019}. For the FORC study, a set of minor loops with various reversal fields was performed. Before each minor loop, the sample was saturated in a positive magnetic field of 5 T. Then the external magnetic field was decreased to the required reversal field  $H_\text{r}$. The first order reversal curve was determined by measuring the magnetic moment when the magnetic field was increased from $H_\text{r}$ to saturation in positive magnetic fields \cite{Gross2019, Gross2019b, Ilse2021}. This procedure was repeated with step-like decreasing of the reversal field until the reversal field reached negative saturation. The FORC density was calculated by the mixed second derivative of the magnetic moment surface:
\begin{equation}
\rho( H, H_r ) = - \frac{1}{2} \frac{\partial^2 m(H, H_r)}{\partial H \partial H_r}
\label{equation_forcdensity}
\end{equation}
The FORC density was then transformed on the axes of the coercive field $H_\text{c}$ and the interaction field $H_\text{u}$ via:
\begin{equation}
 H_u = \frac{1}{2} ( H + H_r )  ; \quad H_c = \frac{1}{2} ( H - H_r )
\label{equation_transformation}
\end{equation}
From the FORC-density, plotted as function of the interaction field and the coercive field, the sign of the magnetic interlayer coupling can be assessed.
\section{Results}
\subsection{Heterostructure with 2 MLs SrIrO$_3$ spacer (RIR2) }
\begin{figure}[!htb]
\includegraphics[width=0.95\linewidth]{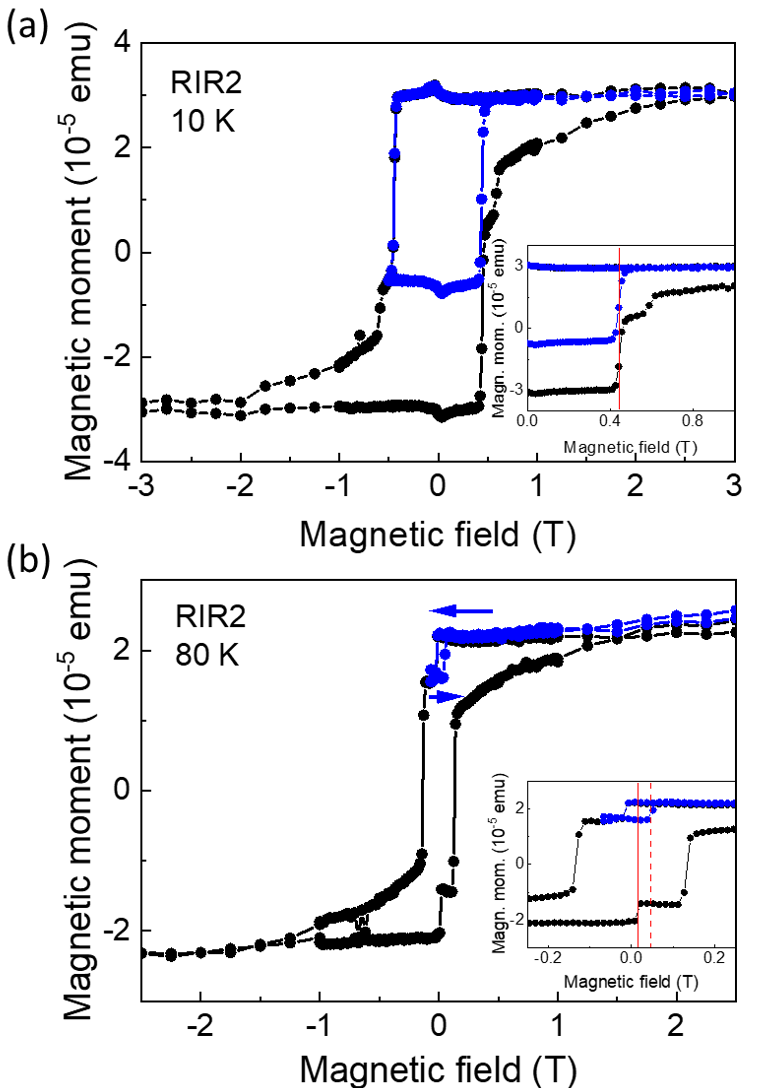}
\caption{\label{fig_overview_RIR2_SQUID} (a) Major (black) and minor (blue) magnetic hysteresis loops for the heterostructure RIR2 with 2 MLs SrIrO$_3$ spacer at 10 K (a) and 80 K (b). The magnetic field was applied perpendicular to the thin film surface. The minor loops were carried out between 5 T and $-0.5$ T (a), and $-0.07$ T (b). The minor loop at 10 K (a) does not show a measurable shift. At 80 K (b), the switching field (during the backward sweep) of the minor loop (red dashed line) is shifted by +30 mT with respect to the reversal field of the magnetically softer layer during the major loop (solid red line).}
\end{figure}

\begin{figure*}[!htb]
\includegraphics[width=0.85\linewidth]{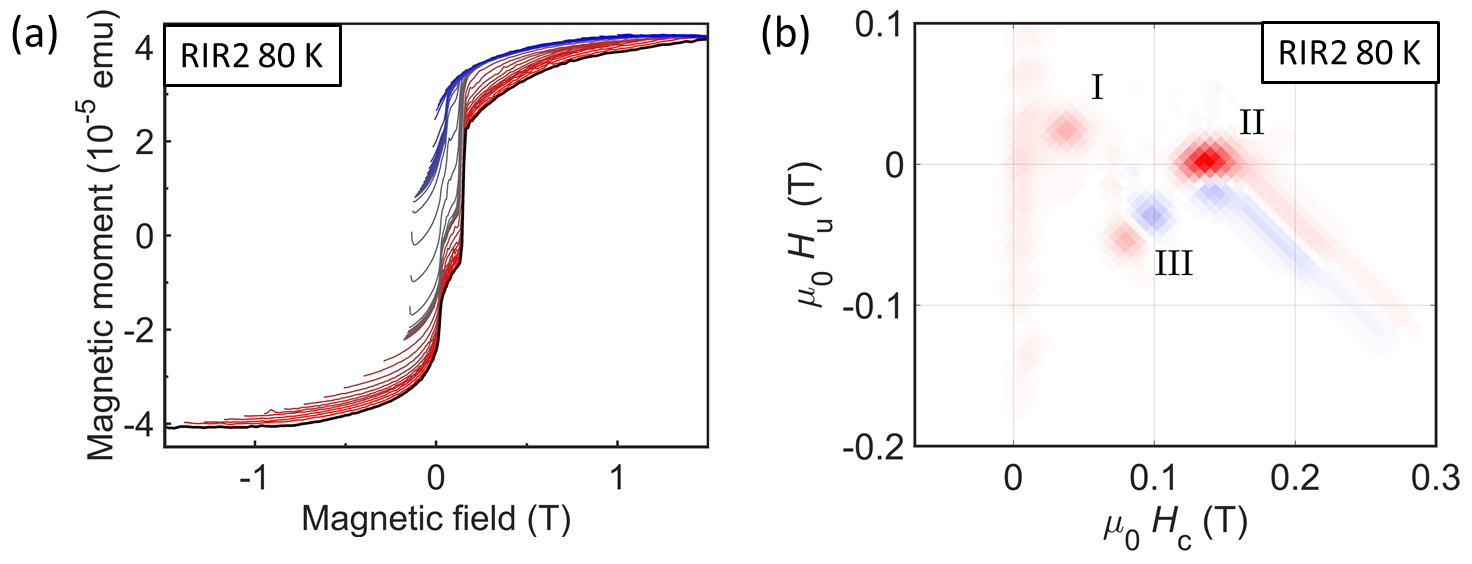}
\caption{\label{fig_overview_RIR2-2} FORC study of the heterotructure RIR2 at 80 K with the magnetic field applied perpendicular to the heterostructure surface. The measured minor loops, corrected for the diamagnetic contribution originating from the substrate, are shown in (a). The color of the respective minor loops changes from red to blue for increasing reversal fields. The FORC density is plotted in (b) as a function of the coercive field $H_\text{c}$ and the interaction field $H_\text{u}$. Positive FORC density peaks are shown in red, negative ones in blue. Feature (I) and (II) correspond to the magnetization switching of the 6 MLs (I) and 18 MLs SrRuO$_3$ (II) layer, respectively. The additional peak pair (III) at 80 K is the interaction peak indicating antiferromagnetic coupling. The additional feature located along $H_\text{c}$ = 0 T is the reversible ridge.}
\end{figure*}

Summarized in \textbf{Figure \ref{fig_overview_RIR2_SQUID}} (a) and (b) are major and minor magnetic hysteresis loops for the heterostructure RIR2 (2 MLs SrIrO$_3$/ 6 MLs SrRuO$_3$ / 2 MLs SrIrO$_3$/ 18 MLs SrRuO$_3$ on SrTiO$_3$) at representative temperatures of 10 K and 80 K. The magnetic field was applied perpendicular to the thin film surface for the presented measurements. The hysteresis loops, acquired by SQUID magnetometry, were corrected by the subtraction of the diamagnetic background of the SrTiO$_3$ substrate and magnetic impurities, following the procedure described in section 2 of the SI (Ref. \cite{supplemental}).\\
The magnetization of the heterostructure RIR2 reverses its orientation in a two-step reversal process indicating at best weak coupling of the two SrRuO$_3$ layers. Since the 18 MLs thick SrRuO$_3$ layer has a larger magnetic moment than the thinner SrRuO$_3$ layer, it can be concluded that the thicker layer is the magnetically softer layer at 10 K. At elevated temperatures, such as 80 K, the thinner 6 MLs SrRuO$_3$ layer is magnetically softer and switches at smaller magnetic fields than the 18 MLs SrRuO$_3$ layer, as it has been shown already in our previous study on similar SrRuO$_3$-based heterostructures \cite{Wysocki2018}. The temperature dependence of the switching fields of the two ferromagnetic layers of this particular heterostructure RIR2 is shown in \textbf{Figure S4b} in the SI (Ref. \cite{supplemental}).
In addition to the sharp two-step magnetization reversal, the magnetic hysteresis loops possess a tail in the high magnetic field range, that can be related most likely to strongly pinned domains in the bottom SrRuO$_3$ layers deposited directly on the SrTiO$_3$ (100) substrate \cite{Wang2020}.\\
The minor loop of heterostructure RIR2, drawn in blue in \textbf{Figure \ref{fig_overview_RIR2_SQUID}(a)}, did not show a measurable shift with respect to the major hysteresis loop at 10 K, showing that the two SrRuO$_3$ layers of the heterostructure are indeed magnetically decoupled.  In contrast, the minor loop of the heterostructure RIR2 is shifted by +30 mT to higher magnetic fields at 80 K (cf. inset in \textbf{Figure \ref{fig_overview_RIR2_SQUID}(b)}). Such a positive shift of the minor loop with respect to the full loop can be an indication for antiferromagnetic coupling of the two SrRuO$_3$ layers (see for instance ref. \cite{Heijden1997,Faure-Vincent2002,Matczak2013}):
As indicated by the red lines in \textbf{Figure \ref{fig_overview_RIR2_SQUID} (b)}, the switching field of the magnetically softer layer increased by + 30 mT with respect to the major loop.
According to the calculation proposed by van der Heijden \textit{et al.}, the magnetic coupling strength is directly proportional to
the difference of the switching fields of the magnetically softer layer of the major loop and of the minor loop \cite{Heijden1997}. As described in detail in section 4 of the SI (Ref. \cite{supplemental}), we estimated a coupling strength of -5 $\mu$J/m$^2$ at 80 K, increasing to -7 $\mu$J/m$^2$ at 40 K. This coupling strength is very weak: In our previous study on asymmetric SrIrO$_3$/SrZrO$_3$ spacers,  we observed a weak ferromagnetic coupling on the order of 35 $\mu$J/m$^2$ for a SrIrO$_3$/SrZrO$_3$ spacers of thickness of about 0.8 nm. As shown in \textbf{Figure S4} in the SI (Ref. \cite{supplemental}), for heterostructure RIR2, the minor loop shift is almost temperature independent between 40 K and 100 K, when the 6 MLs top SrRuO$_3$ layer is the magnetically softer layer of the heterostructure. Thus, the coupling strength, which is directly proportional to the magnetization of the magnetically softer layer \cite{Heijden1997}, decreases for increasing temperature above 40 K, following the temperature dependence of the magnetization of the thinner SrRuO$_3$ layer.
\begin{figure*}[!htb]
\includegraphics[width=0.85\linewidth]{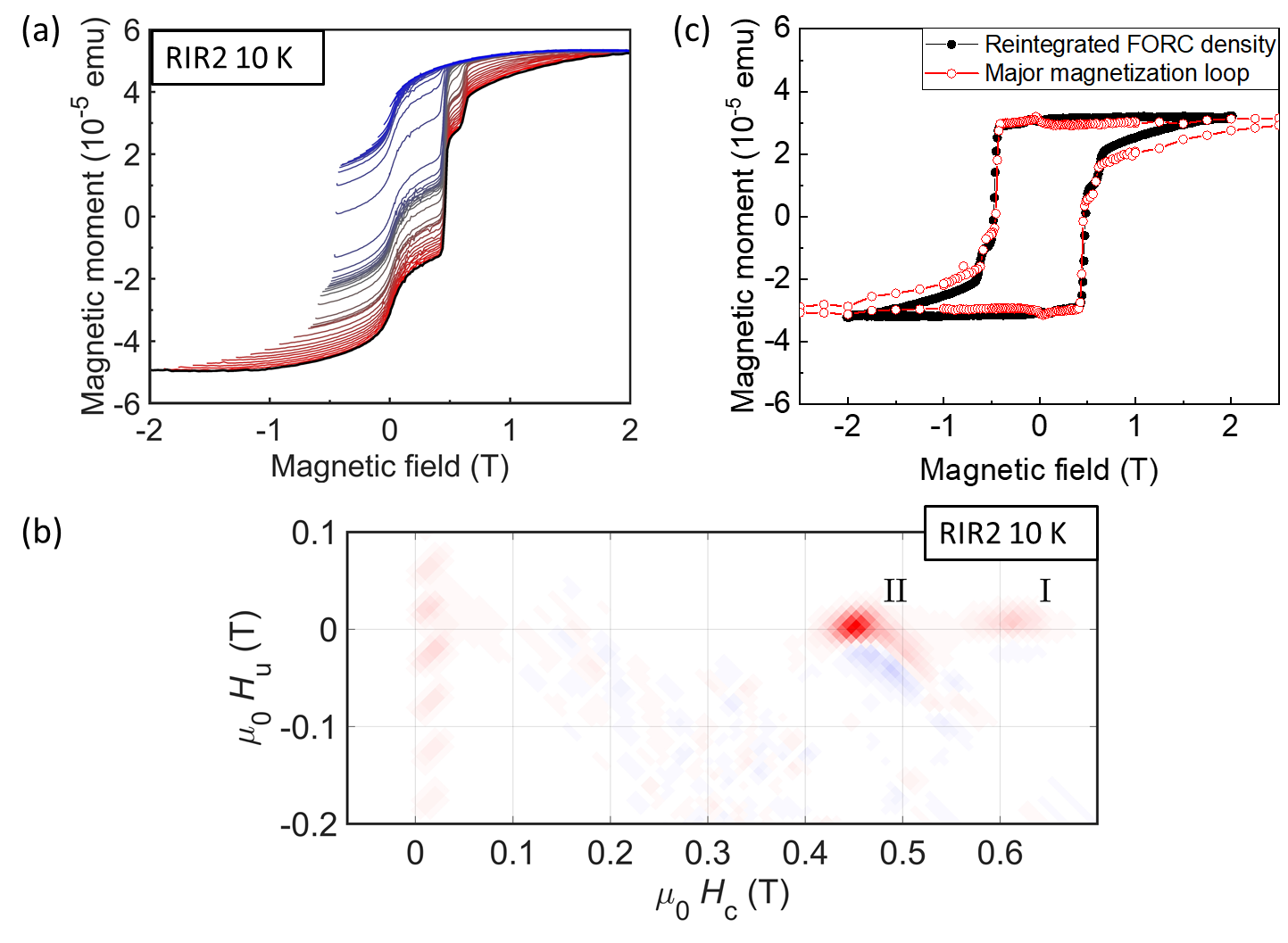}
\caption{\label{fig_overview_RIR2} First order reversal curve study of the heterotructure RIR2 at 10 K. The magnetic field was applied perpendicular to the heterostructure surface. The measured minor loops, corrected for the diamagnetic contribution originating from the substrate, are presented in (a). The color of the respective minor loops changes from red to blue for increasing reversal fields. The FORC densities plotted as function of the coercive field $H_\text{c}$ and the interaction field $H_\text{u}$ at the corresponding temperatures are shown in (b). Positive FORC density peaks are drawn in red, negative ones in blue. Feature (I) and (II) correspond to the magnetization switching of the 6 MLs (I) and 18 MLs SrRuO$_3$ (II) layer, respectively. Shown in (c) is the comparison of the major magnetization loop (red), corrected by the subtraction of the diamagnetic substrate and magnetic impurity contribution (see SI \cite{supplemental}), and the reintegrated FORC density (black) after removal of the soft magnetic contribution of the reversible ridge between -0.05 T $<$ $\mu_0 H_\text{c}$ $<$  0.1 T and -1.5 T$<$ $\mu_0 H_\text{u}$ $<$ 1.5 T.}
\end{figure*}
To confirm the sign and order of magnitude of the minor loop shifts, determined from the magnetometry measurements, Kerr rotation measurements were performed at 10 K and 80 K (see  \textbf{Figure S3} in the SI (Ref. \cite{supplemental})). The Kerr rotation angle, determined in the polar MOKE geometry, scales linearly with the perpendicular component of the magnetization, but is not influenced by magnetic impurities at the backsides or on the edges of the sample for our measurements in reflection geometry and therefore a useful probe of the qualitative interlayer coupling. In agreement with our results from the SQUID investigations, the minor loop at 10 K  (\textbf{Figure S3a} in the SI (Ref. \cite{supplemental} )) is not shifted within the magnetic field accuracy, while the minor loop at 80 K is also shifted by +38 mT ( \textbf{Figure S3b} in the SI (Ref. \cite{supplemental})).\\
To further study the magnetic interlayer coupling in the heterostructure RIR2, FORC measurements were performed at 10 K (\textbf{ Figure \ref{fig_overview_RIR2}}) and 80 K (\textbf{ Figure \ref{fig_overview_RIR2-2}}).
Presented in \textbf{Figure \ref{fig_overview_RIR2}}(a) is the set of minor loops of heterostructure RIR2 at 10 K. All minor loops were corrected by the subtraction of the diamagnetic contribution originating from the SrTiO$_3$ substrate. The soft magnetic contribution visible in the minor loops at small magnetic fields is related to magnetic impurities, often introduced during the sample cutting process, as mentioned earlier. Additionally, the two step-reversal of the magnetization was observed for the minor loops that started close to negative saturation. From these minor loops, the FORC density was calculated according to equation \ref{equation_forcdensity} and is shown in \textbf{Figure \ref{fig_overview_RIR2}}(b). Three general features are present in the FORC density at 10 K. The positive peaks (I) and (II) correspond to the reversal of the two ferromagnetic SrRuO$_3$ layers. The intensity of the peaks is proportional to magnetization of the respective layer. Hence, the more intense peak (II) is related to the switching of the 18 MLs bottom SrRuO$_3$ layer and (I) to the 6 MLs thin SrRuO$_3$ layer. The positions of the center of the peaks at 620 mT (I) and 450 mT (II) are in good agreement with the switching fields determined from the major magnetization hysteresis loops (see \textbf{Figure S4b} in the SI (Ref. \cite{supplemental})). The FORC investigations of heterostructure RIR2 did not show any hints of the coupling of the two ferromagnetic SrRuO$_3$ layers at 10 K.
The additional feature located at tiny magnetic field values is the reversible ridge which is dominated by magnetically soft, reversible contributions originating mainly from magnetic impurities. In case of the SQUID hysteresis loop (\textbf{Figure \ref{fig_overview_RIR2_SQUID}}), these contributions were removed by subtraction of the hysteresis loop measured above the transition temperature of the SrRuO$_3$ layers and therefore related to high $T_\text{c}$ magnetic impurities (see section 2 of the SI (Ref. \cite{supplemental}) for further details). To confirm that the reversible ridge is dominated by the contribution of these magnetic impurities, the FORC density presented in (b) was reintegrated with the exclusion of the contribution between -0.05 T $<$ $\mu_0 H_\text{c}$ $<$  0.1 T and -1.5 T$<$ $\mu_0 H_\text{u}$ $<$ 1.5 T. Such integration of the FORC density was possible, because feature (I) and (II) originating from the magnetization reversal of the layers were sufficiently separated from the reversible ridge. The integration yielded half of the hysteresis loop from -2 T to 2 T and was mirrored at both x- and y-axis in order to reconstruct the full hysteresis loop. Plotted in \textbf{Figure \ref{fig_overview_RIR2}(c) }is the comparison of the reconstructed hysteresis loop of the FORC study (black) and the conventional major magnetization loop (red), which has been corrected for the magnetic impurity contribution. Both hysteresis loops are in good agreement and the switching fields of the two SrRuO$_3$ layers are identical for both techniques within a few mT. The agreement of both hysteresis loops supports our expectation that the reversible ridge is dominated by uncorrelated magnetic impurities that do not influence the switching fields of the magnetic SrRuO$_3$ layers of the heterostructure. This shows that the reintegration of the FORC density without the reversal ridge can be used in this case to obtain a hysteresis loop where the contribution of the soft magnetic impurity is removed, without the need of an additional measurement above the transition temperature of SrRuO$_3$.  \\
The FORC study of heterostructure RIR2 at 80 K is summarized in \textbf{ Figure \ref{fig_overview_RIR2-2}}. Also the minor loops measured at 80 K, presented in\textbf{ Figure \ref{fig_overview_RIR2-2}(a)}, show a two-step magnetization reversal. At 80 K, the 6 MLs thin SrRuO$_3$ switches at smaller magnetic fields than the 18 MLs thick bottom SrRuO$_3$ layer. In the FORC density, shown in (b), feature (I) corresponds again to the reversal of the 6 MLs SrRuO$_3$, while feature (II) originates from the magnetization switching of the 18 MLs thick SrRuO$_3$ layer. In contrast to the FORC density map at 10 K,  an additional positive-negative-peak pair (structure III) is present at finite interaction field (see \textbf{ Figure \ref{fig_overview_RIR2-2}(b)}) at 80 K. According to previous FORC studies on well defined systems of coupled microarrays and on NdFeB samples with components with different coercivities, such additional positive-negative-peak pairs are characteristic for magnetic coupling between two different magnetic sites and denominated as the so called "interaction peak" \cite{Gross2019b, Ilse2021}.
The relative position of the positive and negative part of the interaction peak with respect to each other yields information about the nature of the coupling. As shown in \cite{Gross2019b}, the coupling is antiparallel if the negative FORC density part of the interaction peak is at higher coercive and interaction fields than the positive part of the interaction peak and parallel if it is vice-versa. According to this, the interaction peak in \textbf{Figure \ref{fig_overview_RIR2-2}(b)} shows that the SrRuO$_3$ layers in sample RIR2 are coupled antiparallel at 80 K, which confirms our observation from the conventional SQUID magnetometry.
If exchange bias between a ferromagnet and an antiferromagnet was present, this would lead most likely to a positive peak in the FORC density which is elongated along the interaction field \cite{Graefe2014} rather than a positive-negative peak pair. \\
The FORC density of a SrRuO$_3$-based heterostructure in which the ferromagnetic layers are coupled weakly ferromagnetically is presented in the supplemental material for comparison \cite{supplemental}. In this heterostructure RIZR1, the two SrRuO$_3$ layers were weakly ferromagnetically coupled through a spacer of 1 ML SrIrO$_3$ and 1 ML SrZrO$_3$. The FORC density, plotted in \textbf{Figure S5} in the supplemental material (Ref. \cite{supplemental}), also shows two positive peaks related to the magnetization reversal of the magnetization of the two SrRuO$_3$ layers. The observed interaction peak shows the positive FORC density at higher coercive and smaller interaction field than the negative peak, which indicates the ferromagnetic coupling between the SrRuO$_3$ layers \cite{Gross2019b}.
\subsection{Influence of the SrIrO$_3$ spacer thickness on the interlayer coupling}
\begin{figure*}[!htb]
\includegraphics[width=0.9\linewidth]{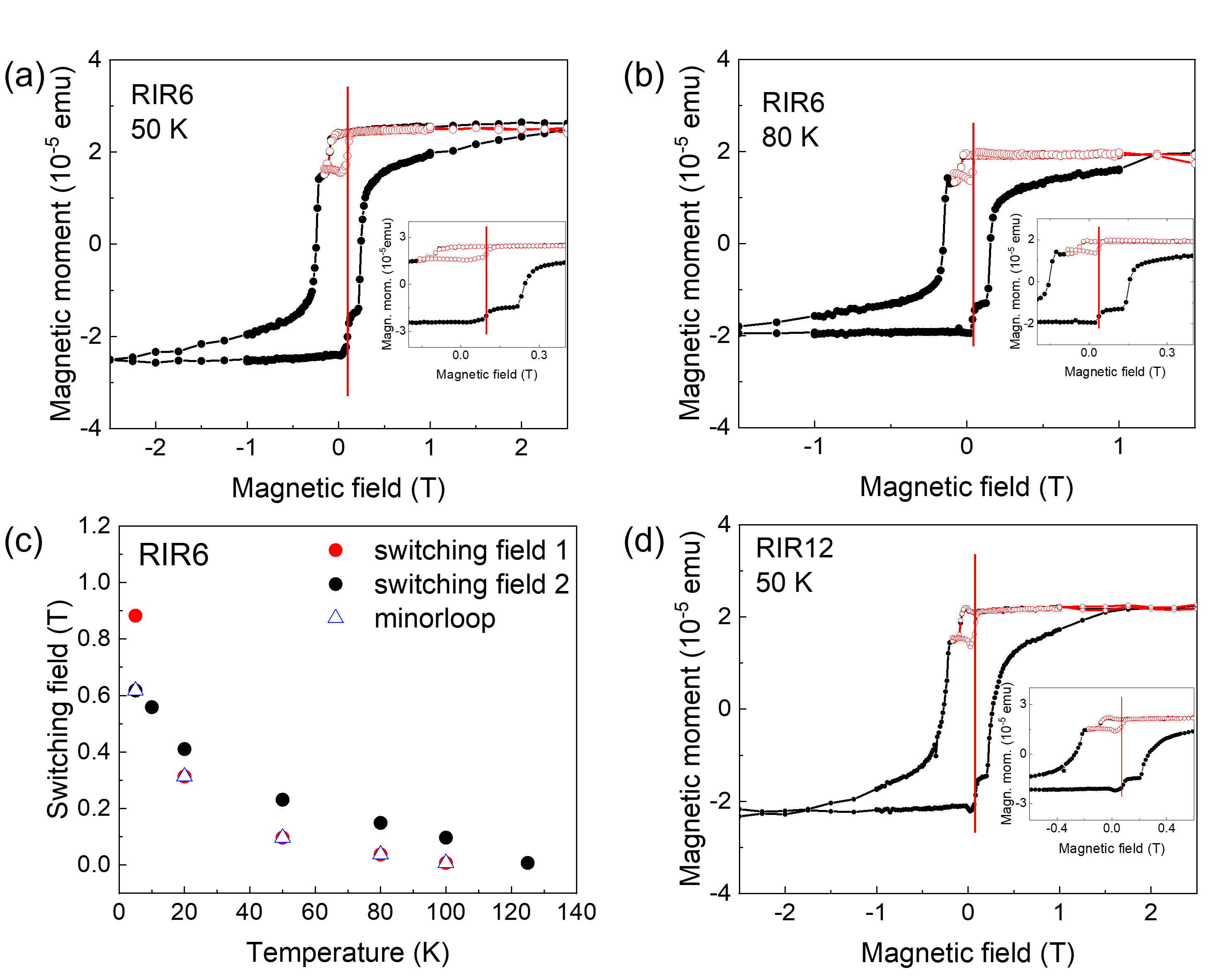}
\caption{\label{fig_RIR6_RIR12} Full and minor hysteresis loops of the magnetic moment of heterostructure RIR6 with 6 MLs SrIrO$_3$ spacer at 50 K (a) and 80 K (b). (c) Temperature dependence of the switching fields of the two SrRuO$_3$ layers with 18 MLs (switching field 2) and 6 MLs (switching field 1) thickness, and the switching fields determined from minor loops experiments for the heterostructure RIR6. (d) Major and minor hysteresis loops of the magnetic moment of heterostructure RIR12 with 12 MLs SrIrO$_3$ spacer at 50 K. The magnetic field was applied perpendicular to the surfaces of the heterostructures.}
\end{figure*}
The possibility of a (semi-)metal-to-insulator transition that was reported by Manca \textit{et al.} to take place between 3 and 4 MLs in bare SrIrO$_3$ thin films \cite{Manca2019} motivated the growth of the two heterostructures RIR6 and RIR12, with SrIrO$_3$ spacers that are considerably thicker than 4 MLs (6 MLs spacer for RIR6 and 12 MLs for RIR12).
If such spacer thickness increase led to a significant change of the SrIrO$_3$ electronic transport properties, a major impact on the interlayer exchange coupling would be expected, as it was achieved in SrRuO$_3$ based heterostructures separated by LaNiO$_3$ spacers \cite{Yang2021}.
Presented in \textbf{Figure \ref{fig_RIR6_RIR12} (a)}  and \textbf{(b)} are the full and minor magnetic hysteresis loops of heterostructure RIR6 (6 MLs SrIrO$_3$ spacer) at 50 K and 80 K, acquired by SQUID magnetometry.
The major hysteresis loops of the heterostructure RIR6 show a two-step reversal of the magnetization, similar to heterostructure RIR2. At 50 K, the switching at 0.1 T originates from the magnetization reversal of the 6 MLs SrRuO$_3$ layer, while the step at 0.25 T is related to the switching of the 18 MLs SrRuO$_3$ layer, which is the magnetically harder layer at 50 K. Such two-step switching process indicates again the decoupling or weak magnetic interlayer coupling. To determine the interlayer coupling strength, the reversal fields of the minor loop were compared to the switching behavior of the magnetically softer layer during the full loop. As highlighted in the inset of \textbf{Figure \ref{fig_RIR6_RIR12}(a)}, the minor loop switching field is equal to the switching field of the major loop. This shows that the minor loop shift and therefore the coupling strength is zero (see section 4 in the SI (Ref. \cite{supplemental}) for further details on the calculation). The two SrRuO$_3$ layers are fully decoupled by 6 MLs SrIrO$_3$ at 50 K.
Also at 80 K, where a weak antiferromagnetic coupling was observed for heterostructure RIR2, the minor loop is not shifted in case of heterostructure RIR6 (see inset of \textbf{Figure \ref{fig_RIR6_RIR12}(b)}). As shown in \textbf{Figure \ref{fig_RIR6_RIR12}(c)}, such equality of the switching fields of minor loop (drawn as blue triangles) and major loop (full symbols) was observed at all temperatures investigated for heterostructure RIR6. The absence of a measurable minor loop shift shows that 6 MLs SrIrO$_3$ spacer decouple the two ferromagnetic SrRuO$_3$ layers fully at all temperatures.\\
Increasing the thickness of the SrIrO$_3$ spacer to 12 MLs was still insufficient to result in a measurable magnetic coupling of the two SrRuO$_3$ layers of heterostructure RIR12. As shown by the hysteresis loop measurements at 50 K in \textbf{Figure \ref{fig_RIR6_RIR12}(d)}, the magnetic hysteresis is consistent with a two step-reversal of the magnetization. The minor loop is not shifted within the magnetic field accuracy, indicating the decoupling of the two ferromagnetic layers. The decoupling of the SrRuO$_3$ layers was confirmed by a magnetotransport study of heterostructure RIR12, presented in the SI (Ref. \cite{supplemental} section 6).\\
The decoupling of the ferromagnetic SrRuO$_3$ layers, observed in a broad temperature range and consistently for the heterostructures RIR6 and RIR12 with spacer thicknesses above the limit for which a MIT was reported, indicates that 
a SrIrO$_3$ spacer may not be a suitable choice for enabling an exchange mediated coupling in these heterostructures.  In contrast to LaNiO$_3$ spacer layers \cite{Yang2021}, the semimetallic SrIrO$_3$ layers may not permit the strong ferromagnetic coupling of SrRuO$_3$ layers, which would be relevant in the context of skyrmion formation in SrRuO$_3$-SrIrO$_3$ multilayers \cite{Yang2020, Esser2021}.
The SrIrO$_3$ resistivity was investigated by reference sample measurements (bare films grown on SrTiO$_3$), given in the SI (Ref. \cite{supplemental}), because the SrIrO$_3$ spacer resistivity could not be measured directly for our heterostructures when the SrIrO$_3$ was sandwiched between the metallic SrRuO$_3$ layers. The resistivities of these 6 MLs and 12 MLs SrIrO$_3$ reference thin films showed a very weak temperature dependence with a small resistivity increase for decreasing temperature, as it was also observed in 20 nm SrIrO$_3$ deposited on SrTiO$_3$ \cite{Gruenewald2014} or when sandwiched between LaMnO$_3$ \cite{Skoropata2020}.\\
Based on the observed semimetallic behavior of the 6 MLs and 12 MLs bare SrIrO$_3$ thin films, also the SrIrO$_3$ spacers of the heterostructures likely did not become metallic.\\
Generally, the coupling of two ferromagnetic layers separated by a nonmagnetic insulator can originate from different mechanisms, such as the direct coupling via pinholes \cite{Fulghum1995}, magnetostatic N\'{e}el's coupling due to correlated surface roughness \cite{Neel1962, Moritz2004}, due to shape induced magnetic poles \cite{Schrag2000a}, or induced by the coupling of magnetic domain walls \cite{Fuller1962,Thomas2000,Platt2000,Baltz2007}. Another coupling mechanism is the magnetic exchange coupling by tunneling of spin-polarized electrons through the insulating barrier \cite{Slonczewski1989, Bruno1995, Faure-Vincent2002}.\\
The coupling via pinholes, which are often present in heterostructures with ultrathin spacers, would lead to trivial ferromagnetic coupling \cite{Fulghum1995,Bobo1999} between the SrRuO$_3$ layers and thus cannot explain the minor loop shift to higher magnetic fields for sample RIR2. On the other hand, we emphasize at this point that the coupling of the two SrRuO$_3$  layers separated by 2 MLs SrIrO$_3$ was found to be very sensitive to the existence of (pin-)holes in the heterostructure. As presented in \textbf{Figure S8} in the SI (Ref. \cite{supplemental}), a second heterostructure where holes of nanometer depth were observed by atomic force microscopy showed weak ferromagnetic coupling. In contrast, atomic force microscopy did not show the existence of holes in any of the heterostructures RIR2, RIR6, RIR12 so that it can be concluded that the density of pinholes connecting the two SrRuO$_3$ layers is most likely small for these samples. The weak antiferromagnetic coupling was observed only in heterostructure RIR2 with 2 MLs SrIrO$_3$ spacer and with a small density of holes seen by AFM. \\
In addition, magnetostatic and interlayer exchange coupling can lead to magnetic coupling of the SrRuO$_3$ layers.  Antiferromagnetic coupling between ferromagnetic layers separated by nonmagnetic, insulating spacers has been previously related also to exchange coupling \cite{Faure-Vincent2002} described by the Slonczweski spin-current model \cite{Slonczewski1989} or the quantum interference model developed by Bruno \cite{Bruno1995}. However, the observed decrease of the coupling strength $J_C$ (calculated with equation (1) in the SI (Ref. \cite{supplemental})) with increasing temperature observed for heterostructure RIR2 cannot be explained within the model of quantum interference effects, which predicts an increase for increasing temperatures in case of insulating spacers \cite{Bruno1995} (for further details on the validity of the approximations made within this model, see Section 4 of the supplemental material \cite{supplemental}).\\
N\'{e}el's theory of the magnetostatic coupling due to magnetic surface charges induced by correlated surface roughness was extended by Moritz \textit{et al.} to magnetic multilayers with perpendicular magnetic anisotropy \cite{Moritz2004}. Depending on the strength of the magnetic anisotropy constant, the magnetostatic orange-peel coupling has been found to be ferromagnetic for a weak perpendicular anisotropy constant and antiferromagnetic for strong perpendicular magnetic anisotropy \cite{Moritz2004}. SrRuO$_3$ thin films deposited on SrTiO$_3$(100) typically have a large magnetic anisotropy with the magnetic easy axis close to the [110]$_\text{orthorh.}$ direction \cite{Ziese2010} so that the orange-peel coupling would be expected to favor antiferromagnetic coupling between the layers. The heterostructures under study were all deposited on SrTiO$_3$ (100) substrates with a step-and-terrace structure of $0.4$ nm height and 250-300 nm width that most likely led to unidirectional interface roughness \cite{Davydenko2015}. However, the orange-peel coupling fields \cite{Schrag2000} expected for the substrate induced roughness would be too small to explain the observed weak antiferromagnetic coupling.\\
One possible coupling mechanism, which is qualitatively consistent with the observed temperature dependence of the weak antiferromagnetic interlayer coupling in heterostructure RIR2, might be the model of domain replication in the hard layer via magnetostatic interactions, as proposed by Nistor \cite{Nistor2011}. When the magnetic field required to reverse the magnetization of the soft layer during the minor loops is close to the nucleation field of the hard layer, inversed domains in the soft layer will generate stray fields that can induce so called replicated domains in the hard layer acting as negative bias field during the second half of the minor loop \cite{Nistor2011,Rodmacq2006}.

\section{Conclusion}
The magnetic interlayer coupling between ferromagnetic SrRuO$_3$ epitaxial layers separated by the strong spin-orbit coupling SrIrO$_3$ was investigated by the combination of conventional SQUID magnetometry and FORC measurements. The minor loops of the heterostructure with 2 MLs of SrIrO$_3$ spacer showed a small shift to higher magnetic field above 40 K, indicating very weak antiferromagnetic coupling of about -7  $\mu$J/m$^2$. The increase of the SrIrO$_3$ layer thickness to 12 MLs did not lead to an increase of the coupling but to rather fully decoupled layers. This is most likely related to the still weak temperature dependence of the SrIrO$_3$ spacer resistivities of our heterostructures and the weak increase of the conductivity upon spacer thickness increase.
Such decoupling or very weak antiferromagnetic coupling of the SrRuO$_3$ by SrIrO$_3$ spacers is undesirable in the context of skyrmion formation. Without ferromagnetic coupling of the magnetic layers, skyrmions would not be coupled through multilayer stacks.\\
Our study further highlights the scientific relevance of first order reversal curve investigations for the study of magnetic interlayer coupling, being capable to detect weak coupling interactions as well as to determine whether the coupling is antiferromagnetic or ferromagnetic. Additionally, FORC measurements have the advantage that a correction for the contribution of magnetic impurities is not necessary, because the peaks representing the various magnetization reversal steps are  well separated in the FORC density maps.  We could also show that reintegrating the FORC density without the reversible ridge can be an alternative method to correct a samples hysteresis loop for a soft magnetic impurity.
\\\\
\section*{Acknowledgment}
We thank Ren\'{e} Borowski and Silvia de Waal for etching the STO substrates and Susanne Heijligen for assistance with SQUID measurements. I.L.-V. acknowledges the financial support from the German Research Foundation (DFG) for Project No. 403504808, within SPP2137, and for Project No. 277146847 within SFB1238 (project A01). We are grateful to DFG for the financing of the PLD-RHEED system (Project No. 407456390). L.Y. thanks China Scholarship Council (File No. 201706750015) for his fellowship.
\nocite{*}
\bibliography{bib}

\begin{thebibliography}{64}%
\makeatletter
\providecommand \@ifxundefined [1]{%
 \@ifx{#1\undefined}
}%
\providecommand \@ifnum [1]{%
 \ifnum #1\expandafter \@firstoftwo
 \else \expandafter \@secondoftwo
 \fi
}%
\providecommand \@ifx [1]{%
 \ifx #1\expandafter \@firstoftwo
 \else \expandafter \@secondoftwo
 \fi
}%
\providecommand \natexlab [1]{#1}%
\providecommand \enquote  [1]{``#1''}%
\providecommand \bibnamefont  [1]{#1}%
\providecommand \bibfnamefont [1]{#1}%
\providecommand \citenamefont [1]{#1}%
\providecommand \href@noop [0]{\@secondoftwo}%
\providecommand \href [0]{\begingroup \@sanitize@url \@href}%
\providecommand \@href[1]{\@@startlink{#1}\@@href}%
\providecommand \@@href[1]{\endgroup#1\@@endlink}%
\providecommand \@sanitize@url [0]{\catcode `\\12\catcode `\$12\catcode
  `\&12\catcode `\#12\catcode `\^12\catcode `\_12\catcode `\%12\relax}%
\providecommand \@@startlink[1]{}%
\providecommand \@@endlink[0]{}%
\providecommand \url  [0]{\begingroup\@sanitize@url \@url }%
\providecommand \@url [1]{\endgroup\@href {#1}{\urlprefix }}%
\providecommand \urlprefix  [0]{URL }%
\providecommand \Eprint [0]{\href }%
\providecommand \doibase [0]{https://doi.org/}%
\providecommand \selectlanguage [0]{\@gobble}%
\providecommand \bibinfo  [0]{\@secondoftwo}%
\providecommand \bibfield  [0]{\@secondoftwo}%
\providecommand \translation [1]{[#1]}%
\providecommand \BibitemOpen [0]{}%
\providecommand \bibitemStop [0]{}%
\providecommand \bibitemNoStop [0]{.\EOS\space}%
\providecommand \EOS [0]{\spacefactor3000\relax}%
\providecommand \BibitemShut  [1]{\csname bibitem#1\endcsname}%
\let\auto@bib@innerbib\@empty
\bibitem [{\citenamefont {Matsuno}\ \emph {et~al.}(2016)\citenamefont
  {Matsuno}, \citenamefont {Ogawa}, \citenamefont {Yasuda}, \citenamefont
  {Kagawa}, \citenamefont {Koshibae}, \citenamefont {Nagaosa}, \citenamefont
  {Tokura},\ and\ \citenamefont {Kawasaki}}]{Matsuno2016}%
  \BibitemOpen
  \bibfield  {author} {\bibinfo {author} {\bibfnamefont {J.}~\bibnamefont
  {Matsuno}}, \bibinfo {author} {\bibfnamefont {N.}~\bibnamefont {Ogawa}},
  \bibinfo {author} {\bibfnamefont {K.}~\bibnamefont {Yasuda}}, \bibinfo
  {author} {\bibfnamefont {F.}~\bibnamefont {Kagawa}}, \bibinfo {author}
  {\bibfnamefont {W.}~\bibnamefont {Koshibae}}, \bibinfo {author}
  {\bibfnamefont {N.}~\bibnamefont {Nagaosa}}, \bibinfo {author} {\bibfnamefont
  {Y.}~\bibnamefont {Tokura}},\ and\ \bibinfo {author} {\bibfnamefont
  {M.}~\bibnamefont {Kawasaki}},\ }\bibfield  {title} {\bibinfo {title}
  {{Interface-driven topological Hall effect in SrRuO$_3$ - SrIrO$_3$
  bilayer}},\ }\href {https://doi.org/10.1126/sciadv.1600304} {\bibfield
  {journal} {\bibinfo  {journal} {Science Advances}\ }\textbf {\bibinfo
  {volume} {2}},\ \bibinfo {pages} {e1600304} (\bibinfo {year}
  {2016})}\BibitemShut {NoStop}%
\bibitem [{\citenamefont {Ohuchi}\ \emph {et~al.}(2018)\citenamefont {Ohuchi},
  \citenamefont {Matsuno}, \citenamefont {Ogawa}, \citenamefont {Kozuka},
  \citenamefont {Uchida}, \citenamefont {Tokura},\ and\ \citenamefont
  {Kawasaki}}]{Ohuchi2018}%
  \BibitemOpen
  \bibfield  {author} {\bibinfo {author} {\bibfnamefont {Y.}~\bibnamefont
  {Ohuchi}}, \bibinfo {author} {\bibfnamefont {J.}~\bibnamefont {Matsuno}},
  \bibinfo {author} {\bibfnamefont {N.}~\bibnamefont {Ogawa}}, \bibinfo
  {author} {\bibfnamefont {Y.}~\bibnamefont {Kozuka}}, \bibinfo {author}
  {\bibfnamefont {M.}~\bibnamefont {Uchida}}, \bibinfo {author} {\bibfnamefont
  {Y.}~\bibnamefont {Tokura}},\ and\ \bibinfo {author} {\bibfnamefont
  {M.}~\bibnamefont {Kawasaki}},\ }\bibfield  {title} {\bibinfo {title}
  {{Electric-field control of anomalous and topological Hall effects in oxide
  bilayer thin films}},\ }\href {https://doi.org/10.1038/s41467-017-02629-3}
  {\bibfield  {journal} {\bibinfo  {journal} {Nature Communications}\ }\textbf
  {\bibinfo {volume} {9}},\ \bibinfo {pages} {213} (\bibinfo {year}
  {2018})}\BibitemShut {NoStop}%
\bibitem [{\citenamefont {Meng}\ \emph {et~al.}(2019)\citenamefont {Meng},
  \citenamefont {Ahmed}, \citenamefont {Ba\'{c}ani}, \citenamefont {Mandru},
  \citenamefont {Zhao}, \citenamefont {Bagu\'{e}s}, \citenamefont {Esser},
  \citenamefont {Flores}, \citenamefont {McComb}, \citenamefont {Hug},\ and\
  \citenamefont {Yang}}]{Meng2019}%
  \BibitemOpen
  \bibfield  {author} {\bibinfo {author} {\bibfnamefont {K.-Y.}\ \bibnamefont
  {Meng}}, \bibinfo {author} {\bibfnamefont {A.~S.}\ \bibnamefont {Ahmed}},
  \bibinfo {author} {\bibfnamefont {M.}~\bibnamefont {Ba\'{c}ani}}, \bibinfo
  {author} {\bibfnamefont {A.-O.}\ \bibnamefont {Mandru}}, \bibinfo {author}
  {\bibfnamefont {X.}~\bibnamefont {Zhao}}, \bibinfo {author} {\bibfnamefont
  {N.}~\bibnamefont {Bagu\'{e}s}}, \bibinfo {author} {\bibfnamefont {B.~D.}\
  \bibnamefont {Esser}}, \bibinfo {author} {\bibfnamefont {J.}~\bibnamefont
  {Flores}}, \bibinfo {author} {\bibfnamefont {D.~W.}\ \bibnamefont {McComb}},
  \bibinfo {author} {\bibfnamefont {H.~J.}\ \bibnamefont {Hug}},\ and\ \bibinfo
  {author} {\bibfnamefont {F.}~\bibnamefont {Yang}},\ }\bibfield  {title}
  {\bibinfo {title} {{Observation of Nanoscale Skyrmions in SrIrO$_3$/SrRuO$_3$
  Bilayers}},\ }\href {https://doi.org/10.1021/acs.nanolett.9b00596} {\bibfield
   {journal} {\bibinfo  {journal} {Nano Letters}\ }\textbf {\bibinfo {volume}
  {19}},\ \bibinfo {pages} {3169} (\bibinfo {year} {2019})}\BibitemShut
  {NoStop}%
\bibitem [{\citenamefont {Nandy}\ \emph {et~al.}(2016)\citenamefont {Nandy},
  \citenamefont {Kiselev},\ and\ \citenamefont {Bl{\"{u}}gel}}]{Nandy2016}%
  \BibitemOpen
  \bibfield  {author} {\bibinfo {author} {\bibfnamefont {A.~K.}\ \bibnamefont
  {Nandy}}, \bibinfo {author} {\bibfnamefont {N.~S.}\ \bibnamefont {Kiselev}},\
  and\ \bibinfo {author} {\bibfnamefont {S.}~\bibnamefont {Bl{\"{u}}gel}},\
  }\bibfield  {title} {\bibinfo {title} {{Interlayer exchange coupling: A
  general scheme turning chiral magnets into magnetic multilayers carrying
  atomic-scale skyrmions}},\ }\href
  {https://doi.org/10.1103/PhysRevLett.116.177202} {\bibfield  {journal}
  {\bibinfo  {journal} {Physical Review Letters}\ }\textbf {\bibinfo {volume}
  {116}},\ \bibinfo {pages} {177202} (\bibinfo {year} {2016})}\BibitemShut
  {NoStop}%
\bibitem [{\citenamefont {Moreau-Luchaire}\ \emph {et~al.}(2016)\citenamefont
  {Moreau-Luchaire}, \citenamefont {Moutafis}, \citenamefont {Reyren},
  \citenamefont {Sampaio}, \citenamefont {Vaz}, \citenamefont {{Van Horne}},
  \citenamefont {Bouzehouane}, \citenamefont {Garcia}, \citenamefont
  {Deranlot}, \citenamefont {Warnicke}, \citenamefont {Wohlh{\"{u}}ter},
  \citenamefont {George}, \citenamefont {Weigand}, \citenamefont {Raabe},
  \citenamefont {Cros},\ and\ \citenamefont {Fert}}]{Moreau-Luchaire2016}%
  \BibitemOpen
  \bibfield  {author} {\bibinfo {author} {\bibfnamefont {C.}~\bibnamefont
  {Moreau-Luchaire}}, \bibinfo {author} {\bibfnamefont {C.}~\bibnamefont
  {Moutafis}}, \bibinfo {author} {\bibfnamefont {N.}~\bibnamefont {Reyren}},
  \bibinfo {author} {\bibfnamefont {J.}~\bibnamefont {Sampaio}}, \bibinfo
  {author} {\bibfnamefont {C.~A.}\ \bibnamefont {Vaz}}, \bibinfo {author}
  {\bibfnamefont {N.}~\bibnamefont {{Van Horne}}}, \bibinfo {author}
  {\bibfnamefont {K.}~\bibnamefont {Bouzehouane}}, \bibinfo {author}
  {\bibfnamefont {K.}~\bibnamefont {Garcia}}, \bibinfo {author} {\bibfnamefont
  {C.}~\bibnamefont {Deranlot}}, \bibinfo {author} {\bibfnamefont
  {P.}~\bibnamefont {Warnicke}}, \bibinfo {author} {\bibfnamefont
  {P.}~\bibnamefont {Wohlh{\"{u}}ter}}, \bibinfo {author} {\bibfnamefont
  {J.~M.}\ \bibnamefont {George}}, \bibinfo {author} {\bibfnamefont
  {M.}~\bibnamefont {Weigand}}, \bibinfo {author} {\bibfnamefont
  {J.}~\bibnamefont {Raabe}}, \bibinfo {author} {\bibfnamefont
  {V.}~\bibnamefont {Cros}},\ and\ \bibinfo {author} {\bibfnamefont
  {A.}~\bibnamefont {Fert}},\ }\bibfield  {title} {\bibinfo {title} {{Additive
  interfacial chiral interaction in multilayers for stabilization of small
  individual skyrmions at room temperature}},\ }\href
  {https://doi.org/10.1038/nnano.2015.313} {\bibfield  {journal} {\bibinfo
  {journal} {Nature Nanotechnology}\ }\textbf {\bibinfo {volume} {11}},\
  \bibinfo {pages} {444} (\bibinfo {year} {2016})}\BibitemShut {NoStop}%
\bibitem [{\citenamefont {Pollard}\ \emph {et~al.}(2017)\citenamefont
  {Pollard}, \citenamefont {Garlow}, \citenamefont {Yu}, \citenamefont {Wang},
  \citenamefont {Zhu},\ and\ \citenamefont {Yang}}]{Pollard2017}%
  \BibitemOpen
  \bibfield  {author} {\bibinfo {author} {\bibfnamefont {S.~D.}\ \bibnamefont
  {Pollard}}, \bibinfo {author} {\bibfnamefont {J.~A.}\ \bibnamefont {Garlow}},
  \bibinfo {author} {\bibfnamefont {J.}~\bibnamefont {Yu}}, \bibinfo {author}
  {\bibfnamefont {Z.}~\bibnamefont {Wang}}, \bibinfo {author} {\bibfnamefont
  {Y.}~\bibnamefont {Zhu}},\ and\ \bibinfo {author} {\bibfnamefont
  {H.}~\bibnamefont {Yang}},\ }\bibfield  {title} {\bibinfo {title}
  {{Observation of stable N{\'{e}}el skyrmions in cobalt/palladium multilayers
  with Lorentz transmission electron microscopy}},\ }\href
  {https://doi.org/10.1038/ncomms14761} {\bibfield  {journal} {\bibinfo
  {journal} {Nature Communications}\ }\textbf {\bibinfo {volume} {8}},\
  \bibinfo {pages} {14761} (\bibinfo {year} {2017})}\BibitemShut {NoStop}%
\bibitem [{\citenamefont {Groenendijk}\ \emph
  {et~al.}(2020{\natexlab{a}})\citenamefont {Groenendijk}, \citenamefont
  {Autieri}, \citenamefont {van Thiel}, \citenamefont {Brzezicki},
  \citenamefont {Hortensius}, \citenamefont {Afanasiev}, \citenamefont
  {Gauquelin}, \citenamefont {Barone}, \citenamefont {van~den Bos},
  \citenamefont {van Aert}, \citenamefont {Verbeeck}, \citenamefont
  {Filippetti}, \citenamefont {Picozzi}, \citenamefont {Cuoco},\ and\
  \citenamefont {Caviglia}}]{Groenendijk}%
  \BibitemOpen
  \bibfield  {author} {\bibinfo {author} {\bibfnamefont {D.~J.}\ \bibnamefont
  {Groenendijk}}, \bibinfo {author} {\bibfnamefont {C.}~\bibnamefont
  {Autieri}}, \bibinfo {author} {\bibfnamefont {T.~C.}\ \bibnamefont {van
  Thiel}}, \bibinfo {author} {\bibfnamefont {W.}~\bibnamefont {Brzezicki}},
  \bibinfo {author} {\bibfnamefont {J.~R.}\ \bibnamefont {Hortensius}},
  \bibinfo {author} {\bibfnamefont {D.}~\bibnamefont {Afanasiev}}, \bibinfo
  {author} {\bibfnamefont {N.}~\bibnamefont {Gauquelin}}, \bibinfo {author}
  {\bibfnamefont {P.}~\bibnamefont {Barone}}, \bibinfo {author} {\bibfnamefont
  {K.~H.~W.}\ \bibnamefont {van~den Bos}}, \bibinfo {author} {\bibfnamefont
  {S.}~\bibnamefont {van Aert}}, \bibinfo {author} {\bibfnamefont
  {J.}~\bibnamefont {Verbeeck}}, \bibinfo {author} {\bibfnamefont
  {A.}~\bibnamefont {Filippetti}}, \bibinfo {author} {\bibfnamefont
  {S.}~\bibnamefont {Picozzi}}, \bibinfo {author} {\bibfnamefont
  {M.}~\bibnamefont {Cuoco}},\ and\ \bibinfo {author} {\bibfnamefont {A.~D.}\
  \bibnamefont {Caviglia}},\ }\bibfield  {title} {\bibinfo {title} {Berry phase
  engineering at oxide interfaces},\ }\href
  {https://doi.org/10.1103/PhysRevResearch.2.023404} {\bibfield  {journal}
  {\bibinfo  {journal} {Physical Review Research}\ }\textbf {\bibinfo {volume}
  {2}},\ \bibinfo {pages} {023404} (\bibinfo {year}
  {2020}{\natexlab{a}})}\BibitemShut {NoStop}%
\bibitem [{\citenamefont {Wysocki}\ \emph
  {et~al.}(2020{\natexlab{a}})\citenamefont {Wysocki}, \citenamefont
  {Sch{\"{o}}pf}, \citenamefont {Ziese}, \citenamefont {Yang}, \citenamefont
  {Kov{\'{a}}cs}, \citenamefont {Jin}, \citenamefont {Versteeg}, \citenamefont
  {Bliesener}, \citenamefont {Gunkel}, \citenamefont {Kornblum}, \citenamefont
  {Dittmann}, \citenamefont {van Loosdrecht},\ and\ \citenamefont
  {Lindfors-Vrejoiu}}]{Wysocki2020}%
  \BibitemOpen
  \bibfield  {author} {\bibinfo {author} {\bibfnamefont {L.}~\bibnamefont
  {Wysocki}}, \bibinfo {author} {\bibfnamefont {J.}~\bibnamefont
  {Sch{\"{o}}pf}}, \bibinfo {author} {\bibfnamefont {M.}~\bibnamefont {Ziese}},
  \bibinfo {author} {\bibfnamefont {L.}~\bibnamefont {Yang}}, \bibinfo {author}
  {\bibfnamefont {A.}~\bibnamefont {Kov{\'{a}}cs}}, \bibinfo {author}
  {\bibfnamefont {L.}~\bibnamefont {Jin}}, \bibinfo {author} {\bibfnamefont
  {R.~B.}\ \bibnamefont {Versteeg}}, \bibinfo {author} {\bibfnamefont
  {A.}~\bibnamefont {Bliesener}}, \bibinfo {author} {\bibfnamefont
  {F.}~\bibnamefont {Gunkel}}, \bibinfo {author} {\bibfnamefont
  {L.}~\bibnamefont {Kornblum}}, \bibinfo {author} {\bibfnamefont
  {R.}~\bibnamefont {Dittmann}}, \bibinfo {author} {\bibfnamefont {P.~H.~M.}\
  \bibnamefont {van Loosdrecht}},\ and\ \bibinfo {author} {\bibfnamefont
  {I.}~\bibnamefont {Lindfors-Vrejoiu}},\ }\bibfield  {title} {\bibinfo {title}
  {{Electronic inhomogeneity influence on the anomalous Hall resistivity loops
  of SrRuO$_3$ epitaxially interfaced with 5\textit{d} perovskites}},\
  }\href@noop {} {\bibfield  {journal} {\bibinfo  {journal} {ACS Omega}\
  }\textbf {\bibinfo {volume} {5}},\ \bibinfo {pages} {5824} (\bibinfo {year}
  {2020}{\natexlab{a}})}\BibitemShut {NoStop}%
\bibitem [{\citenamefont {Kan}\ and\ \citenamefont
  {Shimakawa}(2018)}]{Kan2018}%
  \BibitemOpen
  \bibfield  {author} {\bibinfo {author} {\bibfnamefont {D.}~\bibnamefont
  {Kan}}\ and\ \bibinfo {author} {\bibfnamefont {Y.}~\bibnamefont
  {Shimakawa}},\ }\bibfield  {title} {\bibinfo {title} {{Defect-induced
  anomalous transverse resistivity in an itinerant ferromagnetic oxide}},\
  }\href {https://doi.org/10.1002/pssb.201800175} {\bibfield  {journal}
  {\bibinfo  {journal} {physica status solidi (b)}\ }\textbf {\bibinfo {volume}
  {255}},\ \bibinfo {pages} {1800175} (\bibinfo {year} {2018})}\BibitemShut
  {NoStop}%
\bibitem [{\citenamefont {Kan}\ \emph {et~al.}(2018)\citenamefont {Kan},
  \citenamefont {Moriyama}, \citenamefont {Kobayashi},\ and\ \citenamefont
  {Shimakawa}}]{Kan2018b}%
  \BibitemOpen
  \bibfield  {author} {\bibinfo {author} {\bibfnamefont {D.}~\bibnamefont
  {Kan}}, \bibinfo {author} {\bibfnamefont {T.}~\bibnamefont {Moriyama}},
  \bibinfo {author} {\bibfnamefont {K.}~\bibnamefont {Kobayashi}},\ and\
  \bibinfo {author} {\bibfnamefont {Y.}~\bibnamefont {Shimakawa}},\ }\bibfield
  {title} {\bibinfo {title} {{Alternative to the topological interpretation of
  the transverse resistivity anomalies in SrRuO$_3$}},\ }\href
  {https://doi.org/10.1103/PhysRevB.98.180408} {\bibfield  {journal} {\bibinfo
  {journal} {Physical Review B}\ }\textbf {\bibinfo {volume} {98}},\ \bibinfo
  {pages} {180408(R)} (\bibinfo {year} {2018})}\BibitemShut {NoStop}%
\bibitem [{\citenamefont {Kan}\ \emph {et~al.}(2020)\citenamefont {Kan},
  \citenamefont {Moriyama},\ and\ \citenamefont {Shimakawa}}]{Kan2020}%
  \BibitemOpen
  \bibfield  {author} {\bibinfo {author} {\bibfnamefont {D.}~\bibnamefont
  {Kan}}, \bibinfo {author} {\bibfnamefont {T.}~\bibnamefont {Moriyama}},\ and\
  \bibinfo {author} {\bibfnamefont {Y.}~\bibnamefont {Shimakawa}},\ }\bibfield
  {title} {\bibinfo {title} {{Field-sweep-rate and time dependence of
  transverse resistivity anomalies in ultrathin SrRuO$_3$ films}},\ }\href
  {https://doi.org/10.1103/PhysRevB.101.014448} {\bibfield  {journal} {\bibinfo
   {journal} {Physical Review B}\ }\textbf {\bibinfo {volume} {101}},\ \bibinfo
  {pages} {014448} (\bibinfo {year} {2020})}\BibitemShut {NoStop}%
\bibitem [{\citenamefont {Qin}\ \emph {et~al.}(2019)\citenamefont {Qin},
  \citenamefont {Liu}, \citenamefont {Lin}, \citenamefont {Shu}, \citenamefont
  {Xie}, \citenamefont {Lim}, \citenamefont {Li}, \citenamefont {He},
  \citenamefont {Chow},\ and\ \citenamefont {Chen}}]{Qin2019}%
  \BibitemOpen
  \bibfield  {author} {\bibinfo {author} {\bibfnamefont {Q.}~\bibnamefont
  {Qin}}, \bibinfo {author} {\bibfnamefont {L.}~\bibnamefont {Liu}}, \bibinfo
  {author} {\bibfnamefont {W.}~\bibnamefont {Lin}}, \bibinfo {author}
  {\bibfnamefont {X.}~\bibnamefont {Shu}}, \bibinfo {author} {\bibfnamefont
  {Q.}~\bibnamefont {Xie}}, \bibinfo {author} {\bibfnamefont {Z.}~\bibnamefont
  {Lim}}, \bibinfo {author} {\bibfnamefont {C.}~\bibnamefont {Li}}, \bibinfo
  {author} {\bibfnamefont {S.}~\bibnamefont {He}}, \bibinfo {author}
  {\bibfnamefont {G.~M.}\ \bibnamefont {Chow}},\ and\ \bibinfo {author}
  {\bibfnamefont {J.}~\bibnamefont {Chen}},\ }\bibfield  {title} {\bibinfo
  {title} {{Emergence of topological Hall effect in a SrRuO$_3$ single
  layer}},\ }\href {https://doi.org/10.1002/adma.201807008} {\bibfield
  {journal} {\bibinfo  {journal} {Advanced Materials}\ }\textbf {\bibinfo
  {volume} {31}},\ \bibinfo {pages} {1807008} (\bibinfo {year}
  {2019})}\BibitemShut {NoStop}%
\bibitem [{\citenamefont {Wu}\ \emph {et~al.}(2020)\citenamefont {Wu},
  \citenamefont {Wen}, \citenamefont {Fu}, \citenamefont {Wilson},
  \citenamefont {Liu}, \citenamefont {Zhang}, \citenamefont {Vasiukov},
  \citenamefont {Kareev}, \citenamefont {Pixley},\ and\ \citenamefont
  {Chakhalian}}]{Wu2019}%
  \BibitemOpen
  \bibfield  {author} {\bibinfo {author} {\bibfnamefont {L.}~\bibnamefont
  {Wu}}, \bibinfo {author} {\bibfnamefont {F.}~\bibnamefont {Wen}}, \bibinfo
  {author} {\bibfnamefont {Y.}~\bibnamefont {Fu}}, \bibinfo {author}
  {\bibfnamefont {J.~H.}\ \bibnamefont {Wilson}}, \bibinfo {author}
  {\bibfnamefont {X.}~\bibnamefont {Liu}}, \bibinfo {author} {\bibfnamefont
  {Y.}~\bibnamefont {Zhang}}, \bibinfo {author} {\bibfnamefont {D.~M.}\
  \bibnamefont {Vasiukov}}, \bibinfo {author} {\bibfnamefont {M.~S.}\
  \bibnamefont {Kareev}}, \bibinfo {author} {\bibfnamefont {J.~H.}\
  \bibnamefont {Pixley}},\ and\ \bibinfo {author} {\bibfnamefont
  {J.}~\bibnamefont {Chakhalian}},\ }\bibfield  {title} {\bibinfo {title}
  {{Berry phase manipulation in ultrathin ${\mathrm{SrRuO}}_{3}$ films}},\
  }\href {https://doi.org/10.1103/PhysRevB.102.220406} {\bibfield  {journal}
  {\bibinfo  {journal} {Physical Review B}\ }\textbf {\bibinfo {volume}
  {102}},\ \bibinfo {pages} {220406(R)} (\bibinfo {year} {2020})}\BibitemShut
  {NoStop}%
\bibitem [{\citenamefont {Yang}\ \emph
  {et~al.}(2021{\natexlab{a}})\citenamefont {Yang}, \citenamefont {Wysocki},
  \citenamefont {Sch\"opf}, \citenamefont {Jin}, \citenamefont {Kov\'acs},
  \citenamefont {Gunkel}, \citenamefont {Dittmann}, \citenamefont {van
  Loosdrecht},\ and\ \citenamefont {Lindfors-Vrejoiu}}]{Yang2020}%
  \BibitemOpen
  \bibfield  {author} {\bibinfo {author} {\bibfnamefont {L.}~\bibnamefont
  {Yang}}, \bibinfo {author} {\bibfnamefont {L.}~\bibnamefont {Wysocki}},
  \bibinfo {author} {\bibfnamefont {J.}~\bibnamefont {Sch\"opf}}, \bibinfo
  {author} {\bibfnamefont {L.}~\bibnamefont {Jin}}, \bibinfo {author}
  {\bibfnamefont {A.}~\bibnamefont {Kov\'acs}}, \bibinfo {author}
  {\bibfnamefont {F.}~\bibnamefont {Gunkel}}, \bibinfo {author} {\bibfnamefont
  {R.}~\bibnamefont {Dittmann}}, \bibinfo {author} {\bibfnamefont {P.~H.~M.}\
  \bibnamefont {van Loosdrecht}},\ and\ \bibinfo {author} {\bibfnamefont
  {I.}~\bibnamefont {Lindfors-Vrejoiu}},\ }\bibfield  {title} {\bibinfo {title}
  {{Origin of the hump anomalies in the Hall resistance loops of ultrathin
  SrRuO$_3$/ SrIrO$_3$ multilayers}},\ }\href
  {https://doi.org/10.1103/PhysRevMaterials.5.014403} {\bibfield  {journal}
  {\bibinfo  {journal} {Physical Review Materials}\ }\textbf {\bibinfo {volume}
  {5}},\ \bibinfo {pages} {014403} (\bibinfo {year}
  {2021}{\natexlab{a}})}\BibitemShut {NoStop}%
\bibitem [{\citenamefont {Wysocki}\ \emph
  {et~al.}(2020{\natexlab{b}})\citenamefont {Wysocki}, \citenamefont {Yang},
  \citenamefont {Gunkel}, \citenamefont {Dittmann}, \citenamefont {van
  Loosdrecht},\ and\ \citenamefont {Lindfors-Vrejoiu}}]{Wysocki2020b}%
  \BibitemOpen
  \bibfield  {author} {\bibinfo {author} {\bibfnamefont {L.}~\bibnamefont
  {Wysocki}}, \bibinfo {author} {\bibfnamefont {L.}~\bibnamefont {Yang}},
  \bibinfo {author} {\bibfnamefont {F.}~\bibnamefont {Gunkel}}, \bibinfo
  {author} {\bibfnamefont {R.}~\bibnamefont {Dittmann}}, \bibinfo {author}
  {\bibfnamefont {P.~H.~M.}\ \bibnamefont {van Loosdrecht}},\ and\ \bibinfo
  {author} {\bibfnamefont {I.}~\bibnamefont {Lindfors-Vrejoiu}},\ }\bibfield
  {title} {\bibinfo {title} {{Validity of magnetotransport detection of
  skyrmions in epitaxial SrRuO$_3$ heterostructures}},\ }\href
  {https://doi.org/10.1103/PhysRevMaterials.4.054402} {\bibfield  {journal}
  {\bibinfo  {journal} {Physical Review Materials}\ }\textbf {\bibinfo {volume}
  {4}},\ \bibinfo {pages} {054402} (\bibinfo {year}
  {2020}{\natexlab{b}})}\BibitemShut {NoStop}%
\bibitem [{\citenamefont {Yang}\ \emph
  {et~al.}(2021{\natexlab{b}})\citenamefont {Yang}, \citenamefont {Jin},
  \citenamefont {Wysocki}, \citenamefont {Sch\"opf}, \citenamefont {Jansen},
  \citenamefont {Das}, \citenamefont {Kornblum}, \citenamefont {van
  Loosdrecht},\ and\ \citenamefont {Lindfors-Vrejoiu}}]{Yang2021}%
  \BibitemOpen
  \bibfield  {author} {\bibinfo {author} {\bibfnamefont {L.}~\bibnamefont
  {Yang}}, \bibinfo {author} {\bibfnamefont {L.}~\bibnamefont {Jin}}, \bibinfo
  {author} {\bibfnamefont {L.}~\bibnamefont {Wysocki}}, \bibinfo {author}
  {\bibfnamefont {J.}~\bibnamefont {Sch\"opf}}, \bibinfo {author}
  {\bibfnamefont {D.}~\bibnamefont {Jansen}}, \bibinfo {author} {\bibfnamefont
  {B.}~\bibnamefont {Das}}, \bibinfo {author} {\bibfnamefont {L.}~\bibnamefont
  {Kornblum}}, \bibinfo {author} {\bibfnamefont {P.~H.~M.}\ \bibnamefont {van
  Loosdrecht}},\ and\ \bibinfo {author} {\bibfnamefont {I.}~\bibnamefont
  {Lindfors-Vrejoiu}},\ }\bibfield  {title} {\bibinfo {title} {{Enhancing the
  ferromagnetic interlayer coupling between epitaxial SrRuO$_3$ layers}},\
  }\href {https://doi.org/10.1103/PhysRevB.104.064444} {\bibfield  {journal}
  {\bibinfo  {journal} {Physical Review B}\ }\textbf {\bibinfo {volume}
  {104}},\ \bibinfo {pages} {064444} (\bibinfo {year}
  {2021}{\natexlab{b}})}\BibitemShut {NoStop}%
\bibitem [{\citenamefont {Herranz}\ \emph {et~al.}(2003)\citenamefont
  {Herranz}, \citenamefont {Mart\'{i}nez}, \citenamefont {Fontcuberta},
  \citenamefont {S\'{a}nchez}, \citenamefont {Garc\'{i}a-Cuenca}, \citenamefont
  {Ferrater},\ and\ \citenamefont {Varela}}]{Herranz2003}%
  \BibitemOpen
  \bibfield  {author} {\bibinfo {author} {\bibfnamefont {G.}~\bibnamefont
  {Herranz}}, \bibinfo {author} {\bibfnamefont {B.}~\bibnamefont
  {Mart\'{i}nez}}, \bibinfo {author} {\bibfnamefont {J.}~\bibnamefont
  {Fontcuberta}}, \bibinfo {author} {\bibfnamefont {F.}~\bibnamefont
  {S\'{a}nchez}}, \bibinfo {author} {\bibfnamefont {M.~V.}\ \bibnamefont
  {Garc\'{i}a-Cuenca}}, \bibinfo {author} {\bibfnamefont {C.}~\bibnamefont
  {Ferrater}},\ and\ \bibinfo {author} {\bibfnamefont {M.}~\bibnamefont
  {Varela}},\ }\bibfield  {title} {\bibinfo {title} {{SrRuO$_3$/ SrTiO$_3$/
  SrRuO$_3$ heterostructures for magnetic tunnel junctions}},\ }\href
  {https://doi.org/10.1063/1.1555372} {\bibfield  {journal} {\bibinfo
  {journal} {Journal of Applied Physics}\ }\textbf {\bibinfo {volume} {93}},\
  \bibinfo {pages} {8035} (\bibinfo {year} {2003})}\BibitemShut {NoStop}%
\bibitem [{\citenamefont {Wysocki}\ \emph {et~al.}(2018)\citenamefont
  {Wysocki}, \citenamefont {Mirzaaghayev}, \citenamefont {Ziese}, \citenamefont
  {Yang}, \citenamefont {Sch{\"{o}}pf}, \citenamefont {Versteeg}, \citenamefont
  {Bliesener}, \citenamefont {Engelmayer}, \citenamefont {Kov{\'{a}}cs},
  \citenamefont {Jin}, \citenamefont {Gunkel}, \citenamefont {Dittmann},
  \citenamefont {{van Loosdrecht}},\ and\ \citenamefont
  {Lindfors-Vrejoiu}}]{Wysocki2018}%
  \BibitemOpen
  \bibfield  {author} {\bibinfo {author} {\bibfnamefont {L.}~\bibnamefont
  {Wysocki}}, \bibinfo {author} {\bibfnamefont {R.}~\bibnamefont
  {Mirzaaghayev}}, \bibinfo {author} {\bibfnamefont {M.}~\bibnamefont {Ziese}},
  \bibinfo {author} {\bibfnamefont {L.}~\bibnamefont {Yang}}, \bibinfo {author}
  {\bibfnamefont {J.}~\bibnamefont {Sch{\"{o}}pf}}, \bibinfo {author}
  {\bibfnamefont {R.~B.}\ \bibnamefont {Versteeg}}, \bibinfo {author}
  {\bibfnamefont {A.}~\bibnamefont {Bliesener}}, \bibinfo {author}
  {\bibfnamefont {J.}~\bibnamefont {Engelmayer}}, \bibinfo {author}
  {\bibfnamefont {A.}~\bibnamefont {Kov{\'{a}}cs}}, \bibinfo {author}
  {\bibfnamefont {L.}~\bibnamefont {Jin}}, \bibinfo {author} {\bibfnamefont
  {F.}~\bibnamefont {Gunkel}}, \bibinfo {author} {\bibfnamefont
  {R.}~\bibnamefont {Dittmann}}, \bibinfo {author} {\bibfnamefont {P.~H.~M.}\
  \bibnamefont {{van Loosdrecht}}},\ and\ \bibinfo {author} {\bibfnamefont
  {I.}~\bibnamefont {Lindfors-Vrejoiu}},\ }\bibfield  {title} {\bibinfo {title}
  {{Magnetic coupling of ferromagnetic SrRuO$_3$ epitaxial layers separated by
  ultrathin non-magnetic SrZrO$_3$/ SrIrO$_3$}},\ }\href
  {https://doi.org/10.1063/1.5050346} {\bibfield  {journal} {\bibinfo
  {journal} {Applied Physics Letters}\ }\textbf {\bibinfo {volume} {113}},\
  \bibinfo {pages} {192402} (\bibinfo {year} {2018})}\BibitemShut {NoStop}%
\bibitem [{\citenamefont {Esser}\ \emph {et~al.}(2021)\citenamefont {Esser},
  \citenamefont {Wu}, \citenamefont {Esser}, \citenamefont {Gruhl},
  \citenamefont {Jesche}, \citenamefont {Roddatis}, \citenamefont {Moshnyaga},
  \citenamefont {Pentcheva},\ and\ \citenamefont {Gegenwart}}]{Esser2021}%
  \BibitemOpen
  \bibfield  {author} {\bibinfo {author} {\bibfnamefont {S.}~\bibnamefont
  {Esser}}, \bibinfo {author} {\bibfnamefont {J.}~\bibnamefont {Wu}}, \bibinfo
  {author} {\bibfnamefont {S.}~\bibnamefont {Esser}}, \bibinfo {author}
  {\bibfnamefont {R.}~\bibnamefont {Gruhl}}, \bibinfo {author} {\bibfnamefont
  {A.}~\bibnamefont {Jesche}}, \bibinfo {author} {\bibfnamefont
  {V.}~\bibnamefont {Roddatis}}, \bibinfo {author} {\bibfnamefont
  {V.}~\bibnamefont {Moshnyaga}}, \bibinfo {author} {\bibfnamefont
  {R.}~\bibnamefont {Pentcheva}},\ and\ \bibinfo {author} {\bibfnamefont
  {P.}~\bibnamefont {Gegenwart}},\ }\bibfield  {title} {\bibinfo {title}
  {{Angular dependence of Hall effect and magnetoresistance in
  SrRuO$_3$-SrIrO$_3$ heterostructures}},\ }\href
  {https://doi.org/10.1103/PhysRevB.103.214430} {\bibfield  {journal} {\bibinfo
   {journal} {Physical Review B}\ }\textbf {\bibinfo {volume} {103}},\ \bibinfo
  {pages} {214430} (\bibinfo {year} {2021})}\BibitemShut {NoStop}%
\bibitem [{\citenamefont {Gro{{\ss}}}\ \emph
  {et~al.}(2019{\natexlab{a}})\citenamefont {Gro{{\ss}}}, \citenamefont {Ilse},
  \citenamefont {Sch{\"{u}}tz}, \citenamefont {Gr{\"{a}}fe},\ and\
  \citenamefont {Goering}}]{Gross2019b}%
  \BibitemOpen
  \bibfield  {author} {\bibinfo {author} {\bibfnamefont {F.}~\bibnamefont
  {Gro{{\ss}}}}, \bibinfo {author} {\bibfnamefont {S.~E.}\ \bibnamefont
  {Ilse}}, \bibinfo {author} {\bibfnamefont {G.}~\bibnamefont {Sch{\"{u}}tz}},
  \bibinfo {author} {\bibfnamefont {J.}~\bibnamefont {Gr{\"{a}}fe}},\ and\
  \bibinfo {author} {\bibfnamefont {E.}~\bibnamefont {Goering}},\ }\bibfield
  {title} {\bibinfo {title} {{Interpreting first-order reversal curves beyond
  the Preisach model: An experimental permalloy microarray investigation}},\
  }\href@noop {} {\bibfield  {journal} {\bibinfo  {journal} {Physical Review
  B}\ }\textbf {\bibinfo {volume} {99}},\ \bibinfo {pages} {064401} (\bibinfo
  {year} {2019}{\natexlab{a}})}\BibitemShut {NoStop}%
\bibitem [{\citenamefont {B{\`{e}}ron}\ \emph {et~al.}(2006)\citenamefont
  {B{\`{e}}ron}, \citenamefont {Clime}, \citenamefont {Ciureanu}, \citenamefont
  {Menard}, \citenamefont {Cochrane},\ and\ \citenamefont {Yelon}}]{Beron2006}%
  \BibitemOpen
  \bibfield  {author} {\bibinfo {author} {\bibfnamefont {F.}~\bibnamefont
  {B{\`{e}}ron}}, \bibinfo {author} {\bibfnamefont {L.}~\bibnamefont {Clime}},
  \bibinfo {author} {\bibfnamefont {M.}~\bibnamefont {Ciureanu}}, \bibinfo
  {author} {\bibfnamefont {D.}~\bibnamefont {Menard}}, \bibinfo {author}
  {\bibfnamefont {R.}~\bibnamefont {Cochrane}},\ and\ \bibinfo {author}
  {\bibfnamefont {A.}~\bibnamefont {Yelon}},\ }\bibfield  {title} {\bibinfo
  {title} {First-order reversal curves diagrams of ferromagnetic soft nanowire
  arrays},\ }\href {https://doi.org/10.1109/TMAG.2006.880147} {\bibfield
  {journal} {\bibinfo  {journal} {IEEE Transactions on Magnetics}\ }\textbf
  {\bibinfo {volume} {42}},\ \bibinfo {pages} {3060} (\bibinfo {year}
  {2006})}\BibitemShut {NoStop}%
\bibitem [{\citenamefont {Gr{\"{a}}fe}\ \emph
  {et~al.}(2016{\natexlab{a}})\citenamefont {Gr{\"{a}}fe}, \citenamefont
  {Weigand}, \citenamefont {Stahl}, \citenamefont {Tr{\"{a}}ger}, \citenamefont
  {Kopp}, \citenamefont {Sch{\"{u}}tz}, \citenamefont {Goering}, \citenamefont
  {Haering}, \citenamefont {Ziemann},\ and\ \citenamefont
  {Wiedwald}}]{Graefe2016}%
  \BibitemOpen
  \bibfield  {author} {\bibinfo {author} {\bibfnamefont {J.}~\bibnamefont
  {Gr{\"{a}}fe}}, \bibinfo {author} {\bibfnamefont {M.}~\bibnamefont
  {Weigand}}, \bibinfo {author} {\bibfnamefont {C.}~\bibnamefont {Stahl}},
  \bibinfo {author} {\bibfnamefont {N.}~\bibnamefont {Tr{\"{a}}ger}}, \bibinfo
  {author} {\bibfnamefont {M.}~\bibnamefont {Kopp}}, \bibinfo {author}
  {\bibfnamefont {G.}~\bibnamefont {Sch{\"{u}}tz}}, \bibinfo {author}
  {\bibfnamefont {E.~J.}\ \bibnamefont {Goering}}, \bibinfo {author}
  {\bibfnamefont {F.}~\bibnamefont {Haering}}, \bibinfo {author} {\bibfnamefont
  {P.}~\bibnamefont {Ziemann}},\ and\ \bibinfo {author} {\bibfnamefont
  {U.}~\bibnamefont {Wiedwald}},\ }\bibfield  {title} {\bibinfo {title}
  {{Combined first-order reversal curve and x-ray microscopy investigation of
  magnetization reversal mechanisms in hexagonal antidot lattices}},\
  }\href@noop {} {\bibfield  {journal} {\bibinfo  {journal} {{Physical Review
  B}}\ }\textbf {\bibinfo {volume} {93}},\ \bibinfo {pages} {014406} (\bibinfo
  {year} {2016}{\natexlab{a}})}\BibitemShut {NoStop}%
\bibitem [{\citenamefont {Gr{\"{a}}fe}\ \emph
  {et~al.}(2016{\natexlab{b}})\citenamefont {Gr{\"{a}}fe}, \citenamefont
  {Weigand}, \citenamefont {T{\"{a}}ger}, \citenamefont {Sch{\"{u}}tz},
  \citenamefont {Goering}, \citenamefont {Skripnik}, \citenamefont {Nowak},
  \citenamefont {Haering}, \citenamefont {Ziemann},\ and\ \citenamefont
  {Wiedwald}}]{Graefe2016b}%
  \BibitemOpen
  \bibfield  {author} {\bibinfo {author} {\bibfnamefont {J.}~\bibnamefont
  {Gr{\"{a}}fe}}, \bibinfo {author} {\bibfnamefont {M.}~\bibnamefont
  {Weigand}}, \bibinfo {author} {\bibfnamefont {N.}~\bibnamefont
  {T{\"{a}}ger}}, \bibinfo {author} {\bibfnamefont {G.}~\bibnamefont
  {Sch{\"{u}}tz}}, \bibinfo {author} {\bibfnamefont {E.~J.}\ \bibnamefont
  {Goering}}, \bibinfo {author} {\bibfnamefont {M.}~\bibnamefont {Skripnik}},
  \bibinfo {author} {\bibfnamefont {U.}~\bibnamefont {Nowak}}, \bibinfo
  {author} {\bibfnamefont {F.}~\bibnamefont {Haering}}, \bibinfo {author}
  {\bibfnamefont {P.}~\bibnamefont {Ziemann}},\ and\ \bibinfo {author}
  {\bibfnamefont {U.}~\bibnamefont {Wiedwald}},\ }\bibfield  {title} {\bibinfo
  {title} {{Geometric control of the magnetization reversal in antidot lattices
  with perpendicular magnetic anisotropy}},\ }\href@noop {} {\bibfield
  {journal} {\bibinfo  {journal} {{Physical Review B}}\ }\textbf {\bibinfo
  {volume} {93}},\ \bibinfo {pages} {104421} (\bibinfo {year}
  {2016}{\natexlab{b}})}\BibitemShut {NoStop}%
\bibitem [{\citenamefont {Dobrot\u{a}}\ and\ \citenamefont
  {Stancu}(2013)}]{Dobrota2013}%
  \BibitemOpen
  \bibfield  {author} {\bibinfo {author} {\bibfnamefont {C.-I.}\ \bibnamefont
  {Dobrot\u{a}}}\ and\ \bibinfo {author} {\bibfnamefont {A.}~\bibnamefont
  {Stancu}},\ }\bibfield  {title} {\bibinfo {title} {{What does a first-order
  reversal curve diagram really mean? A study case: Array of ferromagnetic
  nanowires}},\ }\href {https://doi.org/10.1063/1.4789613} {\bibfield
  {journal} {\bibinfo  {journal} {Journal of Applied Physics}\ }\textbf
  {\bibinfo {volume} {113}},\ \bibinfo {pages} {043928} (\bibinfo {year}
  {2013})}\BibitemShut {NoStop}%
\bibitem [{\citenamefont {Ilse}\ \emph {et~al.}(2021)\citenamefont {Ilse},
  \citenamefont {Gro{{\ss}}}, \citenamefont {Sch{\"{u}}tz}, \citenamefont
  {Gr{\"{a}}fe},\ and\ \citenamefont {Goering}}]{Ilse2021}%
  \BibitemOpen
  \bibfield  {author} {\bibinfo {author} {\bibfnamefont {S.~E.}\ \bibnamefont
  {Ilse}}, \bibinfo {author} {\bibfnamefont {F.}~\bibnamefont {Gro{{\ss}}}},
  \bibinfo {author} {\bibfnamefont {G.}~\bibnamefont {Sch{\"{u}}tz}}, \bibinfo
  {author} {\bibfnamefont {J.}~\bibnamefont {Gr{\"{a}}fe}},\ and\ \bibinfo
  {author} {\bibfnamefont {E.}~\bibnamefont {Goering}},\ }\bibfield  {title}
  {\bibinfo {title} {{Understanding the interaction of soft and hard magnetic
  components in NdFeB with first-order reversal curves}},\ }\href
  {https://doi.org/10.1103/PhysRevB.103.024425} {\bibfield  {journal} {\bibinfo
   {journal} {Physical Review B}\ }\textbf {\bibinfo {volume} {103}},\ \bibinfo
  {pages} {024425} (\bibinfo {year} {2021})}\BibitemShut {NoStop}%
\bibitem [{\citenamefont {Muralidhar}\ \emph {et~al.}(2017)\citenamefont
  {Muralidhar}, \citenamefont {Gr{\"{a}}fe}, \citenamefont {Chen},
  \citenamefont {Etter}, \citenamefont {Gregori}, \citenamefont {Ener},
  \citenamefont {Sawatzki}, \citenamefont {Hono}, \citenamefont {Gutfleisch},
  \citenamefont {Kronm{\"{u}}ller}, \citenamefont {Sch\"utz},\ and\
  \citenamefont {Goering}}]{Muralidhar2017}%
  \BibitemOpen
  \bibfield  {author} {\bibinfo {author} {\bibfnamefont {S.}~\bibnamefont
  {Muralidhar}}, \bibinfo {author} {\bibfnamefont {J.}~\bibnamefont
  {Gr{\"{a}}fe}}, \bibinfo {author} {\bibfnamefont {Y.-C.}\ \bibnamefont
  {Chen}}, \bibinfo {author} {\bibfnamefont {M.}~\bibnamefont {Etter}},
  \bibinfo {author} {\bibfnamefont {G.}~\bibnamefont {Gregori}}, \bibinfo
  {author} {\bibfnamefont {S.}~\bibnamefont {Ener}}, \bibinfo {author}
  {\bibfnamefont {S.}~\bibnamefont {Sawatzki}}, \bibinfo {author}
  {\bibfnamefont {K.}~\bibnamefont {Hono}}, \bibinfo {author} {\bibfnamefont
  {O.}~\bibnamefont {Gutfleisch}}, \bibinfo {author} {\bibfnamefont
  {H.}~\bibnamefont {Kronm{\"{u}}ller}}, \bibinfo {author} {\bibfnamefont
  {G.}~\bibnamefont {Sch\"utz}},\ and\ \bibinfo {author} {\bibfnamefont
  {E.~J.}\ \bibnamefont {Goering}},\ }\bibfield  {title} {\bibinfo {title}
  {{Temperature-dependent first-order reversal curve measurements on unusually
  hard magnetic low-temperature phase of MnBi}},\ }\href@noop {} {\bibfield
  {journal} {\bibinfo  {journal} {{Physical Review B}}\ }\textbf {\bibinfo
  {volume} {95}},\ \bibinfo {pages} {024413} (\bibinfo {year}
  {2017})}\BibitemShut {NoStop}%
\bibitem [{\citenamefont {Guo}\ \emph {et~al.}(2020)\citenamefont {Guo},
  \citenamefont {Ji}, \citenamefont {Gu}, \citenamefont {Zhou}, \citenamefont
  {Nie},\ and\ \citenamefont {Pan}}]{Guo2020}%
  \BibitemOpen
  \bibfield  {author} {\bibinfo {author} {\bibfnamefont {W.}~\bibnamefont
  {Guo}}, \bibinfo {author} {\bibfnamefont {D.~X.}\ \bibnamefont {Ji}},
  \bibinfo {author} {\bibfnamefont {Z.~B.}\ \bibnamefont {Gu}}, \bibinfo
  {author} {\bibfnamefont {J.}~\bibnamefont {Zhou}}, \bibinfo {author}
  {\bibfnamefont {Y.~F.}\ \bibnamefont {Nie}},\ and\ \bibinfo {author}
  {\bibfnamefont {X.~Q.}\ \bibnamefont {Pan}},\ }\bibfield  {title} {\bibinfo
  {title} {{Engineering of octahedral rotations and electronic structure in
  ultrathin SrIrO$_3$ films}},\ }\href
  {https://doi.org/10.1103/PhysRevB.101.085101} {\bibfield  {journal} {\bibinfo
   {journal} {Physical Review B}\ }\textbf {\bibinfo {volume} {101}},\ \bibinfo
  {pages} {085101} (\bibinfo {year} {2020})}\BibitemShut {NoStop}%
\bibitem [{\citenamefont {Groenendijk}\ \emph {et~al.}(2017)\citenamefont
  {Groenendijk}, \citenamefont {Autieri}, \citenamefont {Girovsky},
  \citenamefont {Martinez-Velarte}, \citenamefont {Manca}, \citenamefont
  {Mattoni}, \citenamefont {Monteiro}, \citenamefont {Gauquelin}, \citenamefont
  {Verbeeck}, \citenamefont {Otte}, \citenamefont {Gabay}, \citenamefont
  {Picozzi},\ and\ \citenamefont {Caviglia}}]{Groenendijk2017}%
  \BibitemOpen
  \bibfield  {author} {\bibinfo {author} {\bibfnamefont {D.~J.}\ \bibnamefont
  {Groenendijk}}, \bibinfo {author} {\bibfnamefont {C.}~\bibnamefont
  {Autieri}}, \bibinfo {author} {\bibfnamefont {J.}~\bibnamefont {Girovsky}},
  \bibinfo {author} {\bibfnamefont {M.~C.}\ \bibnamefont {Martinez-Velarte}},
  \bibinfo {author} {\bibfnamefont {N.}~\bibnamefont {Manca}}, \bibinfo
  {author} {\bibfnamefont {G.}~\bibnamefont {Mattoni}}, \bibinfo {author}
  {\bibfnamefont {A.~M. R. V.~L.}\ \bibnamefont {Monteiro}}, \bibinfo {author}
  {\bibfnamefont {N.}~\bibnamefont {Gauquelin}}, \bibinfo {author}
  {\bibfnamefont {J.}~\bibnamefont {Verbeeck}}, \bibinfo {author}
  {\bibfnamefont {A.~F.}\ \bibnamefont {Otte}}, \bibinfo {author}
  {\bibfnamefont {M.}~\bibnamefont {Gabay}}, \bibinfo {author} {\bibfnamefont
  {S.}~\bibnamefont {Picozzi}},\ and\ \bibinfo {author} {\bibfnamefont {A.~D.}\
  \bibnamefont {Caviglia}},\ }\bibfield  {title} {\bibinfo {title} {{Spin-Orbit
  Semimetal SrIrO$_{3}$ in the Two-Dimensional Limit}},\ }\href
  {https://doi.org/10.1103/PhysRevLett.119.256403} {\bibfield  {journal}
  {\bibinfo  {journal} {Phys. Rev. Lett.}\ }\textbf {\bibinfo {volume} {119}},\
  \bibinfo {pages} {256403} (\bibinfo {year} {2017})}\BibitemShut {NoStop}%
\bibitem [{\citenamefont {Manca}\ \emph {et~al.}(2018)\citenamefont {Manca},
  \citenamefont {Groenendijk}, \citenamefont {Pallecchi}, \citenamefont
  {Autieri}, \citenamefont {Tang}, \citenamefont {Telesio}, \citenamefont
  {Mattoni}, \citenamefont {Mccollam}, \citenamefont {Picozzi},\ and\
  \citenamefont {Caviglia}}]{Manca2019}%
  \BibitemOpen
  \bibfield  {author} {\bibinfo {author} {\bibfnamefont {N.}~\bibnamefont
  {Manca}}, \bibinfo {author} {\bibfnamefont {D.~J.}\ \bibnamefont
  {Groenendijk}}, \bibinfo {author} {\bibfnamefont {I.}~\bibnamefont
  {Pallecchi}}, \bibinfo {author} {\bibfnamefont {C.}~\bibnamefont {Autieri}},
  \bibinfo {author} {\bibfnamefont {L.~M.~K.}\ \bibnamefont {Tang}}, \bibinfo
  {author} {\bibfnamefont {F.}~\bibnamefont {Telesio}}, \bibinfo {author}
  {\bibfnamefont {G.}~\bibnamefont {Mattoni}}, \bibinfo {author} {\bibfnamefont
  {A.}~\bibnamefont {Mccollam}}, \bibinfo {author} {\bibfnamefont
  {S.}~\bibnamefont {Picozzi}},\ and\ \bibinfo {author} {\bibfnamefont {A.~D.}\
  \bibnamefont {Caviglia}},\ }\bibfield  {title} {\bibinfo {title} {{Balanced
  electron-hole transport in spin-orbit semimetal SrIrO$_3$
  heterostructures}},\ }\href@noop {} {\bibfield  {journal} {\bibinfo
  {journal} {Physical Review B}\ }\textbf {\bibinfo {volume} {97}},\ \bibinfo
  {pages} {081105(R)} (\bibinfo {year} {2018})}\BibitemShut {NoStop}%
\bibitem [{\citenamefont {Biswas}\ \emph {et~al.}(2014)\citenamefont {Biswas},
  \citenamefont {Kim},\ and\ \citenamefont {Jeong}}]{Biswas2014}%
  \BibitemOpen
  \bibfield  {author} {\bibinfo {author} {\bibfnamefont {A.}~\bibnamefont
  {Biswas}}, \bibinfo {author} {\bibfnamefont {K.-S.}\ \bibnamefont {Kim}},\
  and\ \bibinfo {author} {\bibfnamefont {Y.~H.}\ \bibnamefont {Jeong}},\
  }\bibfield  {title} {\bibinfo {title} {{Metal insulator transitions in
  perovskite SrIrO$_3$ thin films}},\ }\href
  {https://doi.org/10.1063/1.4903314} {\bibfield  {journal} {\bibinfo
  {journal} {Journal of Applied Physics}\ }\textbf {\bibinfo {volume} {116}},\
  \bibinfo {pages} {213704} (\bibinfo {year} {2014})}\BibitemShut {NoStop}%
\bibitem [{\citenamefont {Kim}\ \emph {et~al.}(2017)\citenamefont {Kim},
  \citenamefont {Liu},\ and\ \citenamefont {Franchini}}]{Kim2017}%
  \BibitemOpen
  \bibfield  {author} {\bibinfo {author} {\bibfnamefont {B.}~\bibnamefont
  {Kim}}, \bibinfo {author} {\bibfnamefont {P.}~\bibnamefont {Liu}},\ and\
  \bibinfo {author} {\bibfnamefont {C.}~\bibnamefont {Franchini}},\ }\bibfield
  {title} {\bibinfo {title} {{Dimensionality-strain phase diagram of strontium
  iridates}},\ }\href {https://doi.org/10.1103/PhysRevB.95.115111} {\bibfield
  {journal} {\bibinfo  {journal} {Physical Review B}\ }\textbf {\bibinfo
  {volume} {95}},\ \bibinfo {pages} {115111} (\bibinfo {year}
  {2017})}\BibitemShut {NoStop}%
\bibitem [{\citenamefont {Gruenewald}\ \emph {et~al.}(2014)\citenamefont
  {Gruenewald}, \citenamefont {Nichols}, \citenamefont {Terzic}, \citenamefont
  {Cao}, \citenamefont {Brill},\ and\ \citenamefont {Seo}}]{Gruenewald2014}%
  \BibitemOpen
  \bibfield  {author} {\bibinfo {author} {\bibfnamefont {J.~H.}\ \bibnamefont
  {Gruenewald}}, \bibinfo {author} {\bibfnamefont {J.}~\bibnamefont {Nichols}},
  \bibinfo {author} {\bibfnamefont {J.}~\bibnamefont {Terzic}}, \bibinfo
  {author} {\bibfnamefont {G.}~\bibnamefont {Cao}}, \bibinfo {author}
  {\bibfnamefont {J.~W.}\ \bibnamefont {Brill}},\ and\ \bibinfo {author}
  {\bibfnamefont {S.~S.~A.}\ \bibnamefont {Seo}},\ }\bibfield  {title}
  {\bibinfo {title} {{Compressive strain-induced metal–insulator transition
  in orthorhombic SrIrO$_3$ thin films}},\ }\href
  {https://doi.org/10.1557/jmr.2014.288} {\bibfield  {journal} {\bibinfo
  {journal} {Journal of Materials Research}\ }\textbf {\bibinfo {volume}
  {29}},\ \bibinfo {pages} {2491–2496} (\bibinfo {year} {2014})}\BibitemShut
  {NoStop}%
\bibitem [{\citenamefont {Xia}\ \emph {et~al.}(2009)\citenamefont {Xia},
  \citenamefont {Siemons}, \citenamefont {Koster}, \citenamefont {Beasley},\
  and\ \citenamefont {Kapitulnik}}]{Xia2009}%
  \BibitemOpen
  \bibfield  {author} {\bibinfo {author} {\bibfnamefont {J.}~\bibnamefont
  {Xia}}, \bibinfo {author} {\bibfnamefont {W.}~\bibnamefont {Siemons}},
  \bibinfo {author} {\bibfnamefont {G.}~\bibnamefont {Koster}}, \bibinfo
  {author} {\bibfnamefont {M.~R.}\ \bibnamefont {Beasley}},\ and\ \bibinfo
  {author} {\bibfnamefont {A.}~\bibnamefont {Kapitulnik}},\ }\bibfield  {title}
  {\bibinfo {title} {{Critical thickness for itinerant ferromagnetism in
  ultrathin films of SrRuO$_3$}},\ }\href
  {https://doi.org/10.1103/PhysRevB.79.140407} {\bibfield  {journal} {\bibinfo
  {journal} {Physical Review B}\ }\textbf {\bibinfo {volume} {79}},\ \bibinfo
  {pages} {140407(R)} (\bibinfo {year} {2009})}\BibitemShut {NoStop}%
\bibitem [{sup()}]{supplemental}%
  \BibitemOpen
  \href@noop {} {}\bibinfo {note} {{See Supplemental Material at [URL will be
  inserted by publisher] for additional information on the sample deposition,
  the background correction of the SQUID magnetometry data, polar MOKE
  measurements of heterostructure RIR2, the temperature dependence of the
  interlayer coupling strength of RIR2, the FORC study of a heterostructure
  with 1 ML SrIrO$_3$/ 1 ML SrZrO$_3$ spacer, SQUID magnetometry and
  magneto-transport study of the heterostructure RIR12, SQUID magnetometry of a
  second heterostructure with 2 MLs SrIrO$_3$ spacer, and resistance
  measurements of heterostructure RIR12, and 6, and 12 MLs bare SrIrO$_3$
  reference samples }}\BibitemShut {NoStop}%
\bibitem [{\citenamefont {Gr{\"{a}}fe}\ \emph {et~al.}(2021)\citenamefont
  {Gr{\"{a}}fe}, \citenamefont {Gro{\ss}}, \citenamefont {Ilse}, \citenamefont
  {Boltje}, \citenamefont {Muralidhar},\ and\ \citenamefont
  {Goering}}]{Graefe2021}%
  \BibitemOpen
  \bibfield  {author} {\bibinfo {author} {\bibfnamefont {J.}~\bibnamefont
  {Gr{\"{a}}fe}}, \bibinfo {author} {\bibfnamefont {F.}~\bibnamefont
  {Gro{\ss}}}, \bibinfo {author} {\bibfnamefont {S.~E.}\ \bibnamefont {Ilse}},
  \bibinfo {author} {\bibfnamefont {D.~B.}\ \bibnamefont {Boltje}}, \bibinfo
  {author} {\bibfnamefont {S.}~\bibnamefont {Muralidhar}},\ and\ \bibinfo
  {author} {\bibfnamefont {E.~J.}\ \bibnamefont {Goering}},\ }\bibfield
  {title} {\bibinfo {title} {{{LeXtender}: a software package for advanced
  {MOKE} acquisition and analysis}},\ }\href
  {https://doi.org/10.1088/1361-6501/abcdc0} {\bibfield  {journal} {\bibinfo
  {journal} {Measurement Science and Technology}\ }\textbf {\bibinfo {volume}
  {32}},\ \bibinfo {pages} {067002} (\bibinfo {year} {2021})}\BibitemShut
  {NoStop}%
\bibitem [{\citenamefont {Gro{{\ss}}}\ \emph
  {et~al.}(2019{\natexlab{b}})\citenamefont {Gro{{\ss}}}, \citenamefont
  {Mart{\'{i}}nez-Garc{\'{i}}a}, \citenamefont {Ilse}, \citenamefont
  {Sch{\"{u}}tz}, \citenamefont {Goering}, \citenamefont {Rivas},\ and\
  \citenamefont {Gr{\"{a}}fe}}]{Gross2019}%
  \BibitemOpen
  \bibfield  {author} {\bibinfo {author} {\bibfnamefont {F.}~\bibnamefont
  {Gro{{\ss}}}}, \bibinfo {author} {\bibfnamefont {J.~C.}\ \bibnamefont
  {Mart{\'{i}}nez-Garc{\'{i}}a}}, \bibinfo {author} {\bibfnamefont {S.~E.}\
  \bibnamefont {Ilse}}, \bibinfo {author} {\bibfnamefont {G.}~\bibnamefont
  {Sch{\"{u}}tz}}, \bibinfo {author} {\bibfnamefont {E.}~\bibnamefont
  {Goering}}, \bibinfo {author} {\bibfnamefont {M.}~\bibnamefont {Rivas}},\
  and\ \bibinfo {author} {\bibfnamefont {J.}~\bibnamefont {Gr{\"{a}}fe}},\
  }\bibfield  {title} {\bibinfo {title} {{GFORC: A graphics processing unit
  accelerated first-order reversal-curve calculator}},\ }\href@noop {}
  {\bibfield  {journal} {\bibinfo  {journal} {Journal of Applied Physics}\
  }\textbf {\bibinfo {volume} {126}},\ \bibinfo {pages} {163901} (\bibinfo
  {year} {2019}{\natexlab{b}})}\BibitemShut {NoStop}%
\bibitem [{\citenamefont {Wang}\ \emph {et~al.}(2020)\citenamefont {Wang},
  \citenamefont {Li}, \citenamefont {Liu}, \citenamefont {Chen}, \citenamefont
  {Ji}, \citenamefont {Wang}, \citenamefont {Cheng}, \citenamefont {Lu},
  \citenamefont {Rijnders}, \citenamefont {Koster}, \citenamefont {Wu},\ and\
  \citenamefont {Liao}}]{Wang2020}%
  \BibitemOpen
  \bibfield  {author} {\bibinfo {author} {\bibfnamefont {W.}~\bibnamefont
  {Wang}}, \bibinfo {author} {\bibfnamefont {L.}~\bibnamefont {Li}}, \bibinfo
  {author} {\bibfnamefont {J.}~\bibnamefont {Liu}}, \bibinfo {author}
  {\bibfnamefont {B.}~\bibnamefont {Chen}}, \bibinfo {author} {\bibfnamefont
  {Y.}~\bibnamefont {Ji}}, \bibinfo {author} {\bibfnamefont {J.}~\bibnamefont
  {Wang}}, \bibinfo {author} {\bibfnamefont {G.}~\bibnamefont {Cheng}},
  \bibinfo {author} {\bibfnamefont {Y.}~\bibnamefont {Lu}}, \bibinfo {author}
  {\bibfnamefont {G.}~\bibnamefont {Rijnders}}, \bibinfo {author}
  {\bibfnamefont {G.}~\bibnamefont {Koster}}, \bibinfo {author} {\bibfnamefont
  {W.}~\bibnamefont {Wu}},\ and\ \bibinfo {author} {\bibfnamefont
  {Z.}~\bibnamefont {Liao}},\ }\bibfield  {title} {\bibinfo {title} {{Magnetic
  domain engineering in SrRuO$_3$ thin films}},\ }\href
  {https://doi.org/10.1038/s41535-020-00275-5} {\bibfield  {journal} {\bibinfo
  {journal} {npj Quantum Materials}\ }\textbf {\bibinfo {volume} {5}},\
  \bibinfo {pages} {73} (\bibinfo {year} {2020})}\BibitemShut {NoStop}%
\bibitem [{\citenamefont {van~der Heijden}\ \emph {et~al.}(1997)\citenamefont
  {van~der Heijden}, \citenamefont {Bloemen}, \citenamefont {Metselaar},
  \citenamefont {Wolf}, \citenamefont {Gaines}, \citenamefont {van Eemeren},
  \citenamefont {van~der Zaag},\ and\ \citenamefont {de~Jong}}]{Heijden1997}%
  \BibitemOpen
  \bibfield  {author} {\bibinfo {author} {\bibfnamefont {P.~A.~A.}\
  \bibnamefont {van~der Heijden}}, \bibinfo {author} {\bibfnamefont {P.~J.~H.}\
  \bibnamefont {Bloemen}}, \bibinfo {author} {\bibfnamefont {J.~M.}\
  \bibnamefont {Metselaar}}, \bibinfo {author} {\bibfnamefont {R.~M.}\
  \bibnamefont {Wolf}}, \bibinfo {author} {\bibfnamefont {J.~M.}\ \bibnamefont
  {Gaines}}, \bibinfo {author} {\bibfnamefont {J.~T. W.~M.}\ \bibnamefont {van
  Eemeren}}, \bibinfo {author} {\bibfnamefont {P.}~\bibnamefont {van~der
  Zaag}},\ and\ \bibinfo {author} {\bibfnamefont {W.~J.~M.}\ \bibnamefont
  {de~Jong}},\ }\bibfield  {title} {\bibinfo {title} {{Interlayer coupling
  between Fe$_3$O$_4$ layers separated by an insulating nonmagnetic MgO
  layer}},\ }\href {https://doi.org/10.1063/1.2919081} {\bibfield  {journal}
  {\bibinfo  {journal} {Physical Review B}\ }\textbf {\bibinfo {volume} {55}},\
  \bibinfo {pages} {11569} (\bibinfo {year} {1997})}\BibitemShut {NoStop}%
\bibitem [{\citenamefont {Faure-Vincent}\ \emph {et~al.}(2002)\citenamefont
  {Faure-Vincent}, \citenamefont {Tiusan}, \citenamefont {Bellouard},
  \citenamefont {Popova}, \citenamefont {Hehn}, \citenamefont {Montaigne},\
  and\ \citenamefont {Schuhl}}]{Faure-Vincent2002}%
  \BibitemOpen
  \bibfield  {author} {\bibinfo {author} {\bibfnamefont {J.}~\bibnamefont
  {Faure-Vincent}}, \bibinfo {author} {\bibfnamefont {C.}~\bibnamefont
  {Tiusan}}, \bibinfo {author} {\bibfnamefont {C.}~\bibnamefont {Bellouard}},
  \bibinfo {author} {\bibfnamefont {E.}~\bibnamefont {Popova}}, \bibinfo
  {author} {\bibfnamefont {M.}~\bibnamefont {Hehn}}, \bibinfo {author}
  {\bibfnamefont {F.}~\bibnamefont {Montaigne}},\ and\ \bibinfo {author}
  {\bibfnamefont {A.}~\bibnamefont {Schuhl}},\ }\bibfield  {title} {\bibinfo
  {title} {{Interlayer Magnetic Coupling Interactions of Two Ferromagnetic
  Layers by Spin Polarized Tunneling}},\ }\href
  {https://doi.org/10.1103/PhysRevLett.89.107206} {\bibfield  {journal}
  {\bibinfo  {journal} {Physical Review Letters}\ }\textbf {\bibinfo {volume}
  {89}},\ \bibinfo {pages} {107206} (\bibinfo {year} {2002})}\BibitemShut
  {NoStop}%
\bibitem [{\citenamefont {Matczak}\ \emph {et~al.}(2013)\citenamefont
  {Matczak}, \citenamefont {Szymański}, \citenamefont {Urbaniak},
  \citenamefont {Nowicki}, \citenamefont {Głowiński}, \citenamefont
  {Kuświk}, \citenamefont {Schmidt}, \citenamefont {Aleksiejew}, \citenamefont
  {Dubowik},\ and\ \citenamefont {Stobiecki}}]{Matczak2013}%
  \BibitemOpen
  \bibfield  {author} {\bibinfo {author} {\bibfnamefont {M.}~\bibnamefont
  {Matczak}}, \bibinfo {author} {\bibfnamefont {B.}~\bibnamefont {Szymański}},
  \bibinfo {author} {\bibfnamefont {M.}~\bibnamefont {Urbaniak}}, \bibinfo
  {author} {\bibfnamefont {M.}~\bibnamefont {Nowicki}}, \bibinfo {author}
  {\bibfnamefont {H.}~\bibnamefont {Głowiński}}, \bibinfo {author}
  {\bibfnamefont {P.}~\bibnamefont {Kuświk}}, \bibinfo {author} {\bibfnamefont
  {M.}~\bibnamefont {Schmidt}}, \bibinfo {author} {\bibfnamefont
  {J.}~\bibnamefont {Aleksiejew}}, \bibinfo {author} {\bibfnamefont
  {J.}~\bibnamefont {Dubowik}},\ and\ \bibinfo {author} {\bibfnamefont
  {F.}~\bibnamefont {Stobiecki}},\ }\bibfield  {title} {\bibinfo {title}
  {{Antiferromagnetic magnetostatic coupling in Co/Au/Co films with
  perpendicular anisotropy}},\ }\href {https://doi.org/10.1063/1.4819380}
  {\bibfield  {journal} {\bibinfo  {journal} {Journal of Applied Physics}\
  }\textbf {\bibinfo {volume} {114}},\ \bibinfo {pages} {093911} (\bibinfo
  {year} {2013})}\BibitemShut {NoStop}%
\bibitem [{\citenamefont {Gr{\"{a}}fe}\ \emph {et~al.}(2014)\citenamefont
  {Gr{\"{a}}fe}, \citenamefont {Schmidt}, \citenamefont {Audehm}, \citenamefont
  {Sch{\"{u}}tz},\ and\ \citenamefont {Goering}}]{Graefe2014}%
  \BibitemOpen
  \bibfield  {author} {\bibinfo {author} {\bibfnamefont {J.}~\bibnamefont
  {Gr{\"{a}}fe}}, \bibinfo {author} {\bibfnamefont {M.}~\bibnamefont
  {Schmidt}}, \bibinfo {author} {\bibfnamefont {P.}~\bibnamefont {Audehm}},
  \bibinfo {author} {\bibfnamefont {G.}~\bibnamefont {Sch{\"{u}}tz}},\ and\
  \bibinfo {author} {\bibfnamefont {E.}~\bibnamefont {Goering}},\ }\bibfield
  {title} {\bibinfo {title} {{Application of magneto-optical Kerr effect to
  first-order reversal curve measurements}},\ }\href
  {https://doi.org/10.1063/1.4865135} {\bibfield  {journal} {\bibinfo
  {journal} {Review of Scientific Instruments}\ }\textbf {\bibinfo {volume}
  {85}},\ \bibinfo {pages} {023901} (\bibinfo {year} {2014})}\BibitemShut
  {NoStop}%
\bibitem [{\citenamefont {Skoropata}\ \emph {et~al.}(2020)\citenamefont
  {Skoropata}, \citenamefont {Nichols}, \citenamefont {Ok}, \citenamefont
  {Chopdekar}, \citenamefont {Choi}, \citenamefont {Rastogi}, \citenamefont
  {Sohn}, \citenamefont {Gao}, \citenamefont {Yoon}, \citenamefont {Farmer},
  \citenamefont {Desautels}, \citenamefont {Choi}, \citenamefont {Haskel},
  \citenamefont {Freeland}, \citenamefont {Okamoto}, \citenamefont {Brahlek},\
  and\ \citenamefont {Lee}}]{Skoropata2020}%
  \BibitemOpen
  \bibfield  {author} {\bibinfo {author} {\bibfnamefont {E.}~\bibnamefont
  {Skoropata}}, \bibinfo {author} {\bibfnamefont {J.}~\bibnamefont {Nichols}},
  \bibinfo {author} {\bibfnamefont {J.~M.}\ \bibnamefont {Ok}}, \bibinfo
  {author} {\bibfnamefont {R.~V.}\ \bibnamefont {Chopdekar}}, \bibinfo {author}
  {\bibfnamefont {E.~S.}\ \bibnamefont {Choi}}, \bibinfo {author}
  {\bibfnamefont {A.}~\bibnamefont {Rastogi}}, \bibinfo {author} {\bibfnamefont
  {C.}~\bibnamefont {Sohn}}, \bibinfo {author} {\bibfnamefont {X.}~\bibnamefont
  {Gao}}, \bibinfo {author} {\bibfnamefont {S.}~\bibnamefont {Yoon}}, \bibinfo
  {author} {\bibfnamefont {T.}~\bibnamefont {Farmer}}, \bibinfo {author}
  {\bibfnamefont {R.~D.}\ \bibnamefont {Desautels}}, \bibinfo {author}
  {\bibfnamefont {Y.}~\bibnamefont {Choi}}, \bibinfo {author} {\bibfnamefont
  {D.}~\bibnamefont {Haskel}}, \bibinfo {author} {\bibfnamefont {J.~W.}\
  \bibnamefont {Freeland}}, \bibinfo {author} {\bibfnamefont {S.}~\bibnamefont
  {Okamoto}}, \bibinfo {author} {\bibfnamefont {M.}~\bibnamefont {Brahlek}},\
  and\ \bibinfo {author} {\bibfnamefont {H.~N.}\ \bibnamefont {Lee}},\
  }\bibfield  {title} {\bibinfo {title} {{Interfacial tuning of chiral magnetic
  interactions for large topological Hall effects in LaMnO$_3$/SrIrO$_3$
  heterostructures}},\ }\href {https://doi.org/10.1126/sciadv.aaz3902}
  {\bibfield  {journal} {\bibinfo  {journal} {Science Advances}\ }\textbf
  {\bibinfo {volume} {6}},\ \bibinfo {pages} {eaaz3902} (\bibinfo {year}
  {2020})}\BibitemShut {NoStop}%
\bibitem [{\citenamefont {Fulghum}\ and\ \citenamefont
  {Camley}(1995)}]{Fulghum1995}%
  \BibitemOpen
  \bibfield  {author} {\bibinfo {author} {\bibfnamefont {D.~B.}\ \bibnamefont
  {Fulghum}}\ and\ \bibinfo {author} {\bibfnamefont {R.~E.}\ \bibnamefont
  {Camley}},\ }\bibfield  {title} {\bibinfo {title} {{Magnetic behavior of
  antiferromagnetically coupled layers connected by ferromagnetic pinholes}},\
  }\href {https://doi.org/10.1103/PhysRevB.52.13436} {\bibfield  {journal}
  {\bibinfo  {journal} {Physical Review B}\ }\textbf {\bibinfo {volume} {52}},\
  \bibinfo {pages} {13436} (\bibinfo {year} {1995})}\BibitemShut {NoStop}%
\bibitem [{\citenamefont {N\'{e}el}(1962)}]{Neel1962}%
  \BibitemOpen
  \bibfield  {author} {\bibinfo {author} {\bibfnamefont {L.}~\bibnamefont
  {N\'{e}el}},\ }\bibfield  {title} {\bibinfo {title} {{}},\ }\href@noop {}
  {\bibfield  {journal} {\bibinfo  {journal} {Cr. Hebd. Acad. Sci.}\ }\textbf
  {\bibinfo {volume} {255}},\ \bibinfo {pages} {1676} (\bibinfo {year}
  {1962})}\BibitemShut {NoStop}%
\bibitem [{\citenamefont {Moritz}\ \emph {et~al.}(2004)\citenamefont {Moritz},
  \citenamefont {Garcia}, \citenamefont {Toussaint}, \citenamefont {Dieny},\
  and\ \citenamefont {Nozi{\`{e}}res}}]{Moritz2004}%
  \BibitemOpen
  \bibfield  {author} {\bibinfo {author} {\bibfnamefont {J.}~\bibnamefont
  {Moritz}}, \bibinfo {author} {\bibfnamefont {F.}~\bibnamefont {Garcia}},
  \bibinfo {author} {\bibfnamefont {J.~C.}\ \bibnamefont {Toussaint}}, \bibinfo
  {author} {\bibfnamefont {B.}~\bibnamefont {Dieny}},\ and\ \bibinfo {author}
  {\bibfnamefont {J.~P.}\ \bibnamefont {Nozi{\`{e}}res}},\ }\bibfield  {title}
  {\bibinfo {title} {{Orange peel coupling in multilayers with perpendicular
  magnetic anisotropy: Application to (Co/Pt)-based exchange-biased
  spin-valves}},\ }\href {https://doi.org/10.1209/epl/i2003-10063-9} {\bibfield
   {journal} {\bibinfo  {journal} {Europhysics Letters}\ }\textbf {\bibinfo
  {volume} {65}},\ \bibinfo {pages} {123} (\bibinfo {year} {2004})}\BibitemShut
  {NoStop}%
\bibitem [{\citenamefont {Schrag}\ \emph
  {et~al.}(2000{\natexlab{a}})\citenamefont {Schrag}, \citenamefont
  {Anguelouch}, \citenamefont {Xiao}, \citenamefont {Trouilloud}, \citenamefont
  {Lu}, \citenamefont {Gallagher},\ and\ \citenamefont {Parkin}}]{Schrag2000a}%
  \BibitemOpen
  \bibfield  {author} {\bibinfo {author} {\bibfnamefont {B.}~\bibnamefont
  {Schrag}}, \bibinfo {author} {\bibfnamefont {A.}~\bibnamefont {Anguelouch}},
  \bibinfo {author} {\bibfnamefont {G.}~\bibnamefont {Xiao}}, \bibinfo {author}
  {\bibfnamefont {P.}~\bibnamefont {Trouilloud}}, \bibinfo {author}
  {\bibfnamefont {Y.}~\bibnamefont {Lu}}, \bibinfo {author} {\bibfnamefont
  {W.}~\bibnamefont {Gallagher}},\ and\ \bibinfo {author} {\bibfnamefont
  {S.}~\bibnamefont {Parkin}},\ }\bibfield  {title} {\bibinfo {title}
  {{Magnetization reversal and interlayer coupling in magnetic tunneling
  junctions}},\ }\href {https://doi.org/10.1063/1.373129} {\bibfield  {journal}
  {\bibinfo  {journal} {Journal of Applied Physics}\ }\textbf {\bibinfo
  {volume} {87}},\ \bibinfo {pages} {4682} (\bibinfo {year}
  {2000}{\natexlab{a}})}\BibitemShut {NoStop}%
\bibitem [{\citenamefont {Fuller}\ and\ \citenamefont
  {Sullivan}(1962)}]{Fuller1962}%
  \BibitemOpen
  \bibfield  {author} {\bibinfo {author} {\bibfnamefont {H.~W.}\ \bibnamefont
  {Fuller}}\ and\ \bibinfo {author} {\bibfnamefont {D.~L.}\ \bibnamefont
  {Sullivan}},\ }\bibfield  {title} {\bibinfo {title} {{Magnetostatic
  Interactions between Thin Magnetic Films}},\ }\href
  {https://doi.org/10.1063/1.1728600} {\bibfield  {journal} {\bibinfo
  {journal} {Journal of Applied Physics}\ }\textbf {\bibinfo {volume} {33}},\
  \bibinfo {pages} {1063} (\bibinfo {year} {1962})}\BibitemShut {NoStop}%
\bibitem [{\citenamefont {Thomas}\ \emph {et~al.}(2000)\citenamefont {Thomas},
  \citenamefont {Samant},\ and\ \citenamefont {Parkin}}]{Thomas2000}%
  \BibitemOpen
  \bibfield  {author} {\bibinfo {author} {\bibfnamefont {L.}~\bibnamefont
  {Thomas}}, \bibinfo {author} {\bibfnamefont {M.~G.}\ \bibnamefont {Samant}},\
  and\ \bibinfo {author} {\bibfnamefont {S.~S.~P.}\ \bibnamefont {Parkin}},\
  }\bibfield  {title} {\bibinfo {title} {{Domain-Wall Induced Coupling between
  Ferromagnetic Layers}},\ }\href {https://doi.org/10.1103/PhysRevLett.84.1816}
  {\bibfield  {journal} {\bibinfo  {journal} {Physical Review Letters}\
  }\textbf {\bibinfo {volume} {84}},\ \bibinfo {pages} {1816} (\bibinfo {year}
  {2000})}\BibitemShut {NoStop}%
\bibitem [{\citenamefont {Platt}\ \emph {et~al.}(2000)\citenamefont {Platt},
  \citenamefont {McCartney}, \citenamefont {Parker},\ and\ \citenamefont
  {Berkowitz}}]{Platt2000}%
  \BibitemOpen
  \bibfield  {author} {\bibinfo {author} {\bibfnamefont {C.~L.}\ \bibnamefont
  {Platt}}, \bibinfo {author} {\bibfnamefont {M.~R.}\ \bibnamefont
  {McCartney}}, \bibinfo {author} {\bibfnamefont {F.~T.}\ \bibnamefont
  {Parker}},\ and\ \bibinfo {author} {\bibfnamefont {A.~E.}\ \bibnamefont
  {Berkowitz}},\ }\bibfield  {title} {\bibinfo {title} {{Magnetic interlayer
  coupling in ferromagnet/insulator/ferromagnet structures}},\ }\href
  {https://doi.org/10.1103/PhysRevB.61.9633} {\bibfield  {journal} {\bibinfo
  {journal} {Physical Review B}\ }\textbf {\bibinfo {volume} {61}},\ \bibinfo
  {pages} {9633} (\bibinfo {year} {2000})}\BibitemShut {NoStop}%
\bibitem [{\citenamefont {Baltz}\ \emph {et~al.}(2007)\citenamefont {Baltz},
  \citenamefont {Marty}, \citenamefont {Rodmacq},\ and\ \citenamefont
  {Dieny}}]{Baltz2007}%
  \BibitemOpen
  \bibfield  {author} {\bibinfo {author} {\bibfnamefont {V.}~\bibnamefont
  {Baltz}}, \bibinfo {author} {\bibfnamefont {A.}~\bibnamefont {Marty}},
  \bibinfo {author} {\bibfnamefont {B.}~\bibnamefont {Rodmacq}},\ and\ \bibinfo
  {author} {\bibfnamefont {B.}~\bibnamefont {Dieny}},\ }\bibfield  {title}
  {\bibinfo {title} {{Magnetic domain replication in interacting bilayers with
  out-of-plane anisotropy: Application to CoPt multilayers}},\ }\href
  {https://doi.org/10.1103/PhysRevB.75.014406} {\bibfield  {journal} {\bibinfo
  {journal} {Physical Review B}\ }\textbf {\bibinfo {volume} {75}},\ \bibinfo
  {pages} {014406} (\bibinfo {year} {2007})}\BibitemShut {NoStop}%
\bibitem [{\citenamefont {Slonczewski}(1989)}]{Slonczewski1989}%
  \BibitemOpen
  \bibfield  {author} {\bibinfo {author} {\bibfnamefont {J.~C.}\ \bibnamefont
  {Slonczewski}},\ }\bibfield  {title} {\bibinfo {title} {{Conductance and
  exchange coupling of two ferromagnets separated by a tunneling barrier}},\
  }\href {https://doi.org/10.1103/PhysRevB.39.6995} {\bibfield  {journal}
  {\bibinfo  {journal} {Physical Review B}\ }\textbf {\bibinfo {volume} {39}},\
  \bibinfo {pages} {6995} (\bibinfo {year} {1989})}\BibitemShut {NoStop}%
\bibitem [{\citenamefont {Bruno}(1995)}]{Bruno1995}%
  \BibitemOpen
  \bibfield  {author} {\bibinfo {author} {\bibfnamefont {P.}~\bibnamefont
  {Bruno}},\ }\bibfield  {title} {\bibinfo {title} {{Theory of interlayer
  magnetic coupling}},\ }\href {https://doi.org/10.1103/PhysRevB.52.411}
  {\bibfield  {journal} {\bibinfo  {journal} {Physical Review B}\ }\textbf
  {\bibinfo {volume} {52}},\ \bibinfo {pages} {411} (\bibinfo {year}
  {1995})}\BibitemShut {NoStop}%
\bibitem [{\citenamefont {Bobo}\ \emph {et~al.}(1999)\citenamefont {Bobo},
  \citenamefont {Kikuchi}, \citenamefont {Redon}, \citenamefont {Snoeck},
  \citenamefont {Piecuch},\ and\ \citenamefont {White}}]{Bobo1999}%
  \BibitemOpen
  \bibfield  {author} {\bibinfo {author} {\bibfnamefont {J.~F.}\ \bibnamefont
  {Bobo}}, \bibinfo {author} {\bibfnamefont {H.}~\bibnamefont {Kikuchi}},
  \bibinfo {author} {\bibfnamefont {O.}~\bibnamefont {Redon}}, \bibinfo
  {author} {\bibfnamefont {E.}~\bibnamefont {Snoeck}}, \bibinfo {author}
  {\bibfnamefont {M.}~\bibnamefont {Piecuch}},\ and\ \bibinfo {author}
  {\bibfnamefont {R.~L.}\ \bibnamefont {White}},\ }\bibfield  {title} {\bibinfo
  {title} {{Pinholes in antiferromagnetically coupled multilayers: Effects on
  hysteresis loops and relation to biquadratic exchange}},\ }\href
  {https://doi.org/10.1103/PhysRevB.60.4131} {\bibfield  {journal} {\bibinfo
  {journal} {Physical Review B}\ }\textbf {\bibinfo {volume} {60}},\ \bibinfo
  {pages} {4131} (\bibinfo {year} {1999})}\BibitemShut {NoStop}%
\bibitem [{\citenamefont {Ziese}\ \emph {et~al.}(2010)\citenamefont {Ziese},
  \citenamefont {Vrejoiu},\ and\ \citenamefont {Hesse}}]{Ziese2010}%
  \BibitemOpen
  \bibfield  {author} {\bibinfo {author} {\bibfnamefont {M.}~\bibnamefont
  {Ziese}}, \bibinfo {author} {\bibfnamefont {I.}~\bibnamefont {Vrejoiu}},\
  and\ \bibinfo {author} {\bibfnamefont {D.}~\bibnamefont {Hesse}},\ }\bibfield
   {title} {\bibinfo {title} {{Structural symmetry and magnetocrystalline
  anisotropy of SrRuO$_3$ films on SrTiO$_3$}},\ }\href
  {https://doi.org/10.1103/PhysRevB.81.184418} {\bibfield  {journal} {\bibinfo
  {journal} {Physical Review B}\ }\textbf {\bibinfo {volume} {81}},\ \bibinfo
  {pages} {184418} (\bibinfo {year} {2010})}\BibitemShut {NoStop}%
\bibitem [{\citenamefont {Davydenko}\ \emph {et~al.}(2015)\citenamefont
  {Davydenko}, \citenamefont {Pustovalo}, \citenamefont {Ognev}, \citenamefont
  {Kozlov}, \citenamefont {Chetbotkevich},\ and\ \citenamefont
  {Han}}]{Davydenko2015}%
  \BibitemOpen
  \bibfield  {author} {\bibinfo {author} {\bibfnamefont {A.~V.}\ \bibnamefont
  {Davydenko}}, \bibinfo {author} {\bibfnamefont {E.~V.}\ \bibnamefont
  {Pustovalo}}, \bibinfo {author} {\bibfnamefont {A.~V.}\ \bibnamefont
  {Ognev}}, \bibinfo {author} {\bibfnamefont {A.~G.}\ \bibnamefont {Kozlov}},
  \bibinfo {author} {\bibfnamefont {L.~A.}\ \bibnamefont {Chetbotkevich}},\
  and\ \bibinfo {author} {\bibfnamefont {X.~F.}\ \bibnamefont {Han}},\
  }\bibfield  {title} {\bibinfo {title} {{N\'{e}el coupling in Co/Cu/Co stripes
  with unidirectional interface roughness}},\ }\href
  {https://doi.org/https://doi.org/10.1016/j.jmmm.2014.10.143} {\bibfield
  {journal} {\bibinfo  {journal} {Journal of Magnetism and Magnetic Materials}\
  }\textbf {\bibinfo {volume} {377}},\ \bibinfo {pages} {334} (\bibinfo {year}
  {2015})}\BibitemShut {NoStop}%
\bibitem [{\citenamefont {Schrag}\ \emph
  {et~al.}(2000{\natexlab{b}})\citenamefont {Schrag}, \citenamefont
  {Anguelouch}, \citenamefont {Ingvarsson}, \citenamefont {Xiao}, \citenamefont
  {Lu}, \citenamefont {Trouilloud}, \citenamefont {Gupta}, \citenamefont
  {Wanner}, \citenamefont {Gallagher}, \citenamefont {Rice},\ and\
  \citenamefont {Parkin}}]{Schrag2000}%
  \BibitemOpen
  \bibfield  {author} {\bibinfo {author} {\bibfnamefont {B.~D.}\ \bibnamefont
  {Schrag}}, \bibinfo {author} {\bibfnamefont {A.}~\bibnamefont {Anguelouch}},
  \bibinfo {author} {\bibfnamefont {S.}~\bibnamefont {Ingvarsson}}, \bibinfo
  {author} {\bibfnamefont {G.}~\bibnamefont {Xiao}}, \bibinfo {author}
  {\bibfnamefont {Y.}~\bibnamefont {Lu}}, \bibinfo {author} {\bibfnamefont
  {P.~L.}\ \bibnamefont {Trouilloud}}, \bibinfo {author} {\bibfnamefont
  {A.}~\bibnamefont {Gupta}}, \bibinfo {author} {\bibfnamefont {R.~A.}\
  \bibnamefont {Wanner}}, \bibinfo {author} {\bibfnamefont {W.~J.}\
  \bibnamefont {Gallagher}}, \bibinfo {author} {\bibfnamefont {P.~M.}\
  \bibnamefont {Rice}},\ and\ \bibinfo {author} {\bibfnamefont {S.~S.~P.}\
  \bibnamefont {Parkin}},\ }\bibfield  {title} {\bibinfo {title} {{N\'{e}el
  “orange-peel” coupling in magnetic tunneling junction devices}},\ }\href
  {https://doi.org/10.1063/1.1315633} {\bibfield  {journal} {\bibinfo
  {journal} {Applied Physics Letters}\ }\textbf {\bibinfo {volume} {77}},\
  \bibinfo {pages} {2373} (\bibinfo {year} {2000}{\natexlab{b}})}\BibitemShut
  {NoStop}%
\bibitem [{\citenamefont {Nistor}(2011)}]{Nistor2011}%
  \BibitemOpen
  \bibfield  {author} {\bibinfo {author} {\bibfnamefont {L.~E.}\ \bibnamefont
  {Nistor}},\ }\href@noop {} {\bibinfo {title} {{Magnetic tunnel junctions with
  perpendicular magnetization: Anisotropy, magnetoresistance, magnetic coupling
  and spin transfer torque switching}}} (\bibinfo {year} {2011}),\ \bibinfo
  {note} {{PhD Thesis}}\BibitemShut {NoStop}%
\bibitem [{\citenamefont {Rodmacq}\ \emph {et~al.}(2006)\citenamefont
  {Rodmacq}, \citenamefont {Baltz},\ and\ \citenamefont {Dieny}}]{Rodmacq2006}%
  \BibitemOpen
  \bibfield  {author} {\bibinfo {author} {\bibfnamefont {B.}~\bibnamefont
  {Rodmacq}}, \bibinfo {author} {\bibfnamefont {V.}~\bibnamefont {Baltz}},\
  and\ \bibinfo {author} {\bibfnamefont {B.}~\bibnamefont {Dieny}},\ }\bibfield
   {title} {\bibinfo {title} {{Macroscopic probing of domain configurations in
  interacting bilayers with perpendicular magnetic anisotropy}},\ }\href
  {https://doi.org/10.1103/PhysRevB.73.092405} {\bibfield  {journal} {\bibinfo
  {journal} {Physical Review B}\ }\textbf {\bibinfo {volume} {73}},\ \bibinfo
  {pages} {092405} (\bibinfo {year} {2006})}\BibitemShut {NoStop}%
\bibitem [{\citenamefont {Choi}\ \emph {et~al.}(2001)\citenamefont {Choi},
  \citenamefont {Eom}, \citenamefont {Rijnders}, \citenamefont {Rogalla},\ and\
  \citenamefont {Blank}}]{Choi2001}%
  \BibitemOpen
  \bibfield  {author} {\bibinfo {author} {\bibfnamefont {J.}~\bibnamefont
  {Choi}}, \bibinfo {author} {\bibfnamefont {C.~B.}\ \bibnamefont {Eom}},
  \bibinfo {author} {\bibfnamefont {G.}~\bibnamefont {Rijnders}}, \bibinfo
  {author} {\bibfnamefont {H.}~\bibnamefont {Rogalla}},\ and\ \bibinfo {author}
  {\bibfnamefont {D.~H.~A.}\ \bibnamefont {Blank}},\ }\bibfield  {title}
  {\bibinfo {title} {{Growth mode transition from layer by layer to step flow
  during the growth of heteroepitaxial SrRuO$_3$ on (001) SrTiO$_3$}},\ }\href
  {https://doi.org/10.1063/1.1389837} {\bibfield  {journal} {\bibinfo
  {journal} {Applied Physics Letters}\ }\textbf {\bibinfo {volume} {79}},\
  \bibinfo {pages} {1447} (\bibinfo {year} {2001})}\BibitemShut {NoStop}%
\bibitem [{\citenamefont {Mohseni}\ \emph {et~al.}(2011)\citenamefont
  {Mohseni}, \citenamefont {Dumas}, \citenamefont {Fang}, \citenamefont {Lau},
  \citenamefont {Sani}, \citenamefont {Persson},\ and\ \citenamefont
  {\AA{}kerman}}]{Mohseni2011}%
  \BibitemOpen
  \bibfield  {author} {\bibinfo {author} {\bibfnamefont {S.~M.}\ \bibnamefont
  {Mohseni}}, \bibinfo {author} {\bibfnamefont {R.~K.}\ \bibnamefont {Dumas}},
  \bibinfo {author} {\bibfnamefont {Y.}~\bibnamefont {Fang}}, \bibinfo {author}
  {\bibfnamefont {J.~W.}\ \bibnamefont {Lau}}, \bibinfo {author} {\bibfnamefont
  {S.~R.}\ \bibnamefont {Sani}}, \bibinfo {author} {\bibfnamefont
  {J.}~\bibnamefont {Persson}},\ and\ \bibinfo {author} {\bibfnamefont
  {J.}~\bibnamefont {\AA{}kerman}},\ }\bibfield  {title} {\bibinfo {title}
  {{Temperature-dependent interlayer coupling in Ni/Co perpendicular
  pseudo-spin-valve structures}},\ }\href@noop {} {\bibfield  {journal}
  {\bibinfo  {journal} {Physical Review B}\ }\textbf {\bibinfo {volume} {84}},\
  \bibinfo {pages} {174432} (\bibinfo {year} {2011})}\BibitemShut {NoStop}%
\bibitem [{\citenamefont {Pugh}\ and\ \citenamefont
  {Rostoker}(1953)}]{Pugh1953}%
  \BibitemOpen
  \bibfield  {author} {\bibinfo {author} {\bibfnamefont {E.~M.}\ \bibnamefont
  {Pugh}}\ and\ \bibinfo {author} {\bibfnamefont {N.}~\bibnamefont
  {Rostoker}},\ }\bibfield  {title} {\bibinfo {title} {Hall effect in
  ferromagnetic materials},\ }\href {https://doi.org/10.1103/RevModPhys.25.151}
  {\bibfield  {journal} {\bibinfo  {journal} {Rev. Mod. Phys.}\ }\textbf
  {\bibinfo {volume} {25}},\ \bibinfo {pages} {151} (\bibinfo {year}
  {1953})}\BibitemShut {NoStop}%
\bibitem [{\citenamefont {van Thiel}\ \emph {et~al.}(2020)\citenamefont {van
  Thiel}, \citenamefont {Groenendijk},\ and\ \citenamefont
  {Caviglia}}]{van_Thiel_2020}%
  \BibitemOpen
  \bibfield  {author} {\bibinfo {author} {\bibfnamefont {T.~C.}\ \bibnamefont
  {van Thiel}}, \bibinfo {author} {\bibfnamefont {D.~J.}\ \bibnamefont
  {Groenendijk}},\ and\ \bibinfo {author} {\bibfnamefont {A.~D.}\ \bibnamefont
  {Caviglia}},\ }\bibfield  {title} {\bibinfo {title} {{Extraordinary Hall
  balance in ultrathin {SrRuO$_3$} bilayers}},\ }\href
  {https://doi.org/10.1088/2515-7639/ab7a03} {\bibfield  {journal} {\bibinfo
  {journal} {J. Phys.: Materials}\ }\textbf {\bibinfo {volume} {3}},\ \bibinfo
  {pages} {025005} (\bibinfo {year} {2020})}\BibitemShut {NoStop}%
\bibitem [{\citenamefont {Groenendijk}\ \emph
  {et~al.}(2020{\natexlab{b}})\citenamefont {Groenendijk}, \citenamefont
  {Manca}, \citenamefont {Bruijckere}, \citenamefont {Monteiro}, \citenamefont
  {Gaudenzi}, \citenamefont {van~der Zant},\ and\ \citenamefont
  {Caviglia}}]{Groenendijk2020}%
  \BibitemOpen
  \bibfield  {author} {\bibinfo {author} {\bibfnamefont {D.~J.}\ \bibnamefont
  {Groenendijk}}, \bibinfo {author} {\bibfnamefont {N.}~\bibnamefont {Manca}},
  \bibinfo {author} {\bibfnamefont {J.~D.}\ \bibnamefont {Bruijckere}},
  \bibinfo {author} {\bibfnamefont {A.~M. R. V.~L.}\ \bibnamefont {Monteiro}},
  \bibinfo {author} {\bibfnamefont {R.}~\bibnamefont {Gaudenzi}}, \bibinfo
  {author} {\bibfnamefont {H.~S.~J.}\ \bibnamefont {van~der Zant}},\ and\
  \bibinfo {author} {\bibfnamefont {A.~D.}\ \bibnamefont {Caviglia}},\
  }\bibfield  {title} {\bibinfo {title} {{Anisotropic magnetoresistance in
  spin–orbit semimetal SrIrO$_3$}},\ }\href@noop {} {\bibfield  {journal}
  {\bibinfo  {journal} {The European Physical Journal Plus}\ }\textbf {\bibinfo
  {volume} {135}},\ \bibinfo {pages} {627} (\bibinfo {year}
  {2020}{\natexlab{b}})}\BibitemShut {NoStop}%
\bibitem [{\citenamefont {Nie}\ \emph {et~al.}(2015)\citenamefont {Nie},
  \citenamefont {King}, \citenamefont {Kim}, \citenamefont {Uchida},
  \citenamefont {Wei}, \citenamefont {Faeth}, \citenamefont {Ruf},
  \citenamefont {Ruff}, \citenamefont {Xie}, \citenamefont {Pan}, \citenamefont
  {Fennie}, \citenamefont {Schlom},\ and\ \citenamefont {Shen}}]{Nie2015}%
  \BibitemOpen
  \bibfield  {author} {\bibinfo {author} {\bibfnamefont {Y.~F.}\ \bibnamefont
  {Nie}}, \bibinfo {author} {\bibfnamefont {P.~D.~C.}\ \bibnamefont {King}},
  \bibinfo {author} {\bibfnamefont {C.~H.}\ \bibnamefont {Kim}}, \bibinfo
  {author} {\bibfnamefont {M.}~\bibnamefont {Uchida}}, \bibinfo {author}
  {\bibfnamefont {H.~I.}\ \bibnamefont {Wei}}, \bibinfo {author} {\bibfnamefont
  {B.~D.}\ \bibnamefont {Faeth}}, \bibinfo {author} {\bibfnamefont {J.~P.}\
  \bibnamefont {Ruf}}, \bibinfo {author} {\bibfnamefont {J.~P.~C.}\
  \bibnamefont {Ruff}}, \bibinfo {author} {\bibfnamefont {L.}~\bibnamefont
  {Xie}}, \bibinfo {author} {\bibfnamefont {X.}~\bibnamefont {Pan}}, \bibinfo
  {author} {\bibfnamefont {C.~J.}\ \bibnamefont {Fennie}}, \bibinfo {author}
  {\bibfnamefont {D.~G.}\ \bibnamefont {Schlom}},\ and\ \bibinfo {author}
  {\bibfnamefont {K.~M.}\ \bibnamefont {Shen}},\ }\bibfield  {title} {\bibinfo
  {title} {{Interplay of Spin-Orbit Interactions, Dimensionality, and
  Octahedral Rotations in Semimetallic SrIrO$_{3}$}},\ }\href
  {https://doi.org/10.1103/PhysRevLett.114.016401} {\bibfield  {journal}
  {\bibinfo  {journal} {Phys. Rev. Lett.}\ }\textbf {\bibinfo {volume} {114}},\
  \bibinfo {pages} {016401} (\bibinfo {year} {2015})}\BibitemShut {NoStop}%
\end{thebibliography}%

\end{document}


\begin{center}
	\vspace{1cm}
	{\LARGE \bfseries Supplemental Material for\\
Magnetic interlayer coupling between ferromagnetic SrRuO$_3$ layers through a SrIrO$_3$ spacer\par }
	\vspace{1.5cm}
	{\large Lena Wysocki$^1$,  Sven Erik Ilse$^2$, Lin Yang$^1$, Eberhard Goering$^2$, Felix Gunkel$^3$, Regina Dittmann$^3$, Paul H.M. van Loosdrecht,$^1$ \par and Ionela Lindfors-Vrejoiu$^1$ \par}
	\vspace{0.5cm}
	{\large\itshape $^1$ Institute of Physics II, University of Cologne, 50937 Cologne, Germany\newline 
                                     $^2$ Max-Planck Institute for Intelligent Systems, Stuttgart, Germany}\newline
		               $^3$ Peter Grünberg Institut (PGI-7), Forschungszentrum Jülich GmbH, Jülich, Germany
\end{center}
\vspace{0.2cm}

\section{Heterostructure deposition and surface topography investigation}

\begin{figure*}[htp]          
\begin{center}                                                 
\includegraphics[width=\textwidth]{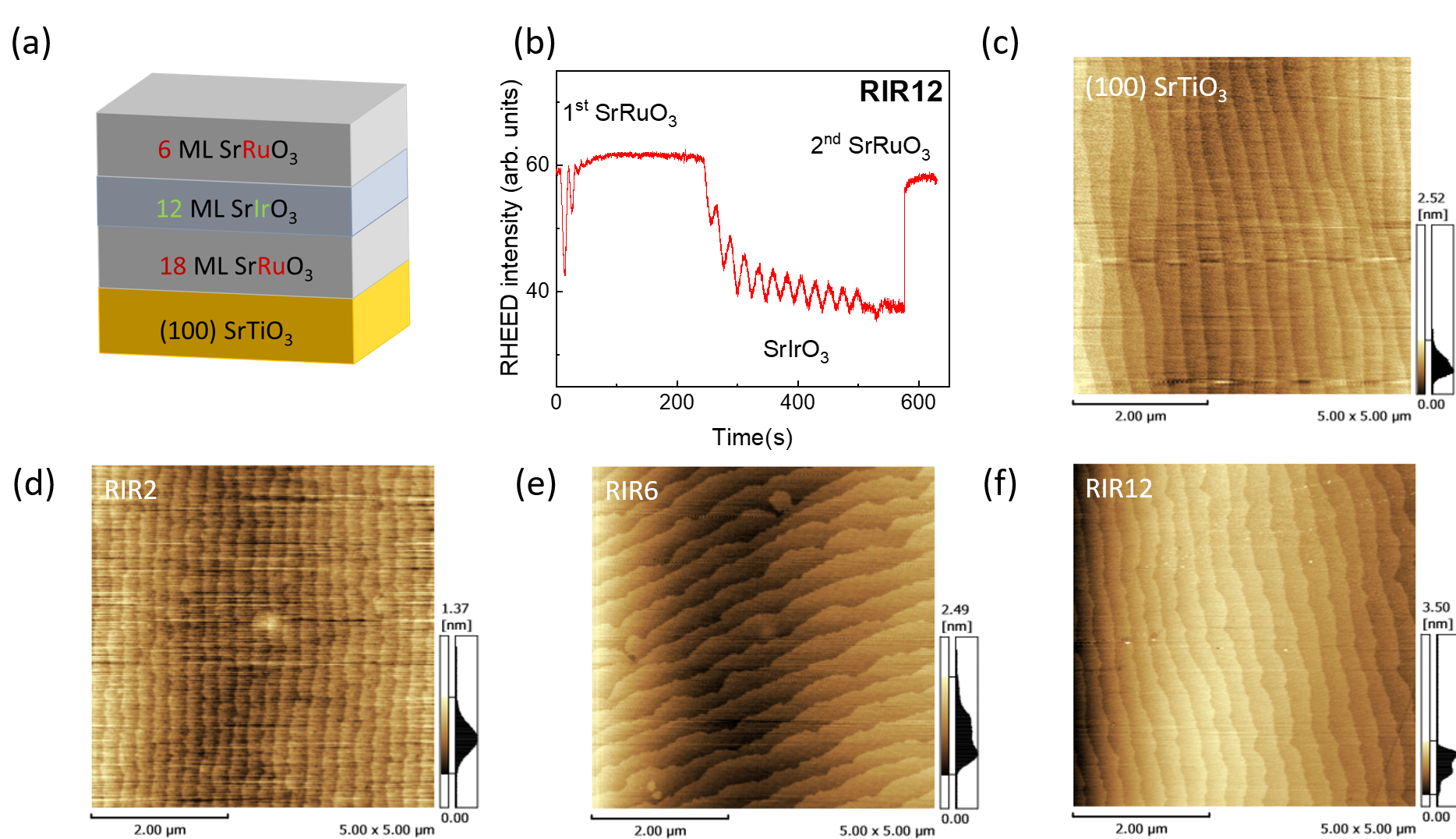}                                      
\caption{\label{fig_RHEEDAFM} (a) Scheme of the heterostructures under study, exemplarily drawn for heterostructure RIR12, with the thickest spacer layer of this study (12 MLs SrIrO$_3$) (b) integrated RHEED intensity plotted as function of deposition time of the heterostructure RIR12. Atomic force microscopy images (5 $\mu $m x 5 $\mu $m) of one SrTiO$_3$ (100) substrate after etching and annealing (c), and AFM images of the heterostructures RIR2 (d), RIR6 (e), and RIR12 (f).} 
\end{center}
\end{figure*}
\noindent
The heterostructures under study have the general design depicted in \textbf{Figure S\ref{fig_RHEEDAFM}} (a). They are composed of two ferromagnetic SrRuO$_3$ layers with deliberately different thicknesses of 18 and 6 monolayers (MLs). The bottom SrRuO$_3$ layer is thicker than the top SrRuO$_3$ layer in each heterostructure and grown directly on top of the SrTiO$_3$ substrate. A spacer of SrIrO$_3$ separates the two ferromagnetic layers. For the heterostructure RIR2, the top SrRuO$_3$ layer is additionally capped by a 2 MLs SrIrO$_3$ layer. \\
In case of decoupled or only weakly coupled layers, the magnetic hysteresis loop of such a heterostructure shows a two-step switching behavior due to the different temperature dependence of the coercive fields of the two SrRuO$_3$ layers. This allows to assess the magnetic interlayer coupling.\\
The deposition was monitored by in-situ reflection high-energy electron diffraction (RHEED). As shown exemplarily for heterostructure RIR12 in \textbf{Figure S\ref{fig_RHEEDAFM}} (b), the SrIrO$_3$ layers grew in a layer-by-layer growth, which enabled the precise control of the layer thickness. The SrRuO$_3$ layers grew in step-flow mode which has been proven to result in smooth thin films [1]. The surface of the heterostructures possessing SrIrO$_3$ spacers, presented in (d)-(f), are smooth, resembling the structure of the SrTiO$_3$ substrate with uniform terrace width and one unit cell step height (see (c)). During the investigation of the surface topography by AFM, we did not observe any etch pitchs on the substrates before deposition, or deep holes in the heterostructures with SrIrO$_3$ spacer (RIR2, RIR6, RIR12).

\section{SQUID magnetometry: Background correction procedure exemplarily shown for heterostructure RIR2 at 10 K}
\begin{figure*}[h]   
\begin{center}                                                         
\includegraphics[width=\textwidth]{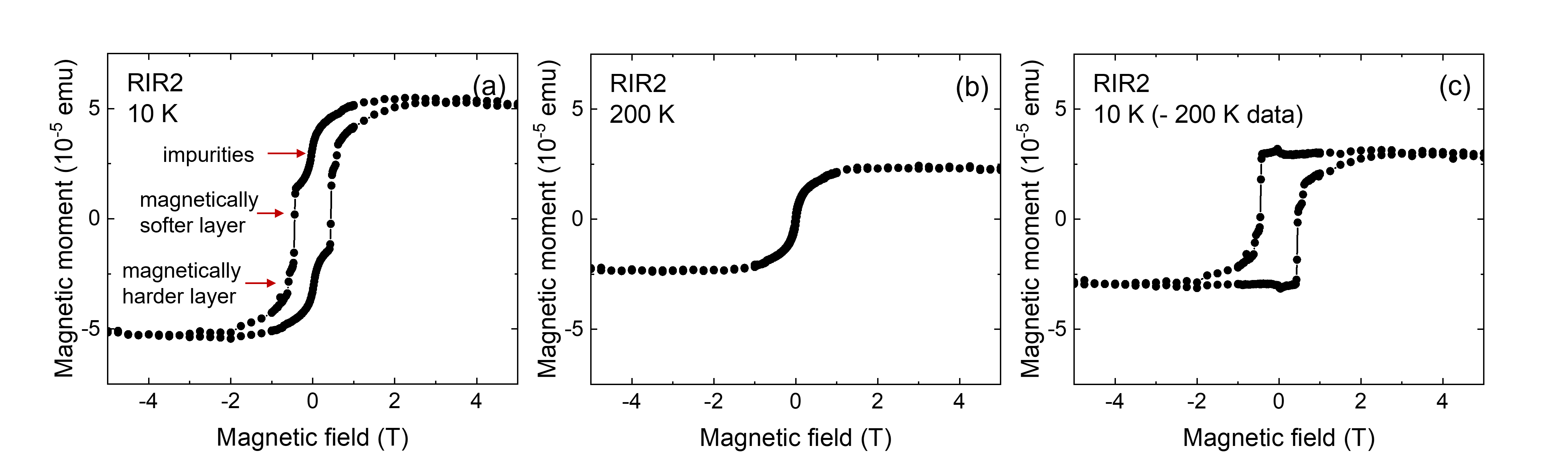}                                      
\caption{\label{fig_backgroundcorrection} Magnetic hysteresis loops of heterostructure RIR2 at 10 K (a) and 200 K (b). The loops have been corrected by subtraction of the diamagnetic contribution from the SrTiO$_3$ substrate. (c) Magnetic hysteresis loop at 10 K after subtraction of the 200 K measurement. } 
\end{center}
\end{figure*}
\noindent
The background correction procedure that has been used for the hysteresis loops measured by SQUID magnetometry is presented in \textbf{Figure S\ref{fig_backgroundcorrection}}, exemplarily for heterostructure RIR2 at 10 K. All magnetic hysteresis loops were corrected by the subtraction of the diamagnetic response from the substrate by the perfomance of linear fitting in the high magnetic field range where the sample is in its saturated state. In order to correct for the magnetically soft contribution of magnetic impurities, visible in the hysteresis loop drawn in (a), a reference hysteresis loop was measured at 200 K, above the transition temperatures of the two SrRuO$_3$ layers of the heterostructures. By subtraction of the 200 K measurement shown in (b), this soft magnetic contribution was removed successfully, as shown in (c). However, small peak-like features around 0 T originating from the imperfect background correction remain.

\newpage \noindent
\section{Heterostructure RIR2: Magneto-optical Kerr measurements}
\begin{figure*}[h]   
\begin{center}                                                         
\includegraphics[width=\textwidth]{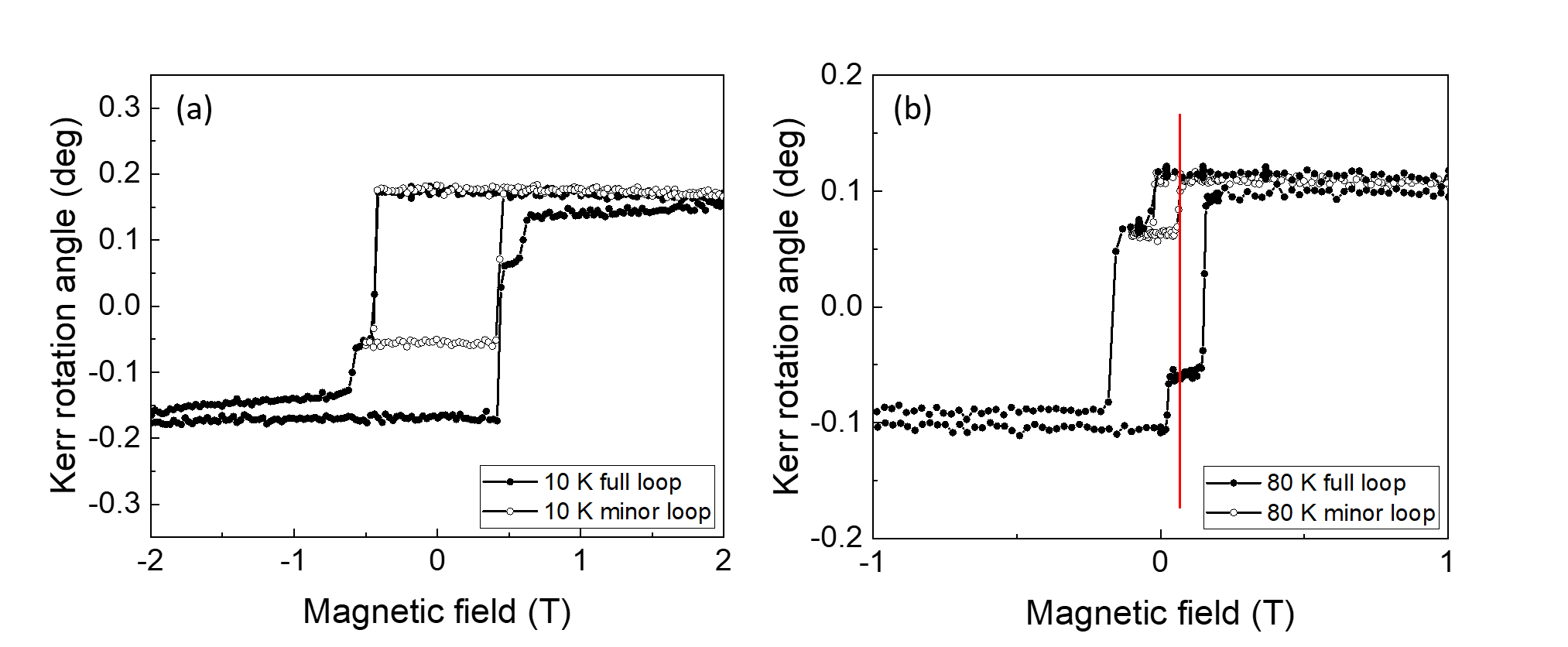}                                      
\caption{\label{fig_RIR2_MOKE} Magneto-optical Kerr rotation measurements of heterostructure RIR2 at 10 K (a) and 80 K (b), measured with incoherent light of 540 nm. Drawn with full symbols are the major hysteresis loops; the open symbols are the minor loops.  } 
\end{center}
\end{figure*}
\noindent
In order to confirm the sign and magnitude of the observed minor loop shift of heterostructure RIR2 with 2 MLs SrIrO$_3$ spacer and capping layer, polar magneto-optical Kerr effect measurements were performed with our home-built polar-MOKE set up based on the well-established double modulation technique with light from a Xe-lamp. At 10 K, the minor loop is not shifted measurably compared to the full loop, as depicted in \textbf{Figure S\ref{fig_RIR2_MOKE}}, which confirms our results from the SQUID magnetometry and FORC study that the SrRuO$_3$ layers are decoupled at low temperatures. The minor loop at 80 K, measured between 2.5 T and -0.1 T is shifted by + 38 mT with respect to the full loop. This shift to higher magnetic fields is similar to the minor loop shift that was observed in our SQUID magnetometry and is also in agreement with the (antiferromagnetic) interaction peak seen in our FORC study of this sample.

\section{Heterostructure RIR2: Temperature dependence of the switching fields and the magnetic interlayer coupling strength}
\begin{figure*}[h]   
\begin{center}                                                         
\includegraphics[width=\textwidth]{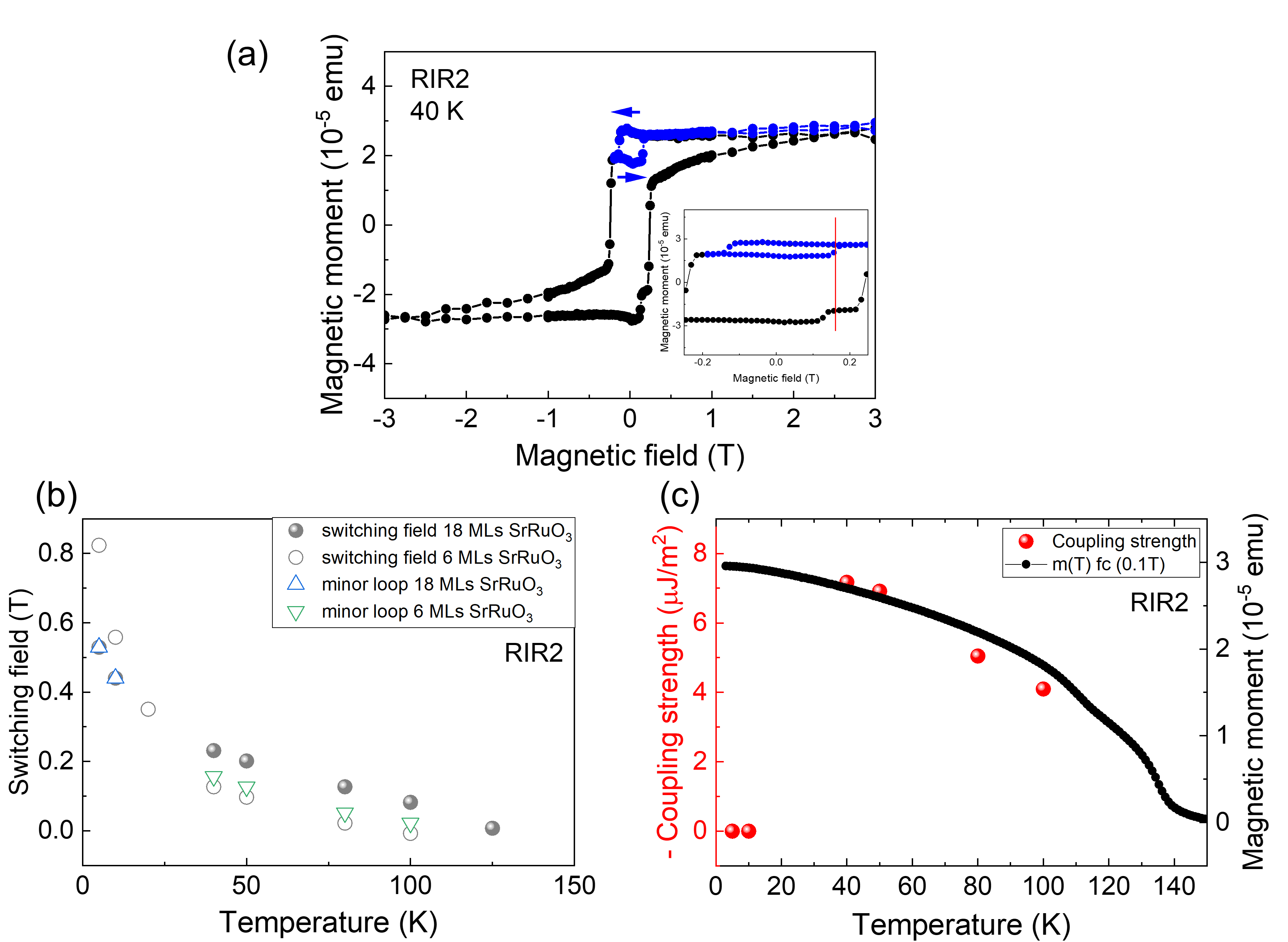}                                      
\caption{\label{fig_RIR2_T} (a) Full and minor magnetic hysteresis loops of heterostructure RIR2 at 40 K. Shown in the inset is the magnetic field range in which the shift of the minor loop to elevated magnetic fields is highlighted. (b) Temperature dependence of the switching fields of the two SrRuO$_3$ layers of heterostructure RIR2 determined from the major hysteresis (circles) and from the minor loops (open triangles). (c) Temperature dependence of the calculated coupling strength (red) and the magnetic moment during warming in 0.1T after field cooling (in 0.1 T). } 
\end{center}
\end{figure*}
\noindent
The full and minor magnetic hysteresis loops of heterostructure RIR2 at 40 K are presented in \textbf{Figure S\ref{fig_RIR2_T}(a)}. The minor loop (blue)  is shifted by +30 mT with respect to the full loop (black). 40 K is the lowest temperature of the current study at which a positive minor loop shift was observed. Shown in \textbf{Figure S\ref{fig_RIR2_T}(b)} are the switching fields of the two SrRuO$_3$ layers of heterostructure RIR2 as function of temperature. The two SrRuO$_3$ layers of the heterostructure have the same switching field at 20 K. Below 20 K, the bottom 18 MLs SrRuO$_3$ switches at smaller magnetic fields than the 6 MLs top SrRuO$_3$, while they behave vice versa above 20 K. Drawn with open triangles are the switching fields of the minor loops at the respective temperatures. The minor loop shift is zero below 20 K, when the bottom, 18 MLs thick SrRuO$_3$ is magnetically softer than the 6 MLs thin SrRuO$_3$ layer. When the thinner layer switches at smaller magnetic fields than the thicker SrRuO$_3$ layer, the minor loop is shifted by 30 mT to elevated fields at all investigated temperatures below the Curie temperature of the thin layer at 110 K (compare the m(T) measurement in (c)).  
\newpage \noindent
From the difference of the switching fields determined from the major loops and the minor loops of the magnetically softer SrRuO$_3$ layer, the coupling strength $J_C$ can be calculated with the following equation introduced by Heijden \textit{et al.} [2]. 
\begin{equation}
J_C = \mu_0 \Delta H_{shift} M_{soft,rev} t _{soft,rev}
\end{equation}
with the minor loop shift $\Delta H_{shift} = H_{major} - H_{minor}$. $t _{soft,rev}$ and $M_{soft,rev}$ are the thickness and the magnetization of the magnetically softer layer, respectively.
Due to the direct proportionality to the magnetization of the magnetically softer layer  $M_{soft,rev}$ and the observed almost temperature independent minor loop shift (above 40 K), the total magnitude of the coupling strength is maximum at 40 K and decreases with increasing temperature following the trend of M(T) of the 6 MLs thin SrRuO$_3$ layer. The temperature dependence of the magnetic moment of the heterostructure RIR2 is shown in \textbf{Figure S\ref{fig_RIR2_T}(c)} for comparison.\\
The change of the magnetic interlayer coupling from decoupling below 20 K to very weak antiferromagnetic coupling above 40 K seems to be correlated to the change of the magnetically softer and harder layer of the heterostructure (cp. \textbf{Figure S\ref{fig_RIR2_T} (b)}). When the 18 MLs thick bottom SrRuO$_3$ layer is the magnetically harder layer of the heterostructure, no shift of the minor loop is observed. Only in the temperature range where the 6 MLs thick top SrRuO$_3$ layer has smaller coercivities, the small shift of the minor loop, indicating weak antiferromagnetic coupling, is seen. 
A similar temperature dependence, namely decoupling at low temperatures and strong coupling at higher temperatures, has been observed in Ni/Co pseudo-spin-valve structures in which the Ni/Co multilayers of different repetition numbers, separated by 4.6 nm Cu, were strongly coupled via magnetic dipolar coupling [3]. Mohnseni \textit{et al.} attributed this dependence and the decoupling at low temperatures to the increase of the difference of the coercivities of the magnetically softer and harder layers at low temperatures so that the stray fields of the magnetically harder layer were insufficient to initiate the reversal of the magnetically softer layer [3]. \\
Maybe also in our case the coupling at low temperatures, where the difference between the coercive fields is more than two times larger than above 40 K, is too weak to initiate the switching in the magnetically harder (18 MLs thick) layer. On the other hand, assuming that the coupling strength was also on the order of a few  $\mu$J/m$^2$ at low temperature, the expected shift of the minor loop would be smaller by a factor of 5-6.5, since the thicker SrRuO$_3$ is the magnetically softer one at low temperature. However, a minor loop shift of 6 mT is close to the detection limit of our experimental set ups and therefore experimentally challenging to observe. \\
The absolute magnitude of the interlayer coupling is maximum at 40 K and decreases upon temperature increase. This indicates that the interlayer exchange mechanism within the quantum interference model, which predicts an increase with temperature in case of insulating spacers [4], cannot explain the observed coupling. The prediction of the temperature-induced increase of the magnitude of the coupling strength is based on approximations that are only valid for temperatures below $T= \hbar^2 k_F / 2 k_b m D$ [4]. In case of heterostructure RIR2, with 0.8 nm thick SrIrO$_3$ spacer, this corresponds to temperatures on the order of 365 K (for Fermi velocities of about 7.6$\cdot 10^{4}$ m/s [5]), which confirms that the approximations should be valid for the temperature range that was studied here. However, due to the dependence of the electronic transport properties of SrIrO$_3$ on the details of the interfacial environment, and the thickness, the Fermi wavevector for the SrIrO$_3$ layers of heterostructure RIR2 might differ from the reported values of bare SrIrO$_3$ films.\\
Orange-peel coupling due to correlated surface roughness, which can cause antiferromagnetic coupling in case of magnetic systems with strong perpendicular magnetic anisotropy [6], is expected to be too small to explain the interlayer coupling in heterostructure RIR2.
One mechanism of magnetostatic origin that is qualitatively consistent with the observed dependence of the (antiferromagnetic) interlayer coupling on temperature, is the model of domain replication in the hard layer via magnetostatic interactions, as proposed by Nistor [7]. When the magnetic field required to reverse the magnetization of the soft layer during the minor loops is close to the nucleation field of the hard layer, inversed domains in the soft layer will generate stray fields that can induce so called replicated domains in the hard layer. The stray fields in the hard layer then act as negative bias field during the second half of the minor loop [7]. 
However, based on our experimental data we cannot prove that this is the only explanation for the weak antiferromagnetic coupling observed for the 2 MLs SrIrO$_3$ spacer.
\section{First order reversal curve study of heterostructure RIZR1 showing weak ferromagnetic coupling of the SrRuO$_3$ layers}
\begin{figure}[h]
\includegraphics[width=\linewidth]{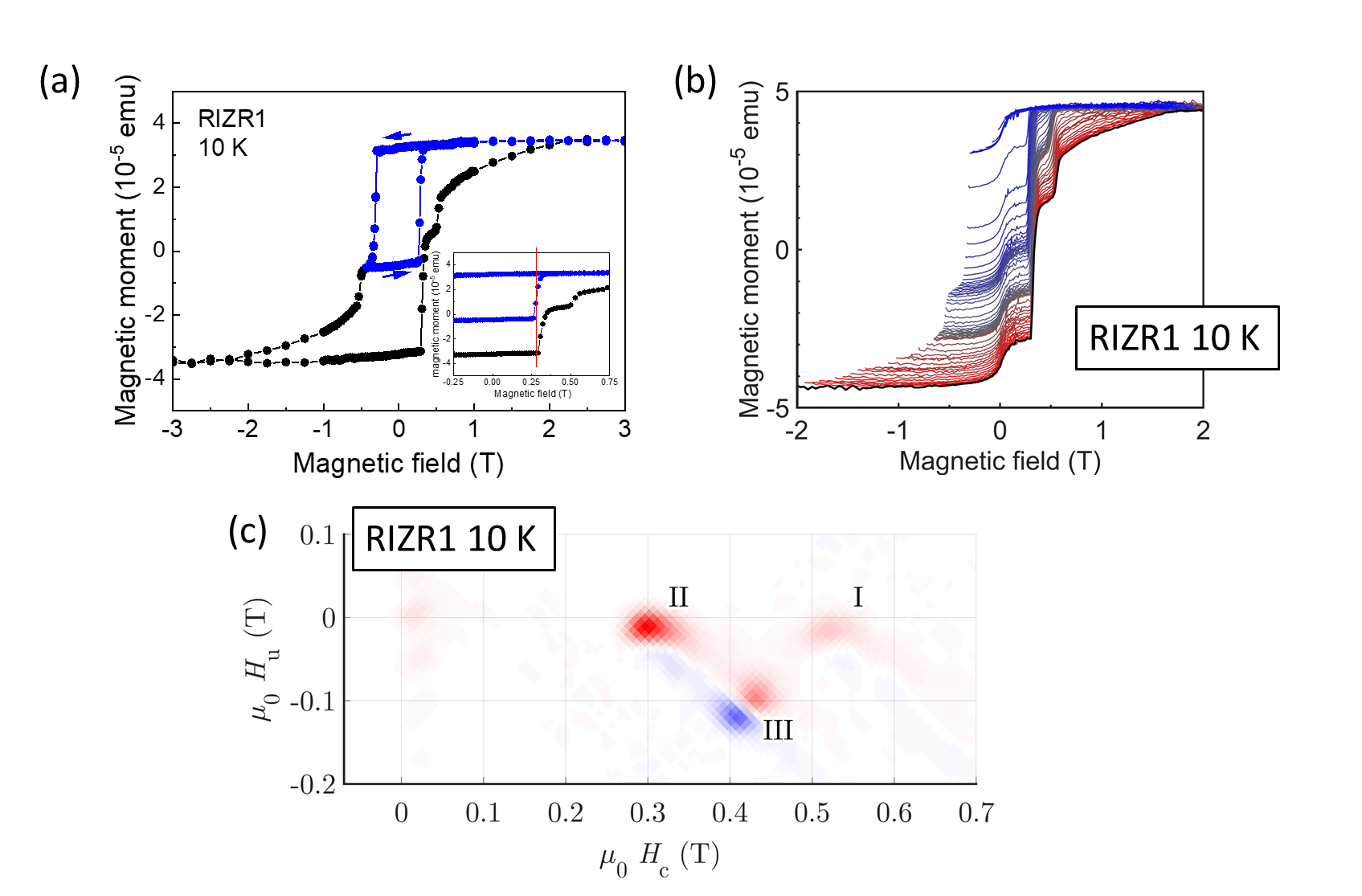}
\caption{\label{fig_FORC}(a) Major (black) and minor (blue) magnetic hysteresis loops of heterostructure RIZR1 at 10 K, measured with conventional SQUID magnetometry. This heterostructure has 1 ML SrIrO$_3$ and 1 ML SrZrO$_3$ as spacer and capping layers. (b) Minor loops of heterostructure RIZR1 measured at 10 K. The color of the respective minor loops changes from red to blue for increasing reversal fields. (c) FORC density plotted as a function of the coercive field H$_c$ and the interaction field H$_u$ for the heterostructure RIZR1 at 10 K. Positive FORC density peaks are shown in red, negative ones in blue. Feature (I) and (II) correspond to the magnetization switching of the 6 MLs (I) and 18 MLs SrRuO$_3$ (II) layer, respectively. (III) is the interaction peak indicating ferromagnetic coupling.}
\end{figure}
\noindent
The major and minor hysteresis loops of heterostructure RIZR1, determined by SQUID magnetometry, are shown exemplarily at 10 K in (a). Similar to heterostructure RIR2, also this heterostructure is composed of two SrRuO$_3$ layers of 18 MLs and 6 MLs thickness. They are separated and capped by 1 ML SrIrO$_3$ /1 ML SrZrO$_3$. As highlighted in the inset of (a), the minor loop is shifted by 30 mT to smaller magnetic fields indicating a weak ferromagnetic coupling on the order of 35 $\mu$J/m$^2$, as we have found in our previous study [8].\\
In order to confirm the relation of the interaction peak with the type of magnetic coupling, FORC studies were also performed at 10 K for the heterostructure RIZR1 with heterogeneous spacer of 1 ML SrIrO$_3$/ 1 ML SrZrO$_3$. \\
Depicted in (b) are the individual minor loops of the first order reversal curve study at 10 K. Only the diamagnetic contribution originating from the SrTiO$_3$ substrate has been corrected by linear fitting in the high magnetic field range where the sample is in its saturated state. The soft magnetic contribution is related to magnetic impurities introduced during the sample cutting and handling.\\
The FORC density at 10 K is plotted in \textbf{Figure S\ref{fig_FORC}} (c) as function of the interaction field H$_u$ and the coercive field H$_c$.\\
Similar to heterostructure RIR2 with 2 MLs SrIrO$_3$ spacer, a reversible ridge was observed at small magnetic field values which is related to purely reversible magnetization switching. Due to the relative intensity of the peaks, the more intense peak (II) can be related to the switching field of the 18 MLs thick SrRuO$_3$ layer.\\ 
Feature (I) corresponds to the irreversable switching of the 6 MLs SrRuO$_3$ layer of the heterostructure. In accordance with the global SQUID magnetometry shown in (a), the 6 MLs SrRuO$_3$ is the magnetically harder layer at 10 K. \\
An additional positive-negative peak pair is present in the FORC density with the orientation opposite to the one observed for heterostructure RIR2. While the orientation of feature (III) in \textbf{Figure 2} of the paper indicated antiferromagnetic coupling for heterostructure RIR2, the reversed orientation of feature (III) in \textbf{Figure S\ref{fig_FORC}} supports the observation of ferromagnetic coupling at 10 K in heterostructure RIZR1.\\
\newpage
\section{Magnetotransport investigations of heterostructure RIR12 with 12 MLs SrIrO$_3$ spacer}
\begin{figure*}[h]   
\begin{center}                                                         
\includegraphics[width=\textwidth]{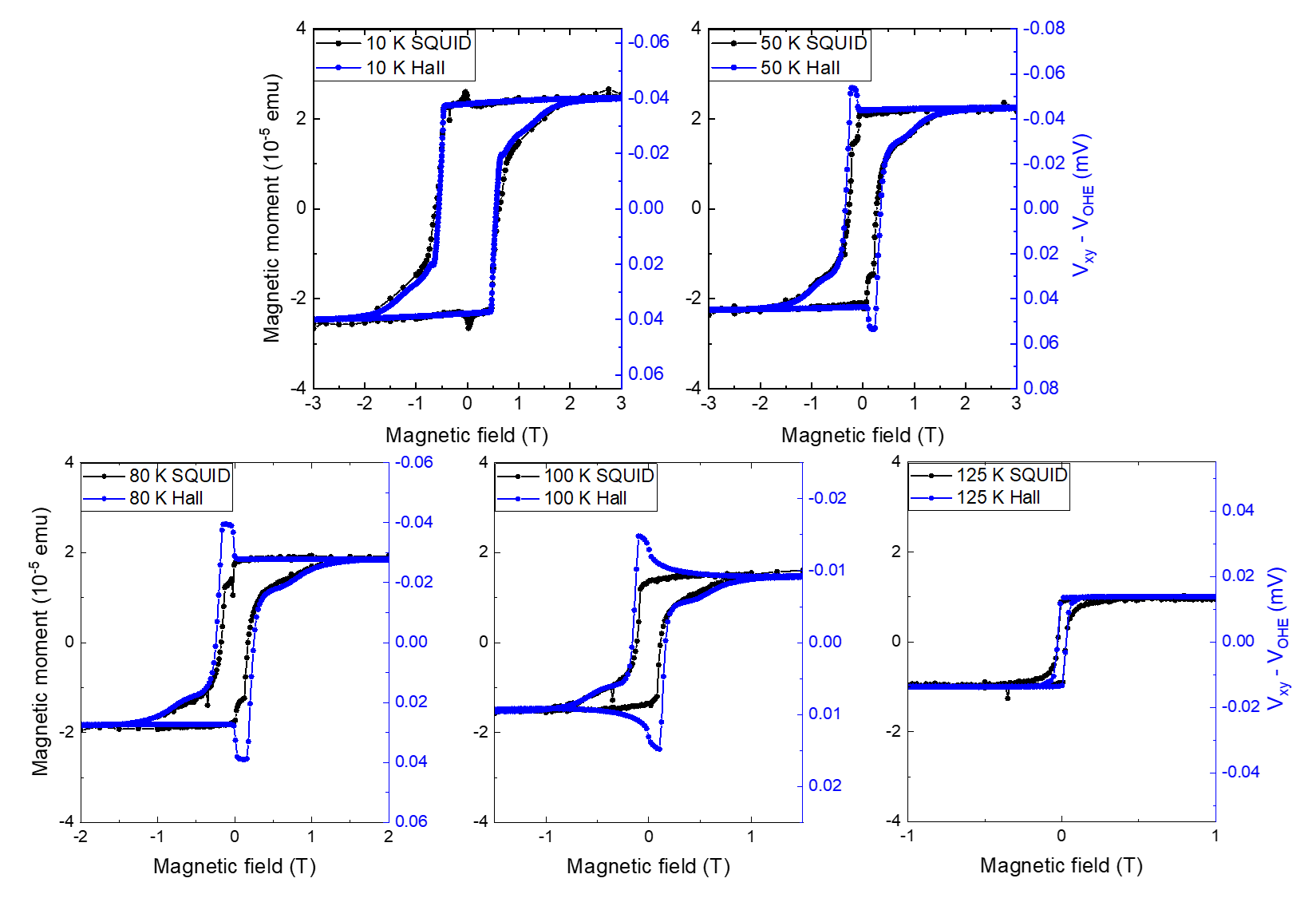}                                      
\caption{\label{fig_RIR12}Comparison of the magnetic hysteresis loops determined by SQUID magnetometry (black) and the Hall voltage (blue), corrected for the ordinary Hall contribution, as function of applied magnetic field for heterostructure RIR12 with 12 MLs SrIrO$_3$ spacer. To increase the comparability of the magnetic field dependencies of Hall effect and magnetization reversal, the anomalous Hall voltage was plotted from positive to negative values for 10 K, 50 K, 80 K, and 100 K. } 
\end{center}
\end{figure*}
\noindent
As described in detail in the previous section 2, all SQUID magnetometry measurements of this study have been corrected by subtraction of the magnetic hysteresis loop at 200 K, above the Curie temperature of the SrRuO$_3$ layers of the heterostructures. However, this correction leads to artifacts, such like the small peaklike features close to zero T at low temperatures. In order to confirm that this correction does not lead to a misinterpretation of the main physical properties of the heterostructures, Hall measurements were performed for sample RIR12, the heterostructure with the thickest SrIrO$_3$ spacer of this study.\\
In a single domain ferromagnet, the anomalous Hall constant is directly proportional to the out-of-plane component of the magnetization [9]. As shown also by van Thiel \textit{et al.}, the measured Hall voltage of a magnetic sample that contains several anomalous Hall conduction channels is given by the sum of the different individual contributions [10]. For heterostructure RIR12, the overall anomalous Hall voltage is given by the sum of the contributions of the two SrRuO$_3$ layers with distinct thicknesses and therefore different temperature dependencies of the anomalous Hall constant.\\
Depicted in \textbf{Figure S\ref{fig_RIR12}} is the comparison of the magnetic hysteresis loops (black) and the Hall voltage (after subtraction of the ordinary Hall effect) of heterostructure RIR12 at several temperatures below the ferromagnetic transition temperature of the 18 MLs SrRuO$_3$ layer of the heterostructure. The observation of hump-like features between 50 K and 100 K confirms our expectation of the different temperature dependencies of the anomalous Hall constant. The anomalous Hall constant of the thin SrRuO$_3$ layer is most likely positive in this temperature range, while the AHE constant of the thicker SrRuO$_3$ layer is still negative up to 100 K. Most relevant for the present study is the magnetic field range in which the two different SrRuO$_3$ layers reverse their magnetization. The comparison of the SQUID and Hall measurements confirms that the hump-like features appear in the same magnetic field range in which the magnetically softer layer reverses its magnetization and disappears when the magnetically harder layer switches its magnetization. This confirms the switching fields determined by SQUID magnetometry. At 100 K, the hump-like feature has an s-shape which is most likely related to the 6 MLs thin layer which is already in its paramagnetic phase with a still measurable contribution to the Hall voltage.  At 125 K, the Hall voltage loop is mainly determined by the anomalous Hall voltage of the 18 MLs SrRuO$_3$, which is still in its ferromagnetic phase, again confirming the SQUID magnetometry. 
\begin{figure*}[h]   
\begin{center}                                                         
\includegraphics[width=0.9\textwidth]{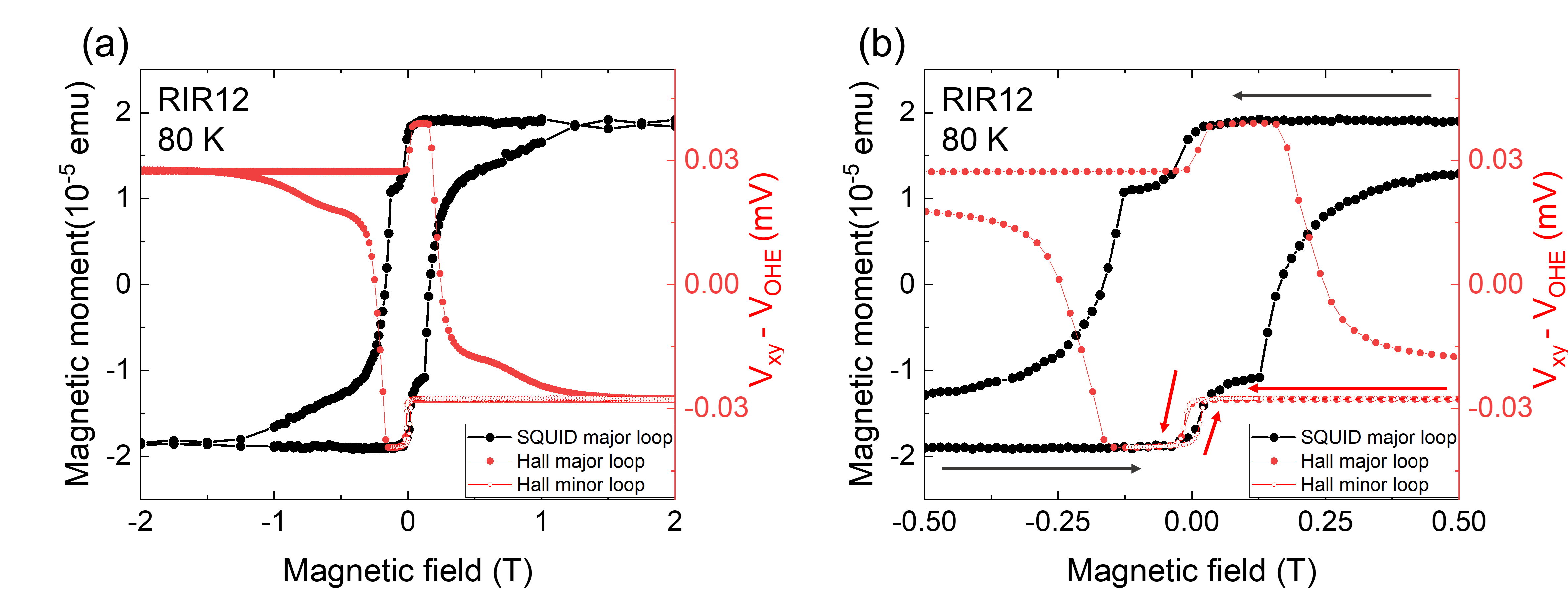}                                      
\caption{\label{fig_RIR12-minor}Comparison of the magnetic hysteresis loops determined by SQUID magnetometry (black) and the anomalous Hall voltage (red) as function of applied magnetic field for heterostructure RIR12 with 12 MLs SrIrO$_3$ spacer at 80 K. To increase the visibility in the magnetic field range of the minor loop switching, a zoom- in of (a) is shown in (b). } 
\end{center}
\end{figure*}
\noindent
Depicted in \textbf{Figure S\ref{fig_RIR12-minor}} are the major and minor loop hystereses of the Hall voltage, after subtraction of the ordinary Hall contribution, and compared to the major magnetic hysteresis loop (in black). As highlighted in (b), the minor loop of the anomalous Hall voltage, which is proportional to the magnetization of the magnetically softer layer, is in good agreement with the magnetic field dependence of the major magnetic hysteresis loop for the reversal of the magnetically softer SrRuO$_3$ layer. This confirms that the minor loop is not shifted with respect to the full loop.

\newpage
\section{Minor loop investigations of a SrRuO$_3$-based heterostructure with 2 MLs SrIrO$_3$ spacer having nanometer-deep holes}
\begin{figure*}[h]   
\begin{center}                                                         
\includegraphics[width=0.9\textwidth]{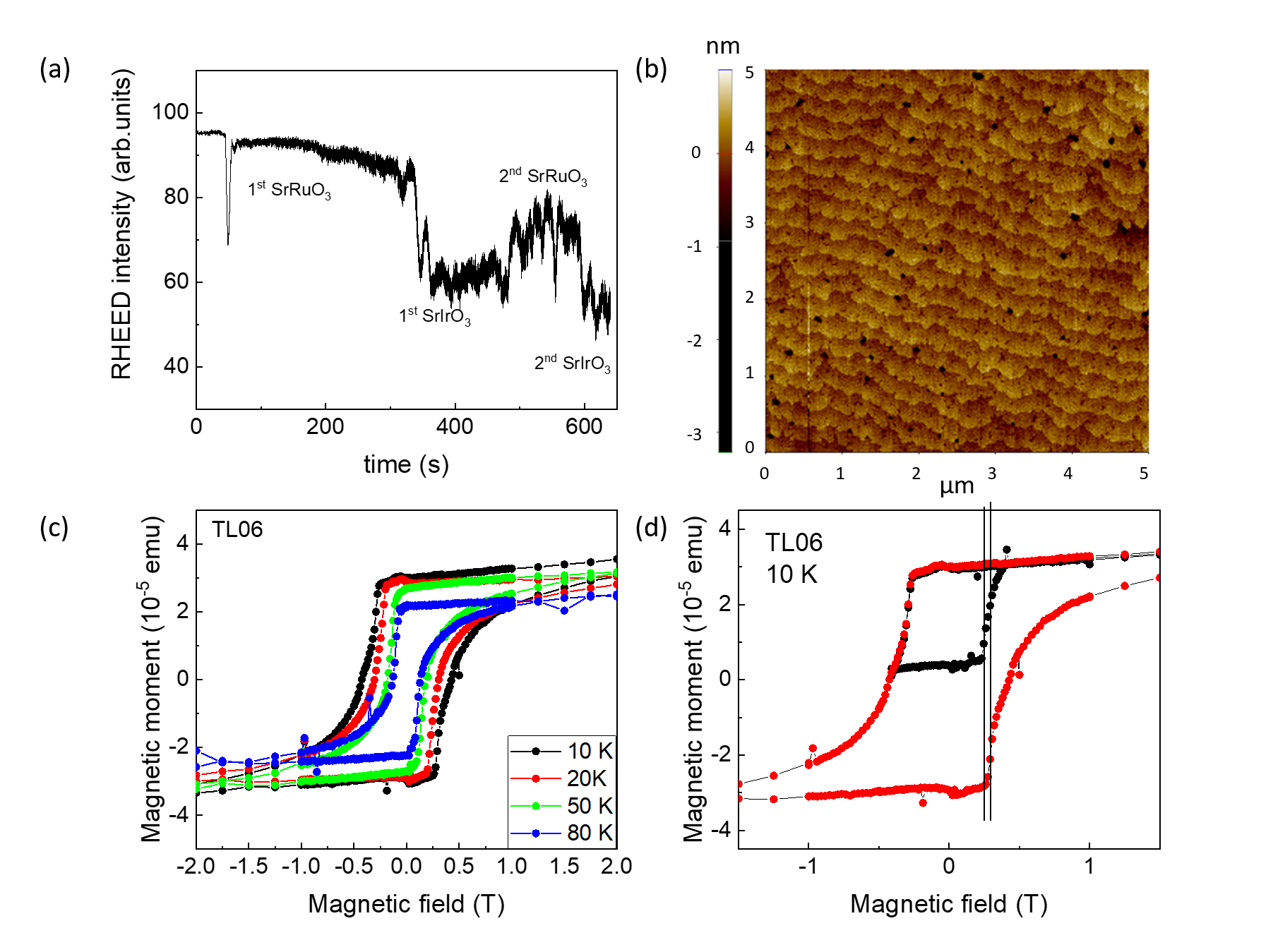}                                      
\caption{\label{fig_TL06} (a) Integrated RHEED intensity of the specular spot during the deposition of a second sample (TL06) with 2 MLs SrIrO$_3$ spacer and capping layer. (b) Atomic force microscopy image (5 $\mu $m x 5 $\mu $m) of the surface topography of TL06. Major (c) and minor (d) magnetic hysteresis loop measurements of the heterostructure determined by SQUID magnetometry. } 
\end{center}
\end{figure*}
\noindent
The influence of holes in the heterostructures on the magnetic interlayer coupling was investigated by the comparison of heterostructure RIR2 with a second sample with 2 MLs SrIrO$_3$ spacer where holes of minimum 1-2 nanometer depth were observed by atomic force microscopy. As shown in \textbf{Figure S\ref{fig_TL06}} by the time-dependent RHEED intensity plot, this sample also has a SrIrO$_3$ spacer and capping layer of 2 MLs thickness and two SrRuO$_3$ layers of distinct thicknesses. In contrast to the other heterostructures of the current study, this sample was deposited by our new pulsed laser deposition set up at the University of Cologne, manufactured by SURFACE Inc. 
As depicted in (b), the heterostructure surface of this sample shows the existence of nanometer deep holes. Because there were no holes observed in the AFM investigations of the heterostructures RIR2, RIR6, or RIR12, it was expected that this heterostructure was influenced more strongly by the existence of ferromagnetic bridges by pinholes. Shown in (c) are major hysteresis loops of this sample at various temperatures. Only at 10 K, the switching fields of the two individual SrRuO$_3$ layers are distinguishable. The performed minor loop at 10 K, drawn in (d), shows a small negative shift of 45 mT, indicating weak ferromagnetic coupling. Thus, both samples with 2 MLs SrIrO$_3$ spacer indicate opposite sign of the coupling of the two SrRuO$_3$ layers. The difference could originate from the different densities of nanometer deep (pin-)holes in the heterostructures, which was increased for sample TL06. Such holes are expected to lead to the formation of ferromagnetic bridges by pinholes connecting the two ferromagnetic SrRuO$_3$ layers. 

\newpage \noindent
\section{Resistance measurements of heterostructure RIR12, and 6 MLs and 12 MLs bare SrIrO$_3$ thin films deposited on SrTiO$_3$(100)}
\begin{figure*}[h]   
\begin{center}                                                         
\includegraphics[width=\textwidth]{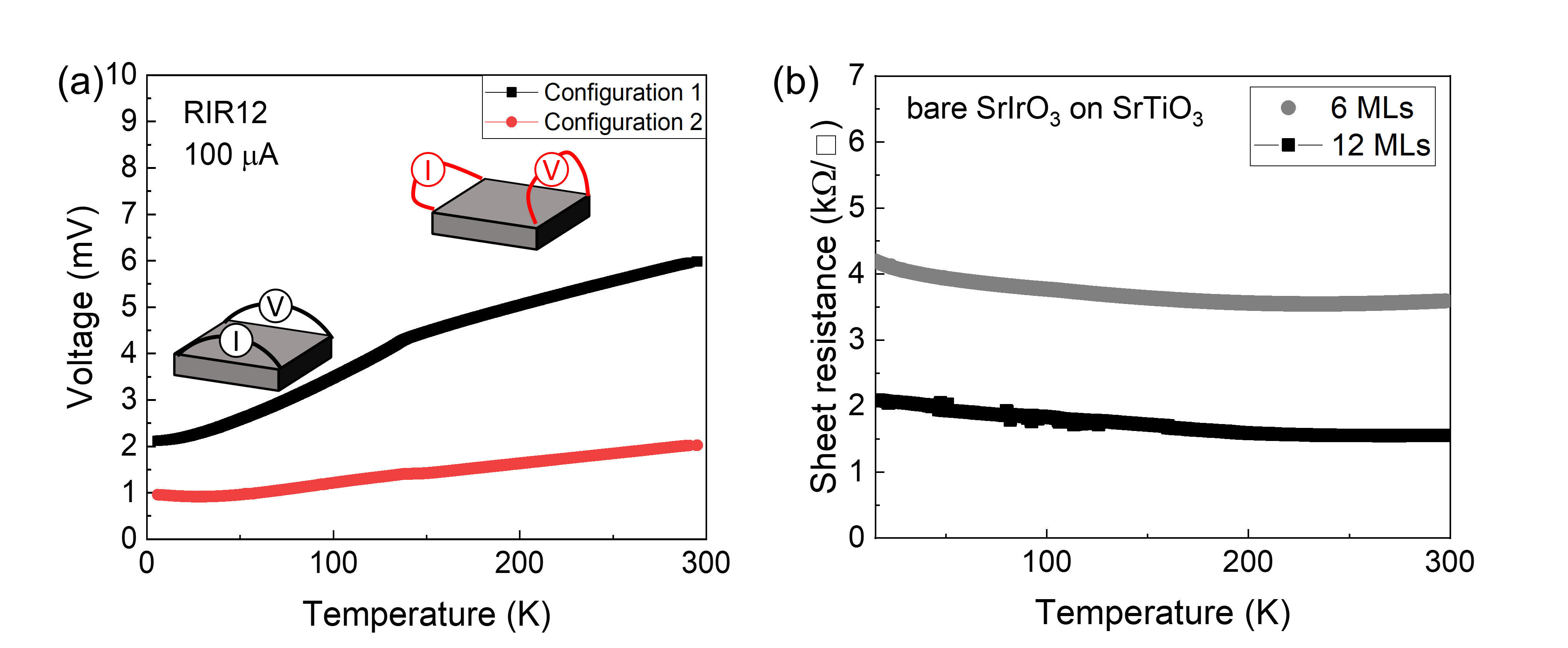}                                      
\caption{\label{fig_bareSIO}(a) Temperature dependence of the voltage measured along the edges of heterostructure RIR12  when 100 $\mu$A were applied. The two configurations that correspond to the current parallel to the two in-plane principal axes of the orthorhombic SrRuO$_3$ layers are sketched in the inset. Temperature dependence of the sheet resistance of  6 (grey) and 12 MLs (black) bare SrIrO$_3$ thin films deposited on SrTiO$_3$ (100).} 
\end{center}
\end{figure*}
\noindent
Because it is sandwiched between two SrRuO$_3$ layers with much lower resistivity (about 0.5 $\mu\Omega$m at 10 K), the resistivity of the SrIrO$_3$ spacer layers cannot be assessed easily [11,12,13]. 
The voltage drop measured along the whole heterostructure will therefore be dominated by the SrRuO$_3$ layers. This is confirmed by a resistance measurement of heterostructure RIR12 in comparison to reference SrIrO$_3$ films (see \textbf{Figure S\ref{fig_bareSIO}}). The voltage drop of this anisotropic heterostructure was measured in van der Pauw geometry as indicated in the inset in \textbf{Figure S\ref{fig_bareSIO}(a)}. The sheet resistances of the 6 and 12 MLs bare SrIrO$_3$ thin films (shown in (b)), deposited on SrTiO$_3$, correspond to a measured voltage drop that is almost two order of magnitudes larger than the measured voltage drop in case of the heterostructure. Thus, our following estimations of the SrIrO$_3$ spacer transport properties are based on the investigations of 6 and 12 MLs bare SrIrO$_3$ reference thin films deposited on SrTiO$_3$ (100) substrates.\\
The sheet resistances of the 6 and 12 MLs SrIrO$_3$ thin films (see \textbf{Figure S\ref{fig_bareSIO}(b)}) show only a very weak temperature dependence of about 20\% variation between 5 K and 300 K, with a small upturn for decreasing temperature.
The classification of the observed transport properties of the SrIrO$_3$ reference films is not straightforward due to the variety of different experimental results of SrIrO$_3$ thin films deposited on SrTiO$_3$ [11,14,15] or when SrIrO$_3$ was interfaced with dissimilar perovskite oxides [16]. A (semi-)metal-to-insulator transition, with a clear change to a positive slope of $\rho$(T) in the thicker films, was observed between 3 and 4 MLs thickness by Manca \textit{et al.} [11] or between 5 and 6 MLs by Groenendijk \textit{et al.} [15]. In contrast, weakly temperature-dependent resistivity was seen in even 20 nm thick SrIrO$_3$ films by Gruenewald \textit{et al.}[14], or when SrIrO$_3$ was interfaced with LaMnO$_3$ [16]. The latter behavior is consistent with the transport behavior of our SrIrO$_3$ thin films. Theoretical band structure calculations as well as experimental ARPES studies [17,5]  showed that the Fermi surface of SrIrO$_3$ in the semimetallic state consists of electron and holelike pockets. Thus, octahedral rotation or slight modifications of the stoichiometry were proposed to change the transport properties [17,5]. Ir deficiency, due to the volatility of IrO$_x$, can lead to an effective hole doping [15], whereas oxygen vacancies can act as electron dopants [18].
 \\\\
\newline
[1] Choi, J., Eom, C. B., Rijnders, G., Rogalla, H., Blank, D. H. A. Growth Mode Transition from Layer by Layer to Step Flow during the Growth of Heteroepitaxial SrRuO$_3$ on (001) SrTiO$_3$. Applied Physics Letters \textbf{79}, 1447 (2001)\newline
[2] P. A. A. van der Heijden, P. J. H. Bloemen, J. M. Metselaar, R. M. Wolf, J. M. Gaines, J. T. W. M. van Eemeren, P. J. van der Zaag, and W. J. M. de Jonge, Physical Review B \textbf{55}, 11569 (1997)  \newline
[3] S. M. Mohseni, R. K. Dumas, Y. Fang, J. W. Lau, S.R. Sani, J. Persson, and J. Akerman, Physical Review B \textbf{84}, 174432 (2011) \newline
[4] P. Bruno, Physical Review B \textbf{52}, 411 (1995) \newline
[5] Y.F. Nie, P.D.C. King, C.H. Kim, M. Uchida, H.I. Wei, B.D. Faeth, J.P. Ruf, J.P.C. Ruff, L. Xie, X.Pan, C.J. Fennie, D.G. Schlom, and K.M. Shen, Phys. Rev. Lett. \textbf{114}, 016401 (2015) \newline
[6] J. Moritz, F. Garcia, J. C. Toussaint, B. Dieny, and J. P.Nozi\'{e}res, Europhysics Letters \textbf{65}, 123 (2004)\newline
[7] L. E. Nistor, Magnetic tunnel junctions with perpendicular magnetization: anisotropy, magnetoresistance, magnetic coupling and spin transfer torque switching (2011), PhD Thesis \newline
[8] L. Wysocki, R. Mirzaaghayev, M. Ziese, L. Yang, J. Sch{\"o}pf, R. B. Versteeg, A. Bliesener, J. Engelmayer,A. Kov\'{a}cs, L. Jin, F. Gunkel, R. Dittmann, P. H. M. van Loosdrecht, and I. Lindfors-Vrejoiu, Applied Physics Letters \textbf{113}, 192402 (2018) \newline
[9] E. M. Pugh and N. Rostoker, Review of Modern Physics \textbf{25}, 151 (1953) \newline
[10] T.C. van Thiel, D. J. Groenendijk, and A. D. Caviglia, J. Phys.: Mater. \textbf{3}, 025005 (2020) \newline
[11] N. Manca, D. J. Groenendijk, I. Pallecchi, C. Autieri, L. M. K. Tang, F. Telesio, G. Mattoni, A. Mccollam, S. Picozzi, and A. D. Caviglia,  Physical Review B \textbf{97}, 081105(R) (2018) \newline
[12] D. J. Groenendijk, N. Manca, J. D. Bruijckere, A. M. R. V. L. Monteiro, R. Gaudenzi, H. S. J. V. D. Zant, and A. D. Caviglia, Eur. Phys. J. Plus \textbf{135}, 627 (2020)\newline
[13] M. Ziese, I. Vrejoiu, and D. Hesse, Physical Review B \textbf{81}, 184418 (2010)  \newline
[14] J. H. Gruenewald, J. Nichols, J. Terzic, G. Cao, J. W. Brill, and S. S. Seo,Journal of Materials Research \textbf{29}, 2491–2496 (2014) \newline
[15] D. J. Groenendijk, C. Autieri, J. Girovsky, M. C. Martinez-Velarte, N. Manca, G. Mattoni, A. M. R. V. L. Monteiro, N. Gauquelin, J.  Verbeeck, A. F. Otte, M. Gabay, S. Picozzi, and A. D. Caviglia, Phys.Rev. Lett. \textbf{119}, 256403 (2017) \newline
[16] E. Skoropata, J. Nichols, J. M. Ok, R. V. Chopdekar, E. S. Choi, A. Rastogi, C. Sohn, X. Gao, S. Yoon, T. Farmer, R. D. Desautels, Y. Choi, D. Haskel, J. W. Freeland, S. Okamoto, M. Brahlek, and H. N. Lee, Science Advances \textbf{6}, eaaz3902 (2020) \newline
[17] W. Guo, D. X. Ji, Z. B. Gu, J. Zhou, Y. F. Nie, and X. Q. Pan, Physical Review B \textbf{101}, 085101 (2020)\newline
[18] Supplemental material of N. Manca, D. J. Groenendijk,I. Pallecchi, C. Autieri, L. M. K. Tang, F. Telesio, G. Mattoni, A. McCollam, S. Picozzi, and A. D. Caviglia, Physical Review B \textbf{97}, 081105(R) (2018)\newline